\documentclass{amsart}

\usepackage{amssymb,amsmath,amscd}
\usepackage{amsfonts}
\usepackage{young,colortbl}
\usepackage{graphicx}

\usepackage{epstopdf}
\usepackage{graphics}
\newtheorem{Theorem}{Theorem}[section]
\newtheorem{Proposition}{Proposition}[section]
\newtheorem{Corollary}{Corollary}[section]

\def\proof{\par{\it Proof}. \ignorespaces}
\def\endproof{{\ \vbox{\hrule\hbox{%
     \vrule height1.3ex\hskip0.8ex\vrule}\hrule }}\par}
\newenvironment{Proof}{\proof}{\endproof}

\theoremstyle{definition}
\newtheorem{Definition}[Theorem]{Definition}
\newtheorem{Example}[Theorem]{Example}

\theoremstyle{remark}
\newtheorem{Remark}[Theorem]{Remark}

\numberwithin{equation}{section}

\numberwithin{figure}{section}

\let\trueint=\int
\let\truesum=\sum
\def\int{\mathop{\textstyle\trueint}\limits}
\def\sum{\mathop{\textstyle\truesum}\limits}

\let\<=\langle
\let\>=\rangle

\def\sech{\mathop{\rm sech}\nolimits}

\def\sf#1#2{{\textstyle\frac{#1}{#2}}}
\def\sech{\mathop{\rm sech}\nolimits} 
\renewcommand\labelitemi{\ifmmode\circ\else$\circ$\fi}

\begin{document}

%\normalbaselineskip=2.0\normalbaselineskip
%\normalbaselines

\title[KP solitons in shallow water]
{KP solitons in shallow water}

%    Information for first author
\author{Yuji Kodama}
%    Address of record for the research reported here
\address{Department of Mathematics, Ohio State University, Columbus, OH 43210}
\email{kodama@math.ohio-state.edu}
%    \thanks will become a 1st page footnote.
%    Information for second author

%\thanks{Partially supported by NSF grant DMS0806219}

\begin{abstract}
The main purpose of the paper is to provide a survey of our recent studies on soliton solutions of the  Kadomtsev-Petviashvili (KP) equation.  The KP equation describes weakly dispersive and small amplitude wave propagation in a quasi-two dimensional framework. Recently a large variety of
exact soliton solutions of the KP equation has been found and classified.
These solutions are localized along  certain lines in a two-dimensional plane and decay
exponentially everywhere else, and are called line-solitons. The classification is based on the far-field patterns of the solutions which consist of a finite number of line-solitons. Each soliton solution
is then defined by a point of the totally non-negative Grassmann variety which can be parametrized 
by a unique derangement of the symmetric group of permutations. Our study also includes certain numerical 
stability problems of those soliton solutions. Numerical simulations of the initial value problems 
indicate that certain class of initial waves asymptotically approach to these exact solutions of
the KP equation.  We then discuss an application of our theory to the Mach reflection problem in shallow water.  This problem describes the resonant interaction of
solitary waves appearing in the reflection of an obliquely incident wave onto a vertical wall, and it predicts an extra-ordinary four-fold amplification of the wave at the wall.
There are several numerical studies confirming the prediction, but all indicate disagreements with the KP theory.  Contrary to those previous numerical studies,  we
find that the KP theory actually provides an excellent model to describe the Mach reflection phenomena
when the higher order corrections are included to the quasi-two dimensional approximation.
We also present laboratory experiments of the Mach reflection recently carried out by Yeh and his colleagues,
and show how precisely the KP theory predicts this wave behavior.

\end{abstract}

\maketitle

\thispagestyle{empty} 
\pagenumbering{roman}\setcounter{page}{1}
\tableofcontents
%\clearpage

\pagenumbering{arabic}
\setcounter{page}{1}
\setcounter{figure}{1}

%%%%%%%%%%%%%%%%%%%%%%%%%%%%%%
\section{Introduction}

It is a quite well-known story that in August 1834 Sir John Scott Russel observed a large solitary wave in a shallow water channel in Scotland.  He noted in his first paper (1838) on the subject  that 
\begin{quote}
 {\it I was observing the motion of a boat which was rapidly drawn along a narrow channel by a pair of horses, when the boat suddenly stopped - not so the mass of water in the channel which it had put in motion; it accumulated round the prow of the vessel in a state of violent agitation, then suddenly leaving it behind, rolled forward with great velocity, assuming the form of a large solitary elevation, a rounded, smooth and well defined heap of water, which continued its course along the channel apparently without change of form or diminution of speed ....}. 
\end{quote}
This solitary wave is now known as an example of a {\it soliton}, and is described by a solution of the Korteweg-de Vries (KdV) equation.   The KdV equation describes one-dimensional wave propagation such as beach waves parallel to the coast line or waves in narrow canal, and is obtained in the leading
order approximation of an asymptotic perturbation theory under the assumptions of weak
nonlinearity (small amplitude) and weak dispersion (long waves). 
The KdV equation has rich mathematical structure including the existence of $N$-soliton
solutions and the Lax pair for the inverse scattering method, and it is a prototype equation of the $1+1$ dimensional integrable systems.  In particular, the initial value problem of the KdV equation has been extensively studied by means of the method of inverse scattering transform (IST).
It is well known that a general initial data decaying rapidly in the spatial variable evolves
to a number of individual solitons and weakly dispersive wave trains separate from the
solitons (see for examples, \cite{AS:81, N:85, NMPZ:84, Wh:74}).

In 1970, Kadomtsev and Petviashvilli \cite{KP:70} proposed a $2+1$ dimensional dispersive wave equation to study the stability 
of the one-soliton solution of the KdV equation under the influence of weak transverse perturbations.
This equation is now referred to as the KP equation.
It turns out that the KP equation has much richer structure than the KdV equation, and might be considered as the most fundamental integrable system in the sense that many known integrable systems can be derived as special reductions of  the so-called KP hierarchy which consists of the KP equation 
together with its
infinitely many symmetries.   The KP equation can be also represented in the Lax form, that is,
there exists a pair of linear equations associated with an eigenvalue problem and an evolution
of the eigenfunction, which enables the method of IST.
However, unlike the case of the KdV equation, the IST for the KP equation does not seem to provide
a practical method of solving the initial value problem for initial waves consisting of  line-solitons in the far field.

It is quite important to recognize that the resonant interaction plays a fundamental role in
multi-dimensional wave phenomenon.  The original description of the soliton interaction for the KP equation
was based on a two-soliton solution found in Hirota bilinear form, which has the shape of ``X'', 
describing the intersection of two lines with oblique angle and a phase shift at the intersection point.
This X-shape solution is referred to as the ``O''-type soliton, where ``O'' stands for {\it original}.
In his study of 1977 on an oblique interaction of two line-solitons, Miles \cite{M:77} pointed out that the O-type solution becomes singular if the angle of the intersection is smaller than certain critical value depending on the amplitudes of the solitons.
Miles then found that at the critical angle, the two line-solitons of the O-type solution interact resonantly, and a third wave is created to make a ``Y-shaped'' wave form.  Indeed, it turns out
that such Y-shaped resonant wave forms are exact solutions of the KP equation (see also \cite{NR:77}).
Miles applied his theory to study the Mach reflection of an incident wave onto a vertical
wall, and predicted that the third wave, called the {\it Mach stem}, created by the resonant interaction
can reach  four-fold amplification of the incidence wave.
Several laboratory and numerical experiments attempted to validate his prediction
of four-fold amplification, but with no definitive success (see for examples
\cite{F:80, KTK:98, T:93} for numerical experiments, and \cite{P:57, Me:80, YCL:10} for laboratory experiments).

After the discovery of the resonant phenomena in the KP equation, several numerical and experimental studies were performed to investigate resonant interactions in other physical
two-dimensional equations such as the ion-acoustic and shallow water wave equations under 
the Boussinesq approximation (see for examples \cite{KY:80, KY:82, FID:80, NN:83, F:80, T:93, MK:96, OT:06, TO:07}).  However, apart from these activities, no significant progress has been made in the
study of the solution space or real applications of the KP equation.  It would appear that the 
general perception was that there were not many new and significant results left to be uncovered
in the soliton solutions of the KP theory.

Over the past several years, we have been working on the classification problem of the soliton solutions
of the KP equation
and their applications to shallow water waves.  Our studies have revealed a large variety of solutions that were totally overlooked in the past \cite{BK:03, K:04, CK:08, CK:08b, CK:09, CK:10}\footnote[1]{The lower dimensional solutions, called $(2,2)$-soliton solutions,  have been found by the
binary Darboux transformation in \cite{BPPP:01}.}, and we found that some of those exact solutions can be applied to study the Mach reflection problem \cite{CK:09, KOT:09, YLK:10, YCL:10}.
Our numerical study \cite{KK:10} indicates that the solution to the initial value problem of the KP equation with certain class of initial waves associated with the Mach reflection problem converges asymptotically to some of these new exact solutions, that is, a separation of dispersive radiations from the soliton solution
similar to the case of the KdV soliton.

The main purpose of this paper is to present a survey of  our  studies on the soliton solutions
of the KP equation. The paper also presents several results for recent laboratory experiments done
by Harry Yeh and his colleagues at Oregon State University.

The paper is organized as follows:

In Section \ref{sec:SWW1}, we present the derivation of the Boussinesq-type equation 
from the three-dimensional 
Euler equation for the irrotational and incompressible fluid under the assumptions of weak nonlinearity and weak dispersion. The purpose of this section is to give
a precise physical meaning to those assumptions and to explain the existence of a solitary wave
solution in the form of the KdV soliton. 

In Section \ref{sec:KP}, we explain the quasi-two dimensional approximation to derive the KP equation, and  discuss physical interpretation of the KP soliton in terms of the KdV soliton.
In order to describe the general soliton solutions, we here
 introduce the $\tau$-function which is expressed 
by a Wronskian determinant for a set of $N$ linearly independent functions $\{f_i:i=1,\ldots,N\}$. 
Each function $f_i$ is  a linear combination of the exponential functions $\{E_j:j=1,\ldots,M\}$, where $E_j=e^{\theta_j}$ with $\theta_j=k_jx+k_j^2y-k_j^3t$ for some $k_j\in\mathbb{R}$. The $\tau$-function in the Wronskian form was found in \cite{Ma:79, Sa:79, FN:83} (see also
\cite{H:04}).  Setting $f_i=\sum_{j=1}^Ma_{ij}E_j$, each solution is parametrized by the $N\times M$
coefficient matrix $A=(a_{ij})$ of rank $N$.  This representation naturally leads to the notion of
the Grassmann variety Gr$(N,M)$, the set of $N$-dimensional subspaces
given by Span$_{\mathbb R}\{f_i:i=1,\ldots,N\}$ of $\mathbb{R}^M=$Span$_{\mathbb R}\{E_j:j=1,\ldots,M\}$, and each point of Gr$(N,M)$ is marked by this $A$-matrix \cite{S:81, K:04,CK:09}.

In Section \ref{sec:Gr}, we provide a brief summary of the totally nonnegative (TNN) Grassmann variety, denoted by Gr$^+(N,M)$, which provides the foundation of the classification theorem for the {\it regular} soliton solutions of the KP equation discussed in the next Section. The main purpose of this section is to show that the $\tau$-function in the Wronskian determinant form described in Section \ref{sec:KP} can be identified as a point of the TNN Grassmannian cell. That is, a classification of the regular soliton solutions of the KP equation is equivalent to a parametrization of TNN Grassmannian cells \cite{CK:08,CK:09,CK:10}.

In Section \ref{sec:CL}, we present a classification theorem which states that  
the $\tau$-function identified as a point of Gr$^+(N,M)$ generates a soliton solution of the KP equation
that has asymptotically
$M-N$ line-solitons for $y\ll 0$ and $N$ line-solitons for $y\gg0$.  Moreover,  these 
solutions can be parametrized by the derangements (the permutations without fixed points) of the symmetric group $S_M$.
This type of solutions is called $(M-N,N)$-soliton solution.
  The derangements then give a parametrization of the TNN Grassmannian cells, and each derangement is expressed by a unique {\it chord diagram}.
The chord diagram is particularly useful to describe the far-field structure of the corresponding
soliton solution. (See \cite{CK:08,CK:09,CK:10}.)

In Section \ref{sec:2-2S}, we explain all the soliton solutions generated by the $\tau$-functions on Gr$^+(2,4)$ (see also \cite{BPPP:01}).  Some of these solutions are useful to describe the Mach reflection problem in shallow water.  We show that the $A$-matrix determine the detailed structure of those solutions,
such as the asymptotic locations of solitons and local interaction patterns. (See \cite{CK:09,KK:10,CK:10}.)

In Section \ref{sec:NS}, we present the numerical study of the KP equation for certain types of initial waves.  In particular, we consider an initial value problem where  the initial wave consists of two semi-infinite line-solitons forming a V-shape
pattern.  Those initial waves were considered in
the study of the generation of large amplitude waves in shallow water
\cite{PTLO:05,TO:07}.  The main result of this section is to show that the solutions of this particular initial value problem  {\it converge} asymptotically to
some of the exact (2,2)-soliton solutions.
These results demonstarte a separation of the (exact) soliton solution from dispersive radiation in the manner similar to the KdV case.  (See \cite{CK:09,KOT:09,KK:10}.)

In Section \ref{sec:SWW2}, we discuss the Mach reflection problem in terms of the KP solitons
which is equivalent to Miles' theory (assuming quasi-two dimensionality).
 We first show that the previous numerical results (see for examples \cite{F:80,T:93}),
 which reported a large discrepancy with the theory, are actually in a good agreement with the predictions given by the KP theory.  However, here one needs to give 
 a proper physical interpretation of the theory when one compares it  with the numerical results.
 We also present some laboratory experiments of  shallow water waves \cite{YCL:10, YLK:10}.
We show that the experimental results are all in good agreement with the predictions of the KP theory which  can describe the evolution of the wave-profile.
Finally we demonstrate that the most complex (2,2)-soliton solution associated with the $\tau$-function on Gr$^+(2,4)$, referred to as T-type solution, can be realized in an experiment.

\bigskip

Sir John Scott Russel continued on to say in his book (1865) that 
\begin{quote}
{\it This is a most beautiful and extraordinary phenomenon:
the first day I saw it was the happiest day of my life. Nobody has ever had the good fortune to see it before or, at all events, to know what it meant. It is now known as the solitary wave of translation. No one before had fancied a solitary wave as a possible thing.}  
\end{quote}
I hope this survey is successful to convince the readers that the two-dimensional wave pattern generated by the soliton solutions of the KP equation {\it is a most beautiful and extraordinary phenomena} of two-dimensional shallow water waves, and one should have no doubt that observing these patterns at a beach will bring the {\it happiest} moment of one's life.

%%%%%%%%%%%%%%%%%%%%%%%%%%%%%%%%%%%%%
\section{Shallow water waves: Basic equations}\label{sec:SWW1}
Let us start with the physical background of the KP equation:  We consider a surface wave
on water which is assumed to be irrotational and incompressible (see for examples \cite{Wh:74, AS:81}).
Then the surface wave may be described by
the three-dimensional Euler equation,
\begin{equation}\label{Euler}
\begin{array}{llll}
~~\tilde\Delta\tilde\phi=0,   \qquad &{\rm for}\quad  0< \tilde z < h_0+\tilde\eta, \\[1.0ex]
~~\tilde\phi_{\tilde z}=0, \qquad &{\rm at}\quad \tilde z=0,\\[1.0ex]
\left.\begin{array}{llll}
\displaystyle{\tilde\phi_{\tilde t}+\frac{1}{2}\left|\tilde\nabla\tilde\phi\right|^2+g\tilde \eta=0,} \\[2.0ex]
\displaystyle{\tilde\eta_{\tilde t}+\tilde\nabla_{\perp}\tilde\phi\cdot\tilde\nabla_{\perp}\tilde\eta=\tilde\phi_{\tilde z},}
\end{array}\right\}\quad&{\rm at}\quad \tilde z=\tilde \eta+h_0,\\
\end{array}
\end{equation}
where  $\tilde \phi$ is
the velocity potential with the Laplacian $\tilde\Delta=\partial_{\tilde x}^2+\partial_{\tilde y}^2+\partial_{\tilde z}^2$, $g=980 \,{\rm cm}/{\rm sec}^2$ the gravitational constant, and
$h_0$  the average depth.
The first two equations of \eqref{Euler} implies,
\[
\tilde\phi(\tilde x,\tilde y,\tilde z,\tilde t)=\cos(\tilde z\sqrt{\tilde\Delta_{\perp}}\,)~\tilde\psi(\tilde x,\tilde y,\tilde t)\qquad
{\rm with}\quad \tilde\Delta_{\perp}=\partial^2_{\tilde x}+\partial^2_{\tilde y},
\]
where $\psi(\tilde x,\tilde y,\tilde t)=\phi(\tilde x, \tilde y,0,\tilde t)$.
In the linear limit, the system can be written in the form,
\[\left\{\begin{array}{llll}
\cos(h_0\sqrt{\tilde\Delta_{\perp}})\,\tilde\psi_{\tilde t}+g\tilde\eta=0\\[1.5ex]
\tilde\eta_{\tilde t}=\sqrt{\tilde\Delta_{\perp}}\,\sin(h_0\sqrt{\tilde\Delta_{\perp}})\,\tilde \psi.
\end{array}\right.
\]
This then gives the dispersion relation,
\begin{align}\label{Drelation}
\omega^2&=gk\,\tanh h_0k
=c_0^2k^2\left(1-\frac{1}{3}h_0^2k^2+\cdots\right),
\end{align}
where $k:=\sqrt{k_x^2+k_y^2}$ and the speed of the surface wave $c_0=\sqrt{gh_0}$ (e.g.
$c_0=70$ cm/sec when $h_0=5$ cm).

Let us denote the following scales:
\begin{align*}
&\lambda_0~\sim~{\rm horizontal~length~scale}\\
&h_0~\sim~{\rm vertical~length~scale=water~depth}\\
&a_0~\sim~{\rm amplitude~scale}
\end{align*}
The non-dimensional variables $\{x,y,z,t,\eta,\phi\}$ are defined as
\begin{equation}\label{Pvariables}
\left\{\begin{array}{llll}
\displaystyle{\tilde x=\lambda_0 x,\quad \tilde y=\lambda_0 y,\quad \tilde z=h_0 z,\quad \tilde t=\frac{\lambda_0}{c_0}t},\\[1.5ex]
\displaystyle{\tilde\eta=a_0\eta,\qquad \tilde\phi=\frac{a_0}{h_0}\lambda_0c_0\phi.}
\end{array}\right.
\end{equation}
Then the shallow water equation in the non-dimesional form is given by
\[
\begin{array}{llll}
~~\phi_{zz}+\beta\Delta_{\perp}\phi=0,   \qquad &{\rm for}\quad  0<  z <1+\alpha\eta, \\[1.0ex]
~~\phi_{ z}=0, \qquad &{\rm at}\quad z=0,\\[1.0ex]
\left.\begin{array}{llll}
\displaystyle{\phi_{ t}+\frac{1}{2}\alpha\left|\nabla_{\perp}\phi\right|^2+\frac{1}{2}\frac{\alpha}{\beta}\phi^2_z+\eta=0,} \\[2.0ex]
\displaystyle{\eta_{t}+\alpha\nabla_{\perp}\phi\cdot\nabla_{\perp}\eta=\frac{1}{\beta}\phi_{ z}},
\end{array}\right\}\quad&{\rm at}\quad z=1+\alpha \eta,\\
\end{array}
\]
where the parameters $\alpha$ and $\beta$ are given by
\[
\alpha=\frac{a_0}{h_0}\qquad {\rm and}\qquad \beta=\left(\frac{h_0}{\lambda_0}\right)^2.
\]
The weak nonlinearity implies $\alpha\ll 1$, and the weak dispersion (or long wave assumption)
implies $\beta\ll 1$. With a small parameter $\epsilon\ll 1$, we assume
\[
\alpha\sim \beta =\mathcal O(\epsilon).
\]
%This implies that the Ursell number $U_r:=\alpha/\beta$ should be of order 1, i.e.
%\begin{equation}\label{Ursell}
%U_r=\frac{a_0\lambda_0^2}{h_0^3}=\mathcal{O}(1).
%\end{equation}
As in the previous manner, $\phi$ can be written formally in the form,
\[
\phi(x,y,z,t)=\cos\left(z\sqrt{\beta\Delta_{\perp}}\right)\psi(x,y,t),
\]
which leads to the expansion,
\[
\phi=\psi-\beta\frac{z^2}{2}\Delta_{\perp}\psi +\mathcal O(\epsilon^2).
\]
Then the equations at the surface gives the following system of equations, sometimes called
the Boussinesq-type equation, 
\begin{equation}\label{Boussinesq}
\left\{
\begin{array}{lll}
\displaystyle{\eta+\psi_t+\frac{\alpha}{2}\left|\nabla\psi\right|^2-\frac{\beta}{2}\Delta\psi_t=\mathcal{O}(\epsilon^2)}\\[1.5ex]
\displaystyle{\eta_t+\Delta\psi-\alpha\nabla\cdot\left(\psi_t\nabla\psi\right)-\frac{\beta}{6}\Delta^2\psi=\mathcal O(\epsilon^2).}
\end{array}\right.
\end{equation}
Here we have omitted $_{\perp}$ sign.  Eliminating $\eta$ in \eqref{Boussinesq}, we obtain the so-called
isotropic Benney-Luke equation \cite{BL:64},
\[
\psi_{tt}-\Delta\psi+\alpha\left(\nabla\psi\cdot\nabla\psi_t+\nabla\cdot(\psi_t\nabla\psi)\right)-\frac{\beta}{2}\left(\Delta\psi_{tt}-\frac{1}{3}\Delta^2\psi\right)=\mathcal O(\epsilon^2).
\]
One can then write this equation in the following form up to the same order,
\begin{equation}\label{eq:B}
\left(1-\frac{\beta}{3}\Delta\right)\psi_{tt}-\Delta\psi+\alpha\left(\nabla\psi\cdot\nabla\psi_t+\nabla\cdot(\psi_t\nabla\psi)\right)=\mathcal O(\epsilon^2).
\end{equation}
This is a regularized form of the two-dimensional Boussinesq-type equation for the shallow water waves.
Note here that the dispersion relation of this equation is given by
\[
\omega^2=\frac{k^2}{1+\frac{\beta}{3}k^2}=k^2\left(1-\frac{\beta}{3}k^2+\ldots\right).
\]
which agrees with \eqref{Drelation} up to $\mathcal{O}(\epsilon^2)$.  

It is also well-known that the Boussinesq-type equation can be reduced to the KdV equation for
a far field with a unidirectional approximation:  Let $\chi$ be the coordinate which is perpendicular
to the wave crest of a linear shape solitary wave, i.e. 
\[
\chi=\tilde{x}\,\cos\Psi_0 +\tilde{y}\,\sin\Psi_0,
\]
where $(\cos\Psi_0,\sin\Psi_0)$ is the unit vector in the propagation direction. Then using the
far-field coordinates in the propgation direction, i.e.
\[
\xi:=\chi-t,\qquad \tau=\epsilon t,
\]
the Boussinesq-type equation \eqref{eq:B} becomes the KdV equation,
\[
\epsilon\psi_{\tau\xi}+\frac{3\alpha}{2}\psi_{\xi}\psi_{\xi\xi}+\frac{\beta}{6}\psi_{\xi\xi\xi}=0.
\]
  From the first equation of \eqref{Boussinesq}, we have $\eta=\psi_{\chi}+\mathcal{O}(\epsilon)$.
Then  the KdV equation
can be expressed in the form with physical coordinates,
\begin{equation}\label{KdV}
\tilde\eta_{\tilde t}+c_0\tilde\eta_{\chi}+\frac{3c_0}{2h_0}\tilde\eta\tilde\eta_{\chi}+\frac{c_0h_0^2}{6}\tilde\eta_{\chi\chi\chi}=0.
\end{equation}
The one-soliton solution of the KdV equation is then given by
\begin{equation}\label{KdVsoliton}
\tilde\eta =\hat{a}_0\sech^2\sqrt{\frac{3\hat{a}_0}{4h_0^3}}\left[\,\chi-c_0\left(1+\frac{\hat{a}_0}{2h_0}\right)\tilde t-\chi_0\,\right],
\end{equation}
where $\hat{a}_0>0$ and $\chi_0$ are arbitrary constants.
One should note that any line-solitary wave in the Euler equation \eqref{Euler} can be (at least locally)
approximated by this soliton under the assumption of weak dispersion and weak nonlinearity.
This remark will be important when we compare any numerical results of the Euler equation or
the Boussinesq-type equation with those of the KP equation.

%%%%%%%%%%%%%%%%%%%%%%%%%%%%%%%%%%%%%%%%%%

\section{The KP equation}\label{sec:KP}
In this section, we give some basic property of the KP equation, and
introduce the soliton solutions relevant to shallow water wave problem.
In particular, we discuss the physical aspect of the KP equation which is derived
by a further assumption called {\it quasi}-two dimensional approximation (see for examples \cite{KP:70, AS:81}).
\subsection{Quasi-two dimensional approximation and one soliton solution}
Let us now assume a quasi-two dimensionality with a weak dependence in the $y$-direction, and we
introduce a small parameter $\gamma$ so that the $y$-coordinate is scaled as
\begin{equation}\label{eq:zeta}
\zeta :=\sqrt{\gamma} y,  \qquad {\rm with}\quad \gamma=\mathcal O(\epsilon).
\end{equation}
Then the system of equations (\ref{Boussinesq}) becomes
\[\left\{
\begin{array}{llll}
\displaystyle{\eta+\psi_t+\frac{\alpha}{2}\psi_{x}^2-\frac{\beta}{2}\psi_{xxt}=\mathcal{O}(\epsilon^2)}\\[1.5ex]
\displaystyle{\eta_t+\psi_{xx}-\alpha\left(\psi_t\psi_x\right)_x-\frac{\beta}{6}\psi_{xxxx}+\gamma\psi_{\zeta\zeta}=\mathcal O(\epsilon^3)}.
\end{array}\right.
\]

Now we consider a far field expressed with the scaling,
\begin{equation}\label{eq:xi}
\xi=x-t\qquad {\rm and}\qquad \tau=\epsilon t.
\end{equation}
Then the above equations have the expansions,
\[\left\{
\begin{array}{lll}
\displaystyle{\eta-\psi_{\xi}+\epsilon\psi_{\tau}+\frac{\alpha}{2}\psi_{\xi}^2+\frac{\beta}{2}\psi_{\xi\xi\xi}=\mathcal O(\epsilon^2)}\\[1.5ex]
\displaystyle{-\eta_\xi+\psi_{\xi\xi}+\epsilon\eta_\tau+\alpha\left(\psi_{\xi}^2\right)_\xi-\frac{\beta}{6}\psi_{\xi\xi\xi\xi}+\gamma\psi_{\zeta\zeta}=\mathcal
O(\epsilon^2).}
\end{array}\right.
\]
Eliminating $\eta$, we obtain
\begin{align}
&2\epsilon\psi_{\tau\xi}+3\alpha\psi_{\xi}\psi_{\xi\xi}+\frac{\beta}{3}\psi_{\xi\xi\xi\xi}+\gamma\psi_{\zeta\zeta}
=\mathcal
O(\epsilon^2)\nonumber
\end{align}
Noting $\eta=\psi_{\xi}+\mathcal{O}(\epsilon)$, we have the KP equation for $\eta$ at the leading order,
\begin{align}\label{KPscale}
\left(2\epsilon \eta_{\tau}+3\alpha \eta\eta_{\xi}+\frac{\beta}{3}\eta_{\xi\xi\xi}\right)_{\xi}+\gamma \eta_{\zeta\zeta}=\mathcal{O}(\epsilon^2)
\end{align}
In terms of physical coordinates, the KP equation is given by
\[
\left(\tilde{\eta}_{\tilde t}+c_0\tilde{\eta}_{\tilde x}+\frac{3c_0}{2h_0}\tilde{\eta}\tilde{\eta}_{\tilde x}+
\frac{c_0h_0^2}{6}\tilde{\eta}_{\tilde x\tilde x\tilde x}\right)_{\tilde x}+\frac{c_0}{2}\tilde{\eta}_{\tilde y\tilde y}=0.
\]
As a particular solution, we have one line-soliton solution in the form,
\begin{equation}\label{Pform}
\tilde \eta = a_0\sech^2\sqrt{\frac{3a_0}{4h_0^3}}\left[\,\tilde x+\tilde y\tan\Psi_0-c_0\left(1+\frac{a_0}{2h_0}+\frac{1}{2}\tan^2\Psi_0\right)\tilde t-\tilde{x}_0\,\right],
\end{equation}
where $a_0>0, \Psi_0$ and $\tilde x_0$ are arbitrary constants.
One should note that the KP equation is derived under the assumption of quasi-two dimensionality,
that is, the angle $\Psi_0$ should be small of order $\mathcal{O}(\epsilon)$, and the solution \eqref{Pform} becomes unphysical for the case with a large angle. This can be seen explicitly
by writing it in the following form in the coordinate perpendicular to the wave crest, i.e.
$\chi=\tilde x\cos\Psi_0+\tilde y\sin\Psi_0$, 
\begin{equation}\label{KPsoliton}
\tilde \eta=a_0\sech^2\sqrt{\frac{3a_0}{4h_0^3\cos^2\Psi_0}}\left[\,\chi-c_0\cos\Psi_0\left(1+\frac{a_0}{2h_0}
+\frac{1}{2}\tan^2\Psi_0\right)\,\tilde t -\chi_0\,\right].
\end{equation}
Noting that $\cos\Psi_0=1-\frac{1}{2}\tan^2\Psi_0+\mathcal{O}(\epsilon^4)$ with $\Psi_0=\mathcal{O}(\epsilon)$, the velocity of the soliton
has the corrected form up to $\mathcal{O}(\epsilon^2)$, i.e.
\[
\cos\Psi_0\left(1+\frac{a_0}{2h_0}+\frac{1}{2}\tan^2\Psi_0\right)=1+\frac{a_0}{2h_0}+\mathcal{O}(\epsilon^3),
\]
which does not depend on the angle up to $\mathcal{O}(\epsilon^2)$. This is consistent with
the assumption of the quasi-two dimensionality.

  We also note that the line-soliton of \eqref{Pform} does not
satisfy the KdV equation \eqref{KdV} except the case with $\Psi_0=0$. Now comparing the KP soliton
\eqref{Pform} with the KdV soliton \eqref{KdVsoliton}, one can find the correction to the quasi-two
dimensional approximation, that is,  the amplitude $a_0$ in \eqref{Pform} is now corrected to
\begin{equation}\label{anglecorrection}
\hat{a}_0=\frac{a_0}{\cos^2\Psi_0}.
\end{equation}
With this correction, the KP soliton \eqref{Pform} now satisfies the KdV equation \eqref{KdV}
up to $\mathcal{O}(\epsilon^2)$.  This correction becomes quite important when we compare
our KP results with numerical results of the Euler or Boussinesq-type equations,
that is, \eqref{anglecorrection} gives the relation between the amplitudes of
the KP soliton and the KdV soliton (see Section \ref{sec:SWW2} for the details).

\subsection{Soliton solutions in the Wronskian determinant}
In order to give a general scheme to discuss the soliton solutions, we first
  put the KP equation (\ref{KPscale}) in the standard form,
\begin{equation}\label{KPS}
\left(4u_T+6uu_X+u_{XXX}\right)_X+3u_{YY}=0,
\end{equation}
where the new variables $(X,Y,T)$ and $u$ are related to the physical ones with
\begin{equation}\label{realvariables}
\tilde \eta=\frac{2h_0}{3}\,u,\quad 
\tilde x-c_0\tilde t=h_0\,X,\quad \tilde y=h_0\,Y,\quad \tilde t=\frac{3h_0}{2c_0}\,T.
\end{equation}

Hereafter we use the lower case letters $(x,y,t)$ for $(X,Y,T)$ (we do not use the non-dimensional 
variables in \eqref{Pvariables}, and the KP variables can be converted to the physical
variables directly through the relations \eqref{realvariables}).
We write the solution of the KP equation \eqref{KPS} in the $\tau$-function form,
\begin{equation}\label{utau}
u(x,y,t)=2\partial_x^2\ln\tau(x,y,t).
\end{equation}
where the $\tau$-function is assumed to be the Wronskian determinant with $N$ functions
$f_i$'s (see for examples \cite{Sa:79, Ma:79, FN:83, H:04}),
\begin{equation}\label{tau}
\tau={\rm Wr}(f_1,f_2,\ldots,f_N):=\left|\begin{matrix}
f_1 & f_1^{(1)} & \cdots & f_1^{(N-1)} \\
f_2 & f_2^{(1)} &\cdots & f_2^{(N-1)} \\
\vdots &\vdots &\ddots &\vdots \\
f_N & f_N^{(1)} &\cdots &f_N^{(N-1)}
\end{matrix}\right|,
\end{equation}
with $f^{(n)}_i=\partial_x^nf_i$.
The functions $\{f_i:i=1,\ldots,N\}$ satisfy the linear equations $\partial_yf_i=\partial_x^2f_i$
 and $\partial_tf_i=-\partial_x^3f_i$, and for the soliton solution, we take 
 \begin{equation}\label{f}
 f_i=\sum_{j=1}^M a_{ij}E_j,\quad {\rm with}\quad E_j=e^{\theta_j}:=\exp\left(k_jx+k_j^2y-k_j^3t\right).
 \end{equation}
 Here $A:=(a_{ij})$ defines an $N\times M$ matrix $A=(a_{ij})$ of rank $N$, and we assume the real parameters $\{k_j:j=1,\ldots,M\}$
 to be ordered,
 \[
 k_1<k_2<\cdots<k_M.
 \]
 One should emphasize here that we have a parametrization of each soliton solution of the KP equation  in terms of the $k$-parameters and the $A$-matrix.
 Then the classification of the soliton solutions is to give a complete characterization of the $\tau$-function \eqref{tau} with the exponential functions in \eqref{f}. This is the main theme in \cite{K:04, BC:06, CK:09, CK:10}
 and will be discussed in the following sections.

In terms of the $\tau$-function, the KP equation \eqref{KPS} is written in the bilinear form,
\begin{equation}\label{bilinear}
4(\tau\tau_{xt}-\tau_x\tau_t)+\tau\tau_{xxxx}-4\tau_x\tau_{xxx}+3\tau_{xx}^2+3(\tau\tau_{yy}
-\tau_{xx}^2)=0.
\end{equation}
To show that the $\tau$-function \eqref{tau}  satisfies this equation, we express the derivatives
of the $\tau$-function using the Young diagram.  Let $Y$ be given by the partition $Y=(\lambda_1\ge\ldots\ge\lambda_n)$ where $\lambda_j$'s represent the numbers of boxes in $Y$,
and $|Y|$ denote the total number of boxes, i.e. $|Y|=\sum_{i=1}^n\lambda_i$.  Denote $\tau$ in the $N$-tablet,
\[
\tau=\tau_{\emptyset}=(0,1,2,\ldots,N-1),
\]
where the numbers describe the orders of the derivative of the column vector
$(f_1,\ldots,f_N)^T$ in the $\tau$-function \eqref{tau}.
Then the number of boxes in each row of $Y$ can be found by counting the missing numbers
which are less than the corresponding number in $\tau_{Y}$.
For example, $\tau_{\mbox{\young[2,1][4][]}}$ represents
\[
\tau_{\mbox{\young[2,1][4][]}}=(0,1,2,\ldots,N-3,N-1,N+1).
\]
For the number $N+1$, two numbers $N-2$ and $N$ are missing, and this gives $\young[2][4][,]$. For the number $N-1$, one number $N-2$ is missing and this gives $\young[1][5][]$. In terms of the determinant, $\tau_{\mbox{\young[2,1][4][,]}}$ represents,
\[
\tau_{\mbox{\young[2,1][4][,]}}=\left|\begin{matrix}
f_1 & \cdots & f_1^{(N-3)}&f_1^{(N-1)}&f_1^{(N+1)}\\
f_2&\cdots & f_2^{(N-3)}&f_2^{(N-1)}&f_2^{(N+1)}\\
\vdots&  & \vdots & \vdots & \vdots \\
f_N&\cdots &f_N^{(N-3)}&f_N^{(N-1)}&f_N^{(N+1)}
\end{matrix}\right|
\]
With this notation and using the equations for $f_i$, i.e. $\partial_yf_i=f^{(2)}_i, \partial_tf_i=-f^{(3)}_i$, the derivatives of the $\tau$-function \eqref{tau} are given by
\begin{align*}
\tau_{\mbox{\young[1][4][ ]}}&=\tau_x,\qquad\tau_{\mbox{\young[2][4][ ,
]}}=\frac{1}{2}(\tau_{xx}+\tau_y),\qquad\tau_{\mbox{\young[1,1][4][ ,
]}}=\frac{1}{2}(\tau_{xx}-\tau_y)\\
\tau_{\mbox{\young[2,2][4][ , , ,
]}}&=\frac{1}{12}(\tau_{xxxx}+3\tau_{yy}-4\tau_{xt}),\qquad\tau_{\mbox{\young[2,1][4][
, , ]}}=\frac{1}{3}(\tau_{xxx}-\tau_t)
\end{align*}
Then the bilinear equation \eqref{bilinear} can be written in the form,
\begin{equation}\label{yp}
\tau_\phi \tau_{\mbox{\young[2,2][4][ , , , ]}} -
\tau_{\mbox{\young[1][4][ ]}}
 \tau_{\mbox{\young[2,1][4][ , , ]}}
+ \tau_{\mbox{\young[2][4][ , ]}}
 \tau_{\mbox{\young[1,1][4][ , ]}} = 0.
\end{equation}
This equation is nothing but the Laplace expansion of the following $2N\times 2N$ determinant identity,
\[
\left|\begin{matrix}
f_1 &\cdots & f_1^{(N-2)} & f_1& \cdots & f_1^{(N-3)}&f_1^{(N-1)}&f_1^{(N)}&f_1^{(N+1)}\\
\vdots&  \ddots    &\vdots &\vdots &  \ddots           &\vdots & \vdots & \vdots &\vdots \\
f_N &\cdots & f_N^{(N-2)} & f_N& \cdots & f_N^{(N-3)}&f_N^{(N-1)}&f_N^{(N)}&f_N^{(N+1)}\\
 0  &\cdots & 0   & f_1& \cdots & f_1^{(N-3)}&f_1^{(N-1)}&f_1^{(N)}&f_1^{(N+1)}\\
\vdots&  \ddots      &\vdots &\vdots &   \ddots          &\vdots & \vdots & \vdots &\vdots \\
 0  &\cdots & 0   & f_N& \cdots & f_N^{(N-3)}&f_N^{(N-1)}&f_N^{(N)}&f_N^{(N+1)}\\
 \end{matrix}\right|\equiv 0\,,
 \]
 so that \eqref{yp} is identically satisfied.
 The relation \eqref{yp} is the simplest example of the Pl\"ucker relations (see below),
 and it can be symbolically written by
 \begin{equation}\label{Gr24Prelation}
 \xi(1,2)\xi(3,4)-\xi(1,3)\xi(2,4)+\xi(1,4)\xi(2,3)=0,
 \end{equation}
 where the Young diagrams are expressed by $Y=(j-2,i-1)$ for the symbol $\xi(i,j)$.
 These symbols $\xi(j_1,\ldots,j_N)$ are the so-called Pl\"ucker coordinates of the
 Granssmann manifold Gr$(N,M)$.
 In the next section, we outline the basic information for the Grassmannian Gr$(N,M)$
 which will provide the foundation of the classification theory for the soliton solutions of the KP equation
 \cite{K:04, CK:09}.

 \begin{Example} Let us express one line-soliton solution \eqref{Pform}  in our setting.  Here we also
 introduce some notations to describe the soliton solutions.
The soliton solution \eqref{Pform} is obtained by the $\tau$-function with $M=2$ and $N=1$, i.e.
 \[ 
 \tau=f_1=a_{11}E_1+a_{12}E_2.
 \]
 Since the solution $u$ is given by \eqref{utau}, one can assume $a_{11}=1$ and denote $a_{12}=a>0$. Then 
 \[
\tau=E_1+aE_2=2\sqrt{a}e^{\frac{1}{2}(\theta_1+\theta_2)}\cosh\frac{1}{2}(\theta_1-\theta_2-\ln a)\,,
\]
with the $1 \times 2$ $A$-matrix of the form $A=(1 \,\, a)$. 
The parameter $a$  in the $A$-matrix must be $a\ge 0$ for a {\it non-singular} solution
and it determines the location of the soliton solution.
Since $a=0$ leads to a trivial solution, we consider only $a>0$.
Then the solution $u=2\partial^2_x(\ln\tau)$ gives 
\[
u=\frac{1}{2}(k_1-k_2)^2\sech^2\frac{1}{2}(\theta_1-\theta_2-\ln a).
\]
Thus the solution is localized along the line $\theta_1-\theta_2=\ln a$, hence we call it {\it line-soliton} solution. 
We emphasize here that the line-soliton appears at the boundary of
two regions where either $E_1$ or $E_2$ is the dominant exponential term,
and because of this we also call this soliton a $[1,2]$-soliton solution.
In Section 5, we will construct more general line-soliton solutions which separates into
a number of one-soliton solutions asymptotically as $|y| \to \infty$. We refer to each of
these asymptotic line-solitons as the $[i,j]$-soliton.
The $[i,j]$-soliton solution with $i<j$ has the same (local) structure as the
one-soliton solution, and can be described as follows
\begin{equation*}
u=A_{[i,j]}\sech^2\frac{1}{2}\left({\bf K}_{[i,j]}\cdot {\bf x}-\Omega_{[i,j]}t+\Theta^0_{[i,j]}\right)
\end{equation*}
with some constant $\Theta^0_{[i,j]}$. 
The amplitude $A_{[i,j]}$, the wave-vector ${\bf K}_{[i,j]}$ and the frequency $\Omega_{[i,j]}$ are defined by
\[\left\{
\begin{array}{lll}
\displaystyle{A_{[i,j]}=\frac{1}{2}(k_j-k_i)^2}\\[1.5ex]
\displaystyle{{\bf K}_{[i,j]}=\left(k_j-k_i, k_j^2-k_i^2\right)=(k_j-k_i)\left(1,k_i+k_j\right),}\\[1.5ex]
\displaystyle{\Omega_{[i,j]}=k_j^3-k_i^3=(k_j-k_i)(k_i^2+k_ik_j+k_j^2).}
\end{array}\right.
\]
The direction of the wave-vector ${\bf K}_{[i,j]}=(K_{[i,j]}^x,K_{[i,j]}^y)$ is 
measured in the counterclockwise sense from the $y$-axis, and
it is given by
\[
\frac{K^y_{[i,j]}}{K^x_{[i,j]}}=\tan\Psi_{[i,j]}=k_i+k_j,
\]
that is, $\Psi_{[i,j]}$ gives the angle between the line ${\bf K}_{[i,j]}\cdot {\bf x} =const$ and the $y$-axis.  Then one line-soliton can be written in the form with three parameters $A_{[i,j]},\Psi_{[i,j]}$ and $x^0_{[i,j]}$,
\begin{equation}\label{Onesoliton}
u=A_{[i,j]}\sech^2\sqrt{\frac{A_{[i,j]}}{2}} \left(x+y\tan\Psi_{[i,j]}-C_{[i,j]}t-x^0_{[i,j]}\right),
\end{equation}
with $C_{[i,j]}=k_i^2+k_ik_j+k_j^2=\frac{1}{2}A_{[i,j]}+\frac{3}{4}\tan^2\Psi_{[i,j]}$.
In  Figure \ref{fig:1soliton}, we illustrate one line-soliton solution of $[i,j]$-type. In the right panel
of this figure, we show a {\it chord diagram} which represents this soliton solution. Here the chord 
diagram indicates the permutation of the dominant exponential terms $E_i$ and $E_j$ 
in the $\tau$-function,  that is, with the ordering
$k_i<k_j$, $E_i$ dominates in $x\ll 0$, while $E_j$ dominates in $x\gg 0$ (see Section \ref{sec:Gr}
for the precise definition of the chord diagram).
%%%%%%%%%%%%%%%%%%%%%%%%%%%
\begin{figure}[t]
\begin{centering}
\includegraphics[scale=0.47]{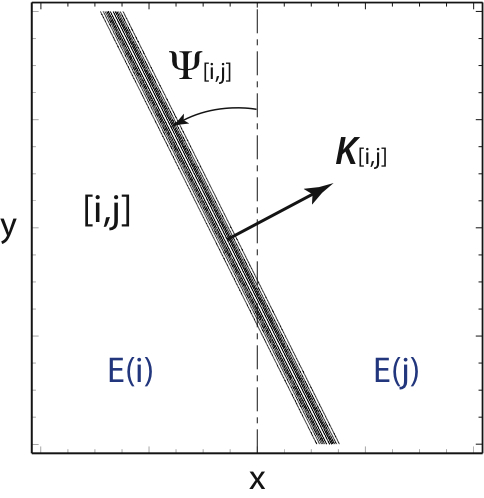}\hskip 1.8cm
\raisebox{0.5cm}{\includegraphics[height=2.8cm]{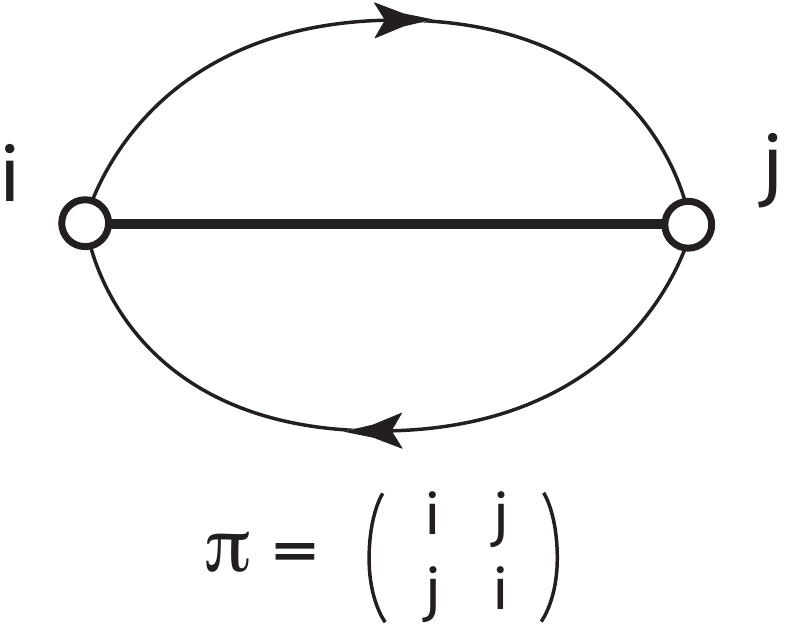}}
\par\end{centering}
\caption{One line-soliton solution of $[i,j]$-type and the corresponding chord diagram. 
The amplitude $A_{[i,j]}$ and the angle $\Psi_{[i,j]}$
are given by $A_{[i,j]}=\frac{1}{2}(k_i-k_j)^2$ and $\tan\Psi_{[i,j]}=k_i+k_j$.
The upper oriented chord represents
the part of $[i,j]$-soliton for $y\gg 0$ and the lower one for $y\ll 0$.
\label{fig:1soliton}}
\end{figure}
%%%%%%%%%%%%%%%%%%%%%%%%%

For each soliton solution of (\ref{Onesoliton}),  the wave vector ${\bf K}_{[i,j]}$ and the 
frequency $\Omega_{[i,j]} $ satisfy the soliton-dispersion relation (see \eqref{Drelation}),
\begin{equation}\label{Sdispersion}
4\Omega_{[i,j]} K_{[i,j]}^x=(K_{[i,j]}^x)^4+3(K_{[i,j]}^y)^2.
\end{equation}
The soliton velocity ${\bf V}_{[i,j]}$ is along the direction of the wave-vector
${\bf K}_{[i,j]}$, and is defined by  ${\bf K}_{[i,j]}\cdot{\bf V}_{[i,j]}=\Omega_{[i,j]}$,
which yields
\[
{\bf V}_{[i,j]}=\frac{\Omega_{[i,j]}}{|{\bf K}_{[i,j]}|^2}{\bf K}_{[i,j]}=
\frac{k_i^2+k_ik_j+k_j^2}{1+(k_i+k_j)^2}\,(1,\,k_i+k_j).
\]
Note in particular that since $C_{[i,j]}=k_i^2+k_ik_j+k_j^2>0$, the $x$-component of 
the soliton velocity is {\it always} positive, i.e., any soliton
propagates in the positive $x$-direction.  In the physical coordinates (see \eqref{realvariables}), this implies that soliton 
propagates in super-sonic (i.e. the speed of soliton is faster than $c_0=\sqrt{gh_0}$,
because of its nonlinear effect with $\tilde\eta>0$, see Section \ref{sec:SWW1}).
 On the other hand, one should note that any 
small perturbation propagates in the negative $x$-direction, i.e., the $x$-component of the 
group velocity is always negative. This can be seen from the dispersion relation of the 
{\it linearized} KP equation for a plane wave $\phi=\exp(i{\bf k}\cdot{\bf x}-i\omega t)$ 
with the wave-vector ${\bf k}=(k_x,k_y)$ and the frequency $\omega$,
\[
\omega=-\frac{1}{4}k_x^3+\frac{3}{4}\,\frac{k_y^2}{k_x},
\]   from which the group velocity of the wave is given by
\[
{\bf v}=\nabla \omega=\left(\frac{\partial\omega}{\partial k_x},\frac{\partial \omega}{\partial k_y}\right)=
\left(-\frac{3}{4}\left(k_x^2+\frac{k_y^2}{k_x^2}\right), \frac{3}{2}\,\frac{k_y}{k_x}\right).
\]
Physically, this means that the radiations disperse with sub-sonic speeds.
This is similar to the case of the KdV equation, and we expect that asymptotically, the 
soliton separates from small radiations.  We further discuss this issue in Section \ref{sec:NS}
where we numerically observe the separation.
 \end{Example}

\begin{Remark}\label{hierarchy}
In the formulas \eqref{f}, if we include the higher times $t_n$ in the exponential functions, i.e.
\[
E_j=\exp\left(\sum_{n=1}^{\infty}k_j^nt_n\right),
\]
then the $\tau$-function \eqref{tau} gives a solution of the KP hierarchy.
The equation for the $t_n$-flow is a symmetry of the KP equation, and the $\tau$-function with
those higher times also satisfies the other Pl\"ucker relations which are expressed with 
the Young diagrams having larger numbers of boxes \cite{MJD:00}.
\end{Remark}

 %%%%%%%%%%%%%%%%%%%%%%%%%%%%%%%%%%%%%%%%%%%

\section{Totally nonnegative Grassmannian Gr$^+(N,M)$}\label{sec:Gr}
In the previous section, we considered a class of solutions which are expressed by the $\tau$-functions
\eqref{tau} with the exponential functions \eqref{f}. Those solutions are determined by the $k$-parameters and the $A$-matrix. Fixing the $k$-parameters, we have a set of $M$ exponentials
$\{E_j=e^{\theta_j}:j=1,\ldots,M\}$ which spans $\mathbb{R}^M$.  Then the set of functions
$\{f_i: i=1,\ldots,N\}$ of \eqref{f} defines an $N$-dimensional subspace of $\mathbb{R}^M$.
This leads naturally to the notion of Grassmannian Gr$(N,M)$, the set of all $N$-dimensional subspaces in $\mathbb{R}^M$, and each point of Gr$(N,M)$ can be parametrized by the $A$-matrix
in \eqref{f}.   Here we give a brief review of the Grassmann manifold $\mathrm{Gr}(N,M)$, in particular, we describe the totally non-negative part of Gr$(N,M)$.  The main purpose of this section is to explain a mathematical background of {\it regular} soliton solutions of the KP equation. 

\subsection{The Grassmannian Gr$(N,M)$}
Recall that the set of the functions $f_i$ spans an $N$-dimensional subspace which is parametrized by
an $N\times M$ matrix $A$ of rank $N$, i.e.
\[
(f_1,f_2,\ldots,f_N)=(E_1,E_2,\ldots,E_M)\,A^T.
\]
Since other set of functions $(g_1,\ldots,g_N)=(f_1,\ldots,f_N)H$ for some $H\in$GL$_N(\mathbb{R})$
gives the same subspace, the $A$-matrix can be canonically chosen in the {\it reduced row echelon
form} (RREF).  This then gives an explicit definition of the Grassmannian,
\[
{\rm Gr}(N,M)={\rm GL}_N(\mathbb{R})\big\backslash \mathcal{M}_{N\times M}(\mathbb{R}),
\]
where  $\mathcal{M}_{N\times M}(\mathbb{R})$
denotes the set of $N\times M$ matrices of rank $N$.  
The canonical form of $A$ is distinguished by a set of {\it pivot} columns
labeled by $I = \{i_1,i_2,\ldots,i_N\}, \,\, 1 \leq i_1 < i_2 < \ldots < i_N \leq M$
such that the $N \times N$ sub-matrix $A_I$ formed by the column set $I$ is the identity 
matrix. Each $N \times M$ matrix $A$ in RREF uniquely determines an $N$-dimensional  subspace, thus providing a coordinate for a point of Gr$(N,M)$.
The set $W_I$ of all points in Gr$(N,M)$ represented by RREF matrices $A$ which have the
same pivot set $I$ is called a Schubert cell which gives the decomposition of the Grassmannian,
the Schubert decomposition,
\begin{equation}
\mathrm{Gr}(N,M)=\bigsqcup_{1 \leq i_1<i_2<\ldots i_N \leq M}\hspace{-0.2 in} W_I \,, 
\qquad I = \{i_1,i_2,\ldots,i_N\}\,.
\label{schubert}
\end{equation}
For example, 
if $I=\{1,2,\ldots,N\}$, then the Schubert cell $W_I$ contains all $A$ matrices 
whose RREF is given by
\begin{equation*}\left(
\begin{array}{ccccccc}
1&0&\cdots&0&*&\cdots&*\\
0&1&\cdots&0&*&\cdots&*\\
\vdots&\vdots&\ddots&\vdots&\vdots&\vdots&\vdots\\
0&0&\cdots&1&*&\cdots&*
\end{array}\right)
\end{equation*}
where the $N(M-N)$ entries of the right-hand block are arbitrary real numbers. 
This particular Schubert cell is often referred to as the {\it top} cell
which has the maximum number of free parameters marked by $*$. 
It follows from this that the dimension of $\mathrm{Gr}(N,M)$ is $N(M-N)$.
The number of free parameters for an $A$-matrix
in RREF with a given pivot set $I = \{i_1,\ldots,i_N\}$ is equal to the 
the dimension of the cell $W_I$, and is given by
$${\rm dim}\, W_I =N(M-N)-\sum_{n=1}^N (i_n-n)\,.$$
Note here that the index set $I=\{i_1,\ldots,i_N\}$ can be expressed by the Young diagram
with $Y_I=(i_N-N,\ldots, i_2-2, i_1-1)$, and then codim $W_I=|Y|$.
\begin{Example} The Schubert decomposition of $\mathrm{Gr}(2,4)$ has the form,
\begin{equation*}
\mathrm{Gr}(2,4)=\bigsqcup_{1\leq i<j\leq 4} W_{\{i,j\}} \,.
\end{equation*}
There are six cells $W_I$ with $\mathrm{dim}\,W_I=7-(i+j)$, and are listed below:
\begin{align*}
&{\rm (a)}\,\, W_{\{1,2\}} =W_{\emptyset}= \left\{\begin{pmatrix} 1&0&*&*\\0&1&*&* \end{pmatrix}\right\},  \qquad  
& {\rm (b)}\,\, W_{\{1,3\}} =W_{\mbox{\young[1][4][]}}= \left\{\begin{pmatrix} 1&*&0&*\\0&0&1&* \end{pmatrix}\right\}, \\
& {\rm (c)}\,\, W_{\{1,4\}}  =W_{\mbox{\young[2][4][]}}= \left\{\begin{pmatrix} 1&*&*&0\\0&0&0&1 \end{pmatrix}\right\}, \qquad  
& {\rm (d)}\,\, W_{\{2,3\}}  =W_{\mbox{\young[1,1][4][]}}= \left\{\begin{pmatrix} 0&1&0&*\\0&0&1&* \end{pmatrix}\right\}, \\
& {\rm (e)}\,\, W_{\{2,4\}}  =W_{\mbox{\young[2,1][4][]}}=\left\{ \begin{pmatrix} 0&1&*&0\\0&0&0&1 \end{pmatrix}\right\}, \qquad  
& {\rm (f)}\,\, W_{\{3,4\}}  =W_{\mbox{\young[2,2][4][]}}= \left\{\begin{pmatrix} 0&0&1&0\\0&0&0&1 \end{pmatrix}\right\}. 
\end{align*}
The top cell $W_{\{1,2\}}$ has four free parameters which gives dim $\mathrm{Gr}(2,4)$, while the bottom
cell $W_{\{3,4\}}$ is 0-dimensional and corresponds to a single point of the Grassmannian. 
\end{Example}

We also note that each cell in the Schubert decomposition can be parametrized by a unique
element of $S_M$, the symmetric group of permutations for $M$ letters. The group $S_M$ is 
generated by the adjacent transpositions $s_j:=(j,j+1)$, i.e.
\[
S_M=\langle s_1,s_2,\ldots,s_{M-1}\rangle,
\]
with $s_i^2=e$, the identity element, $s_is_j=s_js_i$ if $|i-j|>1$ and $(s_is_{i+1})^3=e$.
Let $P_N$ be a maximal parabolic subgroup of $S_M$ generated by $s_j$'s without 
the element $s_{M-N}$, i.e.
\[
P_N:=\langle s_1,\ldots,s_{M-N-1},s_{M-N+1},\ldots, s_{M-1}\rangle \cong S_{M-N}\times S_N.
\]
Then the pivot set
$I=\{i_1,i_2,\ldots,i_N\}$ parametrizing the Schubert cell $W_I$ can be uniquely labeled by 
a minimal length representative of the coset ,
\[
S_M^{(N)}:=\frac{S_M}{P_N}=\{\,{\rm the~reduced~words~ ending~with~}s_{M-N}\,\}.
\] 
Namely, we have the Schubert decomposition of Gr$(N,M)$ in terms of the coset $S_M^{(N)}$,
\[
{\rm Gr}(N,M)=\bigsqcup_{\pi\in S_{M}^{(N)}}W_{\pi}.
\]
where the dimension of the cell $W_{\pi}$ is given by the length of the permutation, i.e.
dim$\,W_{\pi}=\ell (\pi)$.
For example, in the case of Gr$(1,3)$, we have
$S^{(1)}_3=\langle s_1,s_2\rangle/\langle s_1\rangle=\{e,\, s_2, \, s_1s_2\}$,
\[
e=\begin{pmatrix}
1 & 2 & \fbox{3} \\ 1 & 2 & 3
\end{pmatrix}
\quad \overset{s_2}\longrightarrow\quad
\begin{pmatrix}
1 & \fbox{2} & 3 \\ 1 & 3 & 2
\end{pmatrix}
\quad \overset{s_1}\longrightarrow\quad
\begin{pmatrix}
\fbox{1} & 2 & 3 \\ 3 & 1 & 2
\end{pmatrix}
\]
Here $\fbox{i}$ represents a pivot, so that we have
\[
W_e=\{(0,0,1)\},\qquad W_{s_2}=\{(0,1,*)\},\qquad W_{s_1s_2}=\{(1,*,*)\}.
\]
 Also in the case of Gr$(2,3)$, 
we have $S^{(2)}_3=\langle s_1,s_2\rangle/\langle s_2\rangle=\{e, \,s_1,\,s_2s_1\}$,
\[
e=\begin{pmatrix}
1 & \fbox{2} & \fbox{3} \\ 1 & 2 & 3
\end{pmatrix}
\quad \overset{s_1}\longrightarrow\quad
\begin{pmatrix}
\fbox{1} & {2} & \fbox{3} \\ 2 & 1 & 3
\end{pmatrix}
\quad \overset{s_2}\longrightarrow\quad
\begin{pmatrix}
\fbox{1} & \fbox{2} & 3 \\ 2& 3 & 1
\end{pmatrix},
\]
and the Schubert cells $W_{\pi}$ are given by
\[
W_e=\left\{\begin{pmatrix} 0 &1 &0\\ 0&0&1\end{pmatrix}\right\},\qquad
W_{s_1}=\left\{\begin{pmatrix} 1 &*&0\\ 0&0&1\end{pmatrix}\right\},\qquad
W_{s_2s_1}=\left\{\begin{pmatrix} 1 &0 &*\\ 0&1&*\end{pmatrix}\right\}.
\]
Note in particular that the last elements  in the above examples have no fixed points, and they are
called derangements. As we will show that each derangement of $S_M$ parametrizes a unique line-soliton solution generated by the $\tau$-function of the form \eqref{tau}.  It is important for our purposes to remark that each permutation $\pi\in S_M$ with marked pivot positions can be uniquely expressed by the {\it chord diagram}.  This permutation is the decorated permutation defined in \cite{P:06} for a parametrization of the totally non-negative Grassmann cells.  
\begin{Definition}
A chord diagram associated with $\pi\in S_M$ is defined as follows:  Consider a line segment with
$M$ marked points by the numbers $\{1,2,\ldots,M\}$ in the increasing order from the left.
\begin{itemize}
\item[(a)] If $i<\pi(i)$ (excedance), then draw a chord joining $i$ and $\pi(i)$ on the upper part of the line.
\item[(b)] If $j<\pi(j)$ (deficiency), then draw a chord joining $j$ and $\pi(j)$ on the lower part of the line.
\item[(c)] If $l=\pi(l)$ (fixed point), then 
\begin{itemize}
\item[(i)] if $l$ is a pivot, then draw a loop on the upper part of the line at this point.
\item[(ii)] if $l$ is a non-pivot, then draw a loop on the lower part of the line at this point.
\end{itemize}
\end{itemize}
The dimension of each Schubert cell of Gr$(N,M)$ can be also found from the chord diagram, and it is given by
\begin{align*}
{\rm dim}W_{\pi}=& N+\{\#{\rm ~of~crossings}\}+\{\#{\rm ~of ~cusps~ in~the~lower~part}\}\\
&-\{\#{\rm ~of~loops~in ~the ~upper~part}\}.
\end{align*}
Here we say that the point marked by $j$ is a ``cusp'',  if $\pi(j)<j=\pi(k)<k$ or $k<\pi(k)=j<\pi(j)$ for some $k$.
In particular, the point $j$ is a cusp in the lower part of the diagram, if $\pi(j)<j=\pi(k)<k$ (see \cite{C:07,W:05}).
\end{Definition}

\begin{Example}
Consider the case of Gr$(2,4)$.  The Schubert cells $W_{\{i,j\}}$ are marked by the pivots $\{i,j\}$ with $1\le i<j\le 4$, and the permutation representations are given by 
\smallskip
\[
\begin{array}{cccccc}
\begin{pmatrix}
1 & 2 & \fbox{3} & \fbox{4}\\
1 & 2 & 3 & 4
\end{pmatrix} & \overset{s_2}\longrightarrow 
& \begin{pmatrix}
1 & \fbox{2} & 3 & \fbox{4}\\
1& 3 & 2 & 4
\end{pmatrix}  & \overset{s_1}\longrightarrow  &
\begin{pmatrix}
\fbox{1} & 2 & 3 & \fbox{4}\\
3 & 1 & 2 & 4
\end{pmatrix}\\[2.5ex]
&  & s_3\downarrow  & &  s_3 \downarrow  \\[2.5ex]
& &
\begin{pmatrix}
1 &  \fbox{2} & \fbox{3} & {4}\\
1 & 3 & 4 & 2
\end{pmatrix} & \overset{s_1}\longrightarrow 
& \begin{pmatrix}
\fbox{1} & {2} & \fbox{3} & {4}\\
3& 1 & 4 & 2
\end{pmatrix} \\[2.5ex]
&&&& s_2\downarrow\\[2.5ex]
&&&&
\begin{pmatrix}
\fbox{1} & \fbox{2} & {3} & {4}\\
3 & 4 & 1 & 2
\end{pmatrix} 
\end{array}
\]
The chord diagrams are shown below, and the points with filled circle indicate the pivots for those cells:
%%%%%%%%%%%%%%%%%%%%%%%%%%%%%
\begin{figure}[h]
\includegraphics[height=4.5cm]{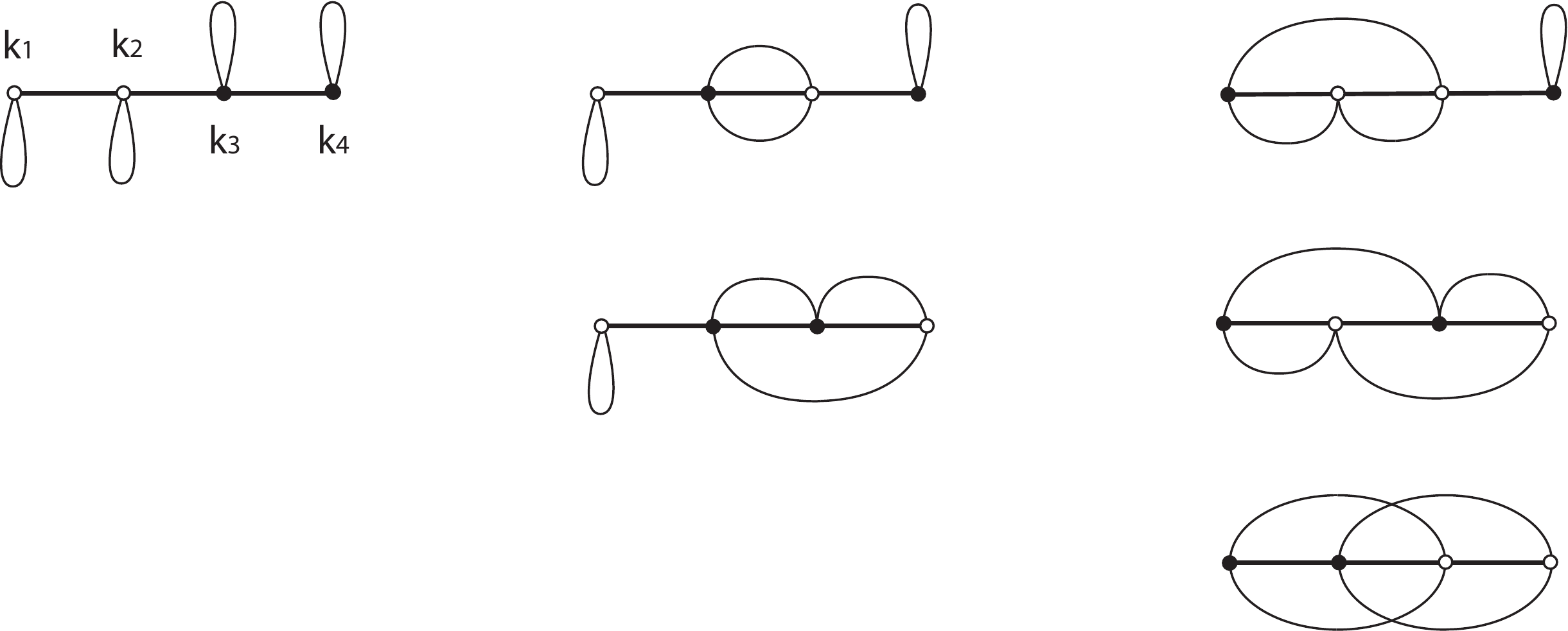}
\end{figure}
%%%%%%%%%%%%%%%%%%%%%%%%%%%%%%
One should note here that each fixed point corresponds to a loop of the diagrams, and
 the diagrams without loops are associated with the derangements of the permutation group.

As we will show, each chord (not loop) identifies a line-soliton for $y\gg0$ (or $y\ll 0$) corresponding
to the location of the chord in the upper (or lower) part of the chord diagram. For example, in the case
$\pi=\binom{1~2~3~4}{3~1~4~2}$, we have $[1,3]$- and $[3,4]$-solitons in $y\gg 0$ and
$[1,2]$- and $[2,4]$-solitons in $y\ll 0$.
\end{Example}

\subsection{The Pl\"ucker coordinates and total non-negativity}
We here describe the totally non-negative (TNN) Grassmannian Gr$^+(N,M)$
as a subspace of Gr$(N,M)$.  Then we will show that the $\tau$-function associated with Gr$^+(N,M)$ is necessary and sufficient conditions for the solution generated by the $\tau$-function to be regular.

We first note that 
the coordinates of Gr$(N,M)$ is given by the Pl\"ucker embedding into the projectivization of the
wedge product space $\wedge^N\mathbb{R}^M$, i.e.
\[
{\rm Gr}(N,M)~ \hookrightarrow~ \mathbb{P}(\wedge^N\mathbb{R}^M),
\]
which maps each frame given by $[f_1,\ldots,f_N]\in$Gr$(N,M)$ to the point on $\mathbb{P}(\wedge^N\mathbb{R}^M)$, i.e.
\begin{equation}\label{embedding}
f_1\wedge \cdots\wedge f_N=\sum_{1\le j_1<\ldots<j_N\le M}\xi(j_1,\ldots,j_N)E_{j_1}\wedge\cdots\wedge E_{j_N}.
\end{equation}
Here the coefficients $\xi(j_1,\ldots,j_N)$ are the $N\times N$ minors of the $A$-matrix defined by
\[
\xi(j_1,\ldots,j_N):={\rm det}|(a_{n,j_n})_{1\le n\le N}|.
\]
Note here that $\xi(j_1,\ldots,j_N)=1$ where the set $\{j_1,\ldots,j_N\}$ is the pivot set.
Those minors $xi(j_1,\ldots,j_N)$ are called the {\it Pl\"ucker coordinates}, which give a coordinate system for the linear space
$\wedge^N\mathbb{R}^M$ with the basis,
\[
\{E_{j_1}\wedge\cdots\wedge E_{j_N}:1\le j_1<\cdots<j_N\le M\}.
\]
Then the Grassmannian structure is determined by certain relations on the Pl\"ucker coordinates,
called the {\it Pl\"ucker relations}:  for any two index sets $\{\alpha_1,\ldots,\alpha_{N-1}\}$
and $\{\beta_1,\ldots,\beta_{N+1}\}$ with $1\le \alpha_i,\beta_j\le M$, they are given by
\begin{equation}\label{Prelation}
\sum_{j=1}^{N+1}\xi(\alpha_1,\ldots,\alpha_{N-1},\beta_j)\,\xi(\beta_1,\ldots,\check{\beta_j},\
\ldots,\beta_{N+1})=0,
\end{equation}
where $\check{\beta}_j$ implies the deletion of $\beta_j$.
The Pl\"ucker relations can be
derived using elementary linear algebra from the Laplace expansion
of the following $2N \times 2N$ determinant formed by the columns $A_i$ of
the matrix $A$, i.e.
\begin{equation*}
\left|\begin{matrix}
A_{\alpha_1} &\cdots & A_{\alpha_{N-1}}  & A_{\beta_1}& \cdots & A_{\beta_{N+1}}\\
 0  &\cdots & 0  & A_{\beta_1}& \cdots & A_{\beta_{N+1}} \\
 \end{matrix}\right| = 0 \,.
\end{equation*}
The Pl\"ucker coordinates, modulo the
Pl\"ucker relations, give the correct dimension of Gr$(N,M)$ which
is typically less than the dimension of $\mathbb{P}(\bigwedge^N\mathbb{R}^M)$.
\begin{Example} For Gr$(2,4)$, the Pl\"{u}cker coordinates are given by the
maximal minors,
\[
\xi(1,2), \quad \xi(1,3), \quad \xi(1,4), \quad\xi(2,3), \quad\xi(2,4), \quad\xi(3,4).
\]
Taking $\{\alpha_1\}=\{1\}$, $\{\beta_1,\beta_2,\beta_3\}=\{2,3,4\}$ in (\ref{Prelation}) gives the
only Pl\"{u}cker relation in this case,
$$\xi(1,2)\xi(3,4)-\xi(1,3)\xi(2,4)+\xi(1,4)\xi(2,3)=0,$$
which is the same as (\ref{Gr24Prelation}).
Since $\dim(\bigwedge^2\mathbb{R}^4) = 6$, and the projectivization 
gives dim($\mathbb{P}(\bigwedge^2\mathbb{R}^4)) = 6-1=5$. Then with one Pl\"ucker relation, 
the dimension of Gr$(2,4)$ turns out be 4, which is consistent with the
dimension of the top cell $W_{\{1,2\}}$ as shown in Example 3.2 (case (a)).
\end{Example}

Since each point of Gr$(N,M)$ is expressed by \eqref{embedding}, the TNN Grassmannian Gr$^+(N,M)$ is defined by the set of $N\times M$ matrices of rank $N$ whose minors, the Pl\"ucker coordinates, are all non-negative, i.e.
\[
{\rm Gr}^+(N,M):=\left\{A\in {\rm Gr}(N,M): \xi(j_1,\ldots,j_N)\ge 0,~\forall~1\le j_1<\cdots<j_M\le M\right\}.
\]
Then the most interesting question is to find a parametrization of all the cells in Gr$^+(N,M)$.
This question has been solved by Postnikov and his colleagues (see \cite{P:06,W:05}), and
our classification theorem of the soliton solutions provides an alternative proof based on
a simple asymptotic analysis as described in Section \ref{sec:CL} (see also \cite{CK:08, CK:09}).

\subsection{The $\tau$-function as a point on Gr$^+(N,M)$}
Expanding the $\tau$-function in the Wronskian determinant (\ref{tau}) by Binet-Cauchy formula,
we have
\begin{equation}\label{tauexpansion}
\tau=\mathrm{Wr}(f_1,f_2,\ldots,f_N)
=\sum_{1\le i_1<\cdots<i_N\le M} \hspace{-0.2 in}
\xi(i_1,i_2,\ldots,i_N)E(i_1,i_2,\ldots,i_N)\,,
\end{equation}
where $\xi(i_1,i_2,\ldots,i_N)$ are the Pl\"ucker coordinates given by the maximal minors of 
the $A$-matrix, and $E(i_1,i_2,\ldots,i_N)={\rm Wr}(E_{i_1},E_{i_2},\ldots,E_{i_N})$.
Here each $E(i_1,i_2,\ldots,i_N)$ can be
identified with $E_{i_1}\wedge E_{i_2}\wedge\cdots E_{i_N}$ which is the basis element 
for $\bigwedge^N\mathbb{R}^M$, i.e.
\[
{\rm Span}_\mathbb{R}\left\{E(i_1,i_2,\ldots,i_N):1\le i_1<i_2<\cdots<i_N\le M\right\}\cong 
\bigwedge^N\mathbb{R}^M\,.
\]
 Note here that the sum
$k_{i_1}+k_{i_2}+\ldots+k_{i_N}$ should be distinct for distinct sets
$\{i_1,i_2,\ldots,i_N\}$ in order for the functions $\{E(i_1,i_2,\ldots,i_N)\}$ 
to be linearly independent.  It is then clear that the $\tau$-function given by the Wronskian determinant
\eqref{tau} can be identified with a point of Gr$(N,M)$, and the Wronskian map Wr$ :[f_1,\ldots,f_N]
\mapsto \tau$ gives the Pl\"ucker embedding.  With the ordering $k_1<\cdots<k_M$, the Wronskian
Wr$(E_{i_1},\ldots,E_{i_N})>0$ for $i_1<\cdots<i_N$.
  Then $\tau\in$Gr$^+(N,M)$ implies that  $\tau$-function is positive definite and the solution
$u(x,y,t)=2\partial^2_x(\ln\tau)$ is regular for all $(x,y,t)\in\mathbb{R}^3$.
In order to prove a converse of this statement, we first show the following: Let $(t_1,t_2,\ldots,t_M)$ be the higer times
for the KP equation (see Remark \ref{hierarchy}, and here the first three times $t_1=x,t_2=y$ and $t_3=-t$ give the KP variables).
\begin{Proposition}
Suppose that the $\tau$-function is regular for all $(t_1,t_2,\ldots,t_M)$. Then $\tau\in$Gr$^+(N,M)$.
\end{Proposition}
\begin{Proof}
Let us first write the exponential terms,
 \[
 E_j=\exp\left(\sum_{n=1}^Mk_j^nt_n +\theta_j^0\right)=: \hat{E}_je^{\theta_j^0} \qquad{\rm for}\quad j=1,\ldots,M,
 \]
 with ${\theta_j^0}\in \mathbb{R}$, i.e. the shifts of $t_n$'s in the exponential functions.
Because the $k$-parameters are all distinct, one can take the coordinates $(\theta_1^0,\ldots,\theta_M^0)$ instead of $(t_1,\ldots,t_M)$.  The $\tau$-function is then given by
\[
\tau=\sum_{1\le j_1<\cdots<j_N\le M}\xi(j_1,\ldots,j_N)\hat{E}(j_1,\ldots,j_N)\prod_{k=1}^Ne^{\theta_{j_k}^0},
\]
where $\hat{E}(j_1,\ldots,j_N)={\rm Wr}(\hat{E}_{j_i},\ldots,\hat{E}_{j_N})$.  Then one can choose
the parameters $(\theta_{j_1}^0,\ldots,\theta_{j_N}^0)$ so that 
\[
\sum_{k=1}^N\theta_{j_k}^0 ~\gg~\sum_{k=1}^N\theta_{l_k}^0,
\]
for any other choice of the parameters $(\theta_{l_1}^0,\ldots,\theta_{l_N}^0)$.
This means that the exponential term having this index set $\{j_1,\ldots,j_N\}$ 
is the dominant one in the $\tau$-function, while all other parameters are of 
$\mathcal{O}(1)$.  Suppose the minor $\xi(j_1,\ldots,j_N)$
associated with this index set is negative, that is, $\tau\approx \xi(j_1,\ldots,j_N)E(j_1,\ldots,j_N)<0$.  
Now note that for $x\ll0$, the dominant exponential in the $\tau$-function is $E(e_1,\ldots,e_N)>0$ with the pivot set $\{e_1,\ldots,e_N\}$ and the ordering $k_1<\cdots<k_M$, so that $\tau\approx E(e_1,\ldots,e_N)>0$.
This implies that the $\tau$-function vanishes at some point in $(x,y)$-plane, and
therefore the solution $u(x,y,t)$ is not regular.
\end{Proof}
It is then clear from the proof that the total non-negativity is not only sufficient but necessary
for the regularity of the solution.  Namely we have the following.
\begin{Corollary}
The solution of the KP equation generated by the $\tau$-function in the form \eqref{tau}
with \eqref{f} is non-singular for any initial data if and only if $\tau\in$Gr$^+(N,M)$.
\end{Corollary}
Thus the classification of the regular soliton solutions
is equivalent  to a study of the totally non-negative Grassmannian.  

\begin{Remark}\label{momentpolytope}
Since each $\tau$-function can be identified as a point on Gr$(N,M)$, one can define 
a moment map, $\mu: {\rm Gr}(N,M) \to \mathfrak{h}_{\mathbb R}^*$ \cite{KS:08},
\[
\mu(\tau)=\frac{\sum_{1\le j_1<\cdots<j_N\le M}|\xi(j_1,\ldots,j_N)E(j_1,\ldots,j_N)|^2(L_{j_1}+\cdots+L_{j_N})}{\sum_{1\le j_<\cdots<j_N\le M}|\xi(j_1,\ldots,j_N)E(j_1,\ldots,j_N)|^2},
\]
where $L_j$ are the weights of the standard representation of SL$(M)$, and $\mathfrak{h}_{\mathbb R}^*$ is the real part of the dual of the Cartan subalgebra of $\mathfrak{sl}(M)$ defined by
\[
\mathfrak{h}_{\mathbb R}^*={\rm Span}_{\mathbb R}\left\{L_1,\ldots,L_M: \sum_{j=1}^M L_j=0\right\}\cong \mathbb{R}^{M-1}.
\]
Then the closure of the image of the moment map is a convex polytope whose vertices are the fixed points of the $S_M$ orbit, that is, the dominant exponentials. 
\end{Remark}

%%%%%%%%%%%%%%%%%%%%%%%%%%%%%%%%%%%%%%%%%%%%%%%%

\section{Classification of soliton solutions}\label{sec:CL}
In this section, we now show the asymptotic behavior of the 
$\tau$-function in \eqref{tauexpansion} and then present a classification
scheme for the regular line-soliton solutions of KP based on the $\tau$-function
asymptotics (see also \cite{BC:06}). 

\subsection{Asymptotic line-solitons}
The $\tau$-function of \eqref{tauexpansion} is given explicitly by the sum of exponential terms
with the Wronskians,
\begin{align}
\tau(x,y,t)
&= \sum_{1\le m_1<\dots<m_N\le M}\hspace{-0.3 in} \xi(m_1,\dots,m_N)
  E(m_1,\ldots,m_N)\,,
\label{tauexp}
\end{align}
where $E(m_1,\ldots,m_N)={\rm Wr}(E_{m_1}, \ldots, E_{m_N})$
with $E_m=e^{\theta_m}$, and $\theta_m(x,y,t)=k_mx+k_m^2y-k_m^3t$.
Since $u=2\partial_x^2(\ln \tau)$, the regularity condition 
on the line-soliton solutions requires that the
$\tau$-function does not vanish for all values of $x, y$ and $t$. To ensure that
it is sign-definite, the following necessary and sufficient conditions are imposed
on the $\tau$-function in \eqref{tauexp}.
\begin{itemize}
\item[(i)] The parameters $k_1,k_2, \ldots k_M$ are ordered as 
$k_1 < k_2 < \ldots < k_M$, and the sums $k_i+k_j$ are all distinct.

%\vspace{-0.1 in}

\item[(ii)]$A$ is a TNN matrix, that is, all its maximal minors
are $\xi(m_1,\dots,m_N) \geq 0$.
\end{itemize}

The asymptotic spatial structure of the solution $u(x,y,t)$ is determined
from the consideration of dominant exponentials $E(m_1,\ldots,m_N)$ in 
the $\tau$-function at different regions of the $(x,y)$-plane for 
large $|y|$. The solution $u = 2 \partial_x^2(\ln\tau)$ is localized at
the {\it boundaries} of two distinct regions where a balance exists between 
two dominant exponentials in the $\tau$-function (\ref{tauexp}), whereas
the solution is exponentially small in the interior of each of these regions 
where only one exponential $E(m_1,\ldots,m_N)$ with a specific index set 
$\{m_1,\ldots,m_N\}$ is dominant.  Before discussing a general theorem, let us first consider the following simple examples
which illustrate  the resonant interactions among the line-solitons.
As discussed in Introduction, the resonant interaction is one of the most important 
features of the KP equation (see e.g. \cite{M:77, NR:77, KY:80}).

\begin{Example}\label{12soliton}
We consider the case with $N=1$ and $M=3$, where 
the $1 \times 3$ coefficient matrix $A$ is given by
 \[
 A=\begin{pmatrix} 1 & a & b \end{pmatrix}.
 \]
 The parameters $a, b$ in the matrix are positive constants; meaning that this $A$-matrix
 marks a point on Gr$^+(1,3)$, and the positivity implies the regularity
 of the KP solution. The $\tau$-function is simply given by
 \[
\tau=f= E_1+aE_2+bE_3.
\]
Now let us determine the dominant
exponentials and analyze the structure of the solution in the $xy$-plane. First we
consider the function $f$ along the line $x=-cy$ with $c=\tan\Psi$ where
$\Psi$ is the angle measured counterclockwise from the $y$-axis (see Figure \ref{fig:1soliton}). 
Then along $x=-cy$, we have the exponential function $E_j =\exp[\eta_j(c)y-k_j^3t]$ with
\begin{equation}\label{eta}
\eta_j(c)=k_j(k_j-c).
\end{equation}
%%%%%%%%%%%%%%%%%%%%%%%%%%%
\begin{figure}[t]
\begin{center}
\includegraphics[height=1.3in]{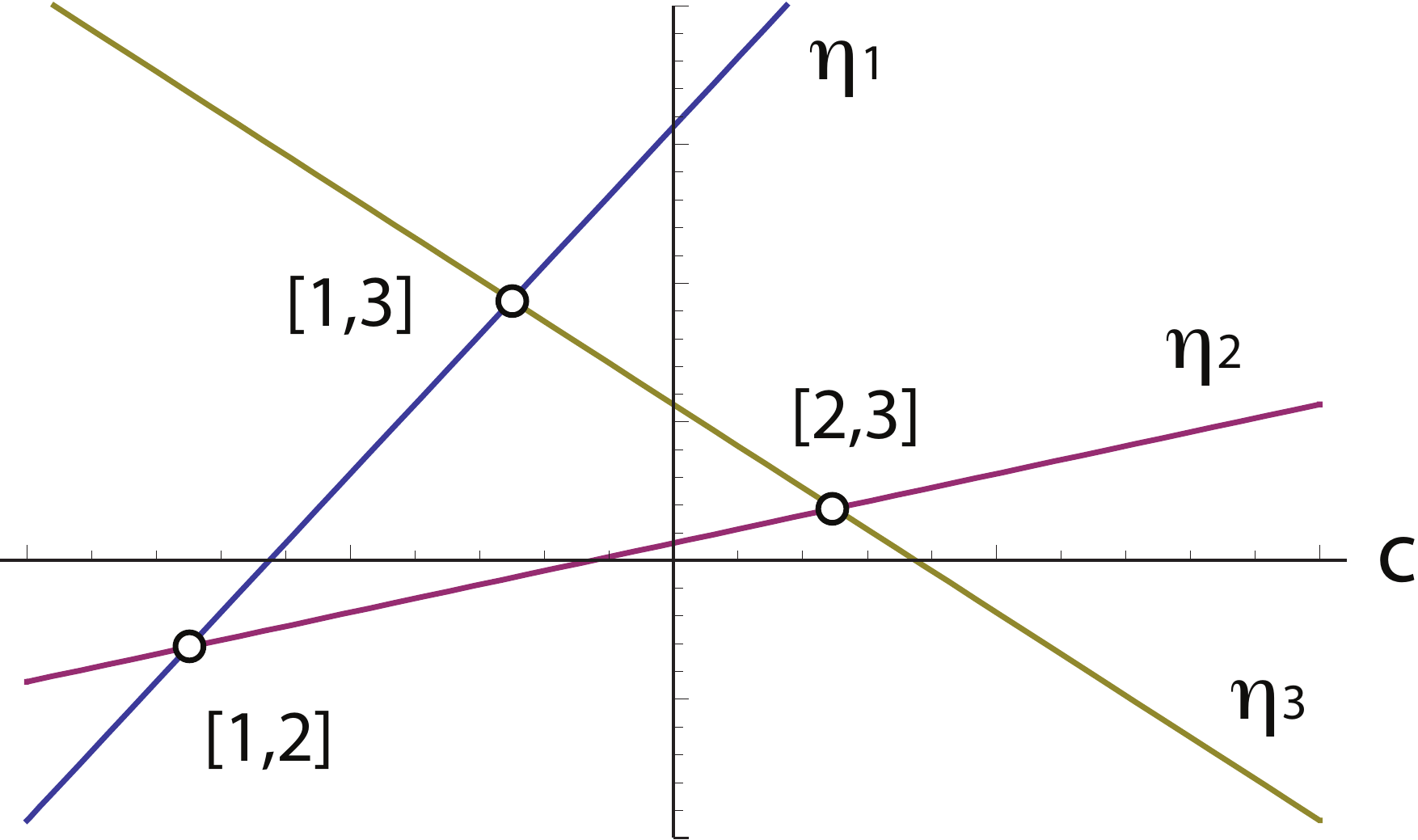}
\end{center}
\caption{The graphs of $\eta_j(c)=k_j(k_j-c)$. Each $[i,j]$ represents the exchange of the order
between $\eta_i$ and $\eta_j$.
The parameters are given by $(k_1,k_2,k_3)=
(-\sf{5}{4},-\sf{1}{4},\sf{3}{4})$.\label{fig:1}}
\end{figure}
%%%%%%%%%%%%%%%%%%%%%%%%%
It is then seen from Figure \ref{fig:1}  that for $y\gg 0$ and a fixed $t$, the exponential term
$E_1$ dominates when $c$ is large positive ($\Psi\approx \frac{\pi}{2}$, i.e. $x\to-\infty$).
Decreasing the value of $c$ (rotating the line clockwise), the dominant term changes to $E_3$.
Thus we have
\begin{equation*}\label{ypositive}
w:=\partial_x\,\ln f\longrightarrow\left\{\begin{array}{lll}
k_1\quad& {\rm as}\quad& x\to -\infty,\\
k_3\quad &{\rm as}\quad &x\to\infty.
\end{array}\right. 
\end{equation*}
The transition of the dominant exponentials $E_1\to E_3$ 
is characterized by the condition $\eta_1 = \eta_3$, which corresponds the 
direction parameter value $c=\tan\Psi_{[1,3]}=k_1+k_3$. In the neighborhood of this line, 
the function $f$ can be approximated as
\[
 f\approx E_1+bE_3,
 \]
which implies that there exists a $[1,3]$-soliton for $y\gg 0$. The constant $b$ can be used to choose
a specific location of this soliton.

Next consider the case of $y\ll 0$. The dominant exponential corresponds to the
{\it least} value of $\eta_j$ for any given value of $c$.
For large positive $c$ ($\Psi\approx\frac{\pi}{2}$, i.e. $x\to \infty$),
$E_3$ is the dominant term. Decreasing the value of $c$ (rotating the line $x=-cy$ clockwise),
the dominant term changes to $E_2$ when $k_2+k_3>c>k_1+k_2$,
and $E_1$ becomes dominant for $c< k_1+k_2$. Hence, we have for $y\ll 0$
\begin{equation*}\label{ynegative}
w=\partial_x\ln f\longrightarrow\left\{\begin{array}{llll}
k_1\quad &{\rm as}\quad &x\to-\infty,\\
k_2\quad &{\rm for}\quad  & -(k_1+k_2)y<x<-(k_2+k_3)y, \\
k_3\quad &{\rm as}\quad &x\to\infty.
\end{array}\right.
\end{equation*}
In the neighborhood of the line $x+(k_1+k_2)y=$constant, 
\[
f\approx E_1+aE_2,
\]
which corresponds to a $[1,2]$-soliton and its location is fixed by the constant $a$.
The solution in $y\ll 0$ also consists of a $[2,3]$-soliton in the neighborhood
of the line $x+(k_2+k_3)y=$constant, and whose location is determined by the 
locations of other line-solitons. Therefore, we need only two parameters $a,\,b$
(besides the $k$-parameters) to specify the solution uniquely, and those parameters fix
 the locations of line $x+y\tan\Psi_{[i,j]}-C_{[i,j]}t=x_{[i,j]}^0$ for $[i,j]$-soliton.  For $[i,j]=[1,3]$ and $[2,3]$ in $x\gg 0$, we have
 \[
 x_{[1,3]}=-\frac{1}{k_3-k_1}\ln b,\qquad x_{[2,3]}=-\frac{1}{k_3-k_2}\ln\frac{b}{a}.
 \]
The shape of solution generated by $f= E_1+aE_2+bE_3$ with $a=b=1$ (i.e. at $t=0$ three
line-solitons meet at the origin)
is illustrated via the contour plot in Figure \ref{fig:2a}.
%%%%%%%%%%%%%%%%%%%%%%%%%%%
\begin{figure}[t]
%\begin{center}
\includegraphics[scale=0.45]{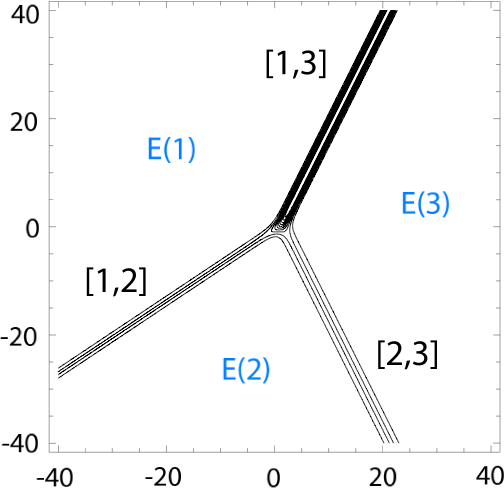}  \hskip 1.5cm
\raisebox{0.45cm}{\includegraphics[height=2.4cm]{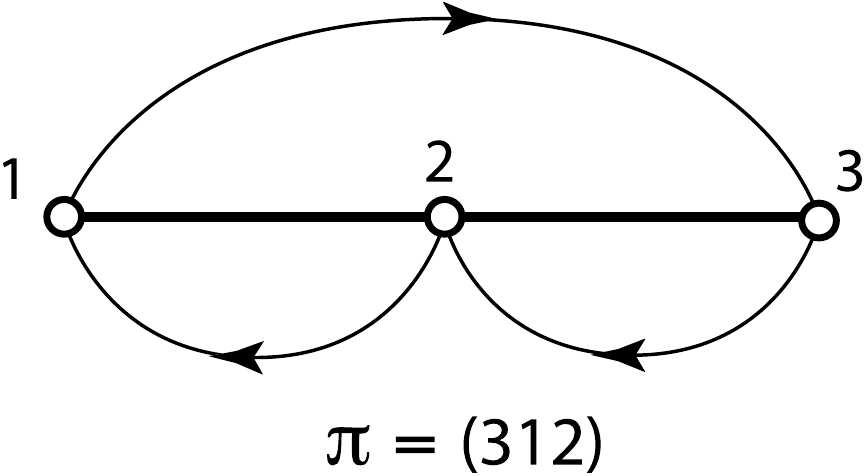}} 
\caption{Example of $(2,1)$-soliton solution and the chord diagram.
The $k$-parameters are chosen as $(k_1,k_2,k_3)=(-\frac{5}{4},-\frac{1}{4},\frac{3}{4})$.
The right panel is the corresponding chord diagrams.
We take the $A$-matrix $A=(1~1~1)$
so that at $t=0$ three line-solitons meet at the origin.
Each $E(j)$ with $j=1,2$ or $3$ indicates the dominant exponential
term $E_j$ in that region. The boundaries of any two adjacent regions give 
the line-solitons indicating the transition
of the dominant terms $E_j$. The $k$-parameters are the same as those
in Figure \ref{fig:1}, and the line-solitons are determined from
the intersection points of the $\eta_j(c)$'s in Figure \ref{fig:1}. 
Here $a=b=1$ (i.e. $\tau=E_1+E_2+E_3$) so that the three solitons 
meet at the origin at $t=0$.\label{fig:2a}}
\end{figure}
%%%%%%%%%%%%%%%%%%%%%%%%%%%%%%%%
%\end{Example}
In this Figure, one can see that the line-soliton in $y\gg 0$ labeled by $[1,3]$,
is localized along the line $\theta_1 = \theta_3$ 
with direction parameter $c = k_1+k_3$; two other line-solitons in $y\ll 0$ labeled by 
$[1,2]$ and $[2,3]$ are localized respectively, along the phase transition lines 
with $c=k_1+k_2$ and $c=k_2+k_3$.
This solution represents a resonant solution of three line-solitons. The resonant condition among those three line-solitons is
given by
\begin{align*}
{\bf K}_{[1,3]}={\bf K}_{[1,2]}+{\bf K}_{[2,3]},\qquad \Omega_{[1,3]}=\Omega_{[1,2]}+\Omega_{[2,3]},
\end{align*}
which are trivially satisfied with 
${\bf K}_{[i,j]}=(k_j-k_i,k_j^2-k_i^2)$ and $\Omega_{[i,j]}=
k_j^3-k_i^3$.
The resonant condition may be symbolically written as
\[
[1,3]=[1,2]+[2,3].
\]
One can also represent this line-soliton solution by a permutation of
three indices: $\{1, 2, 3\}$ which is illustrated by a (linear) {\it chord diagram}
shown below.
Here, the upper chord represents the $[1,3]$-soliton in $y\gg 0$ and 
the lower two chords represent $[1,2]$ and $[2,3]$-solitons in $y\ll 0$. 
Following the arrows in the chord diagram, one recovers the permutation,
\[
\pi=\begin{pmatrix}1 & 2& 3\\3 &1&2\end{pmatrix} \qquad {\rm or~simply}\quad \pi=(312).
\]
In general, each line-soliton solution of the KP equation can be parametrized 
by a {\it unique} permutation corresponding to a chord diagram (see the next subsection). 
\end{Example}

The results described in this example can be easily extended to the
general case where $f$ has arbitrary number of exponential terms (see also \cite{Me:02, BK:03}). 
\begin{Proposition} \label{Mm11sol}
If $f=a_1E_1+a_2E_2+\cdots+a_ME_M$ with $a_j > 0$ for $j=1,2,\ldots,M$, then the 
solution $u$ consists of $M-1$ line-solitons for $y\ll 0$ and one 
line-soliton for $y\gg 0$.
\end{Proposition} 
Such solutions are referred to as the $(M-1,1)$-soliton solutions; meaning
that $(M-1)$ line-solitons for $y\ll 0$ and one line-soliton for $y\gg 0$. Note that the 
line-soliton for $y\gg 0$ is labeled by $[1,M]$, whereas the other 
line-solitons in $y\ll 0$ are labeled by $[k,k+1]$ for $k=1,2,\ldots,M-1$,
counterclockwise from the negative to the positive $x$-axis, i.e.
increasing $\Psi$ from $-\sf{\pi}{2}$ to $\sf{\pi}{2}$. As in the previous
examples one can set $a_1 =1$ without any loss of generality, then the remaining
$M-1$ parameters $a_2, \ldots, a_M$ determine the locations of the $M$ line-solitons.
Also note that the $xy$-plane is divided into $M$ sectors for the asymptotic region with $x^2+y^2\gg0$,
and the boundaries of those sectors are given by the asymptotic line-solitons.
This feature is common even for the general case.

%%%%%%%%%%%%%%%%%%%%%%%%%%%%%%
\begin{figure}[t]
\begin{center}
\includegraphics[height=6.5cm]{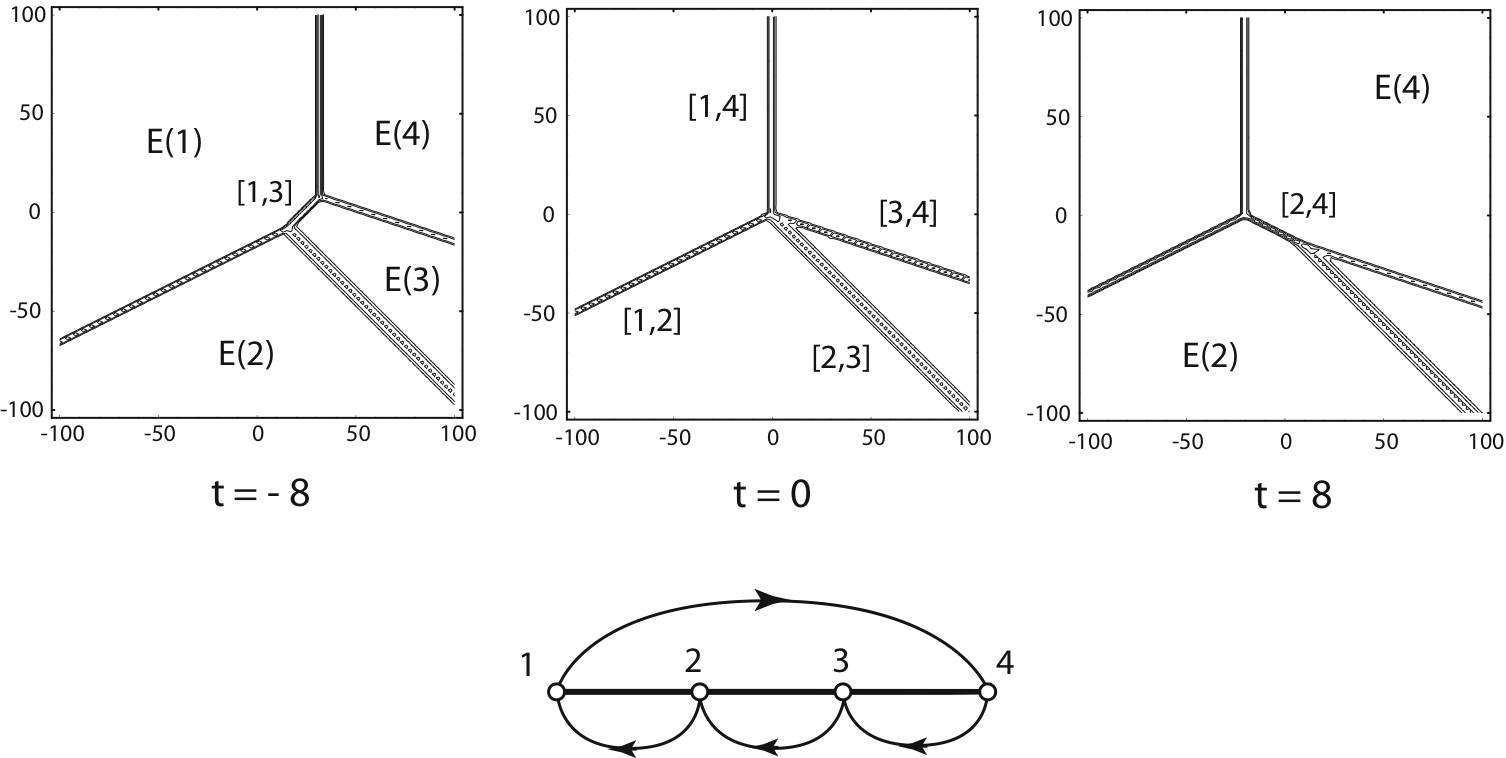}
\caption{The time evolution of a $(3,1)$-soliton solution and the corresponding chord diagram.
The upper chord represents the $[1,4]$-soliton, and the lower ones 
represent $[1,2]$-, $[2,3]$- and $[3,4]$-solitons.
The chord diagram shows $\pi=(4123)$.
The intermediate solitons are $[1,3]$-soliton at $t=-8$ and
[2,4]-soliton at $t=8$, respectively. These solitons appear
as the resonant $(2,1)$-type solutions.\label{fig:31soliton}}
\end{center}
\end{figure}
%%%%%%%%%%%%%%%%%%%%%%%%%%%%%%%%%%
Figure \ref{fig:31soliton} illustrates the case for a $(3,1)$-soliton solution
with $f= E_1+E_2+E_3+E_4$. The chord diagram for this solution
represents the permutation $\pi = (4123)=s_4s_3s_2\in S_4^{(1)}$.

\begin{Example}\label{21soliton}
Let us now consider the case with $N=2$ and $M=3$: We take the $A$-matrix in 
\eqref{f} of the form,
\[
A=\begin{pmatrix}
1 &0 &-b \\ 0& 1 & a
\end{pmatrix}.
\]
where $a$ and $b$ are positive constants, that is, $A$ marks a point on Gr$^+(2,3)$.
Then the $\tau$-function is given by

\[
\tau=E(1,2)+aE(1,3)+bE(2,3).
\]
In order to carry out the asymptotic analysis in this case one needs to 
consider the sum of two $\eta_j(c)$, i.e. $\eta_{i,j}=\eta_i+\eta_j$ for 
$1\le i<j\le 3$. This can still be done using Figure \ref{fig:1},
but a more effective way is described below (see the graph of $\eta(k,c)$
in Figure \ref{fig:eta24}).

 For $y\gg 0$, the transitions of the dominant exponentials are given by following scheme:
$$
E(1,2)\longrightarrow E(1,3) \longrightarrow E(2,3),
$$ 
as $c$ varies from large positive (i.e. $x\to-\infty$) to large negative values (i.e. $x\to\infty$). 
The boundary between the regions with the dominant exponentials $E(1,2)$ and $E(1,3)$
defines the $[2,3]$-soliton solution since here the $\tau$-function can be approximated as
\begin{align*}
\tau&\approx E(1,2)+aE(1,3)=2(k_2-k_1)e^{\theta_1+\sf{1}{2}(\theta_2+\theta_3-\theta_{23})}\cosh\sf{1}{2}(\theta_2-\theta_3+\theta_{23}),
\end{align*}
so that we have
\[
u=2\partial_x^2\ln \tau \approx \sf{1}{2}(k_2-k_3)^2 
\mathrm{sech}^2\sf{1}{2}(\theta_2-\theta_3+\theta_{23})\,,
\]
where $\theta_{23}$ is related to the parameter of the $A$-matrix (see below).
A similar computation
as above near the transition boundary of the dominant exponentials $E(1,3)$ and $E(2,3)$
yields
\begin{align*}
\tau &\approx 2(k_3-k_1)a e^{\theta_3+\sf{1}{2}(\theta_1+\theta_2-\theta_{12})}\cosh
\sf{1}{2}(\theta_1-\theta_2+\theta_{12}).
\end{align*}
The phases $\theta_{12}$ and $\theta_{23}$ are related to the parameters of the $A$-matrix,
\[
a=\frac{k_2-k_1}{k_3-k_1}e^{-\theta_{23}},\qquad  b=\frac{k_2-k_1}{k_3-k_2}e^{-\theta_{12}}.
\]
%%%%%%%%%%%%%%%%%%%%%%%%%%%%%%%
\begin{figure}
\begin{center}
\includegraphics[scale=0.45]{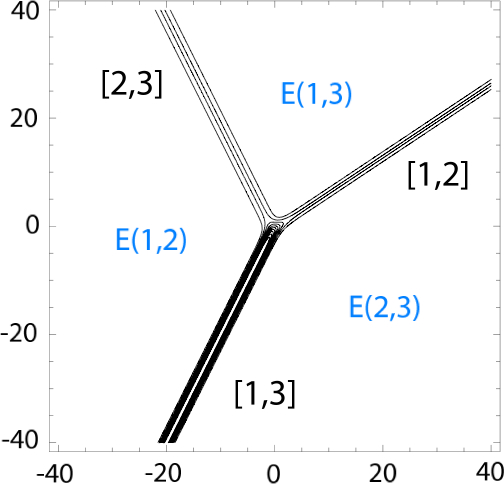} \hskip 1.5cm
\raisebox{0.45cm}{\includegraphics[height=2.4cm]{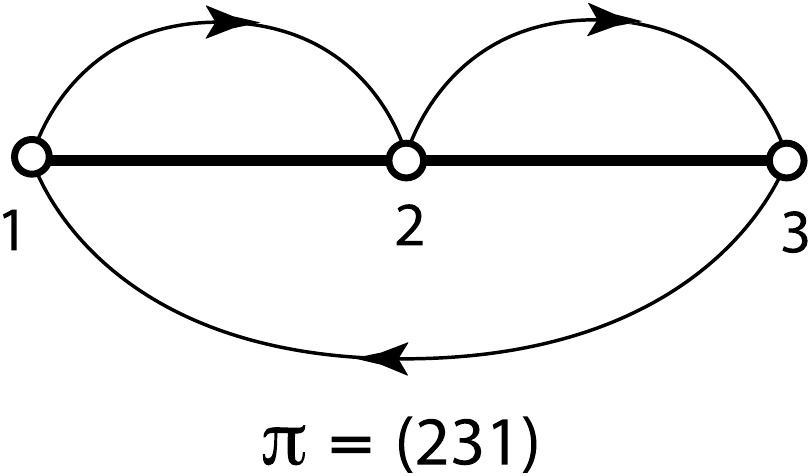}} 
\caption{Example of $(1,2)$-soliton solution and the chord diagram.
The $k$-parameters are the same as the $(2,1)$-soliton in the previous figure.
The right panels are the corresponding chord diagrams.
The parameters in the $A$-matrices are chosen  as $a=\frac{1}{2}$ and $b=1$
so that at $t=0$ three line-solitons meet at the origin.}\label{fig:2}
\end{center}
\end{figure}
%%%%%%%%%%%%%%%%%%%%%%%%%%%%%%%%
For $y\ll 0$,
there is only one transition, namely
$$
E(2,3) \longrightarrow E(1,2),
$$
as $c$ varies from large positive value (i.e. $x\to\infty$) to large negative 
value (i.e. $x\to-\infty$). 
In this case, a $[1,3]$-soliton is formed for $y\ll 0$ at the boundary of the
dominant exponentials $E(2,3)$ and $E(1,2)$.  The contour plot
of the line-soliton solution is shown in Figure \ref{fig:2}. 
Notice that this figure can be obtained from Figure \ref{fig:2a} by changing $(x,y)\to(-x,-y)$.
This solution can be represented by the chord diagram corresponding to the permutation 
$\pi=(231)$ shown below. 
Note that this diagram is the {\it $\pi$-rotation} of the chord diagram in Example \ref{12soliton}
whose permutation $\pi=(312)$ is the inverse of $\pi=(231)$.

\end{Example}

   As shown in those examples, it is now clear that each line-soliton appears as a boundary of 
   two dominant exponentials, and with the condition that $k_i+k_j$ are all distinct for $i\ne j$,
   we have the following Proposition:
\begin{Proposition}\label{TwoDominant}
Two dominant exponentials of the $\tau$-function in adjacent
regions of the $xy$-plane are of the form $E(i,m_2,\dots,m_N)$
and $E(j,m_2,\dots,m_N)$ for some
$N-1$~common indices ${m_2}, \ldots,{m_N}$.
\label{ij}
\end{Proposition}
As a consequence of Proposition \ref{ij}, the KP solution behaves asymptotically
like a single line-soliton 
\begin{equation}
u(x,y,t) \simeq \sf12(k_j-k_i)^2 \mathrm{sech}^2\sf12(\theta_j-\theta_i+\theta_{ij}) \,,
\label{uasymp}
\end{equation}
in the neighborhood of the line $x+(k_i+k_j)y = \mathrm{constant}$, which forms
the boundary between the regions of dominant exponentials $E(i,m_2,\ldots,m_N)$ and 
$E(j,m_2,\ldots,m_N)$.
Equation (\ref{uasymp}) defines an {\it asymptotic} line-soliton, i.e.
$[i,j]$-soliton, as a result of those two dominant exponentials. 
 In order to identify the set of asymptotic line-solitons associated
with a given solution, we need to determine which exponential terms
$E(m_1,m_2,\ldots,m_N)$ are actually dominant along each line 
$[i,j]:\, x = -(k_i+k_j)y$ as 
$|y| \to \infty$. For this purpose, first note that along a line $x=-cy$ 
each exponential term $E(m_1,m_2,\ldots,m_N)$ has the form,
$$
E(m_1,m_2,\ldots,m_N) \propto \exp\left(\sum_{n=1}^N\eta_{m_n}(c)y \right)\,, 
\qquad \eta_m(c) = k_m(k_m-c)\,.
$$
Thus for $y\gg 0$ (or $\ll 0$), 
the dominant exponential corresponds to the largest (or least) value of the
sum of $\eta_{m_n}(c)$ for each $c$. When two dominant exponentials $E(i,m_2,\dots,m_N)$
and $E(j,m_2,\dots,m_N)$ are in balance along the direction of the $[i,j]$-soliton,
we have $\eta_i(c)=\eta_j(c)$ which implies that $c=k_i+k_j$.
Since $\eta_m(c)-\eta_i(c) = (k_m-k_i)(k_m+k_i-c)$ and the $k$-parameters are ordered 
as $k_1<k_2<\cdots<k_M$, we have
the following order relations among the other $\eta_m(c)$'s along $c=k_i+k_j$,
\begin{equation}\label{dominant}
\left\{\begin{array}{lll}
\eta_i=\eta_j<\eta_m \quad {\rm if}\quad  m<i~{\rm or}~ j<m,\\[1.0ex]
\eta_i=\eta_j>\eta_m\quad {\rm if}\quad i<m<j.
\end{array}\right.
\end{equation}
The relations among the phases $\eta_j(c)$ can be seen easily from
the plots of $\eta_j(c)$ versus $c$ as well as $\eta(k,c)=k(k-c)$ versus $k$ 
for a fixed value of $c$ illustrated by Figure \ref{fig:eta24}.
Proposition \ref{ij} and the relations \eqref{dominant} are particularly useful in order to find 
the asymptotic line-solitons from a given KP $\tau$-function as demonstrated by the
example below.
%%%%%%%%%%%%%%%%%%%%%%%%%%%%
\begin{figure}
%\begin{center}
\includegraphics[height=4cm]{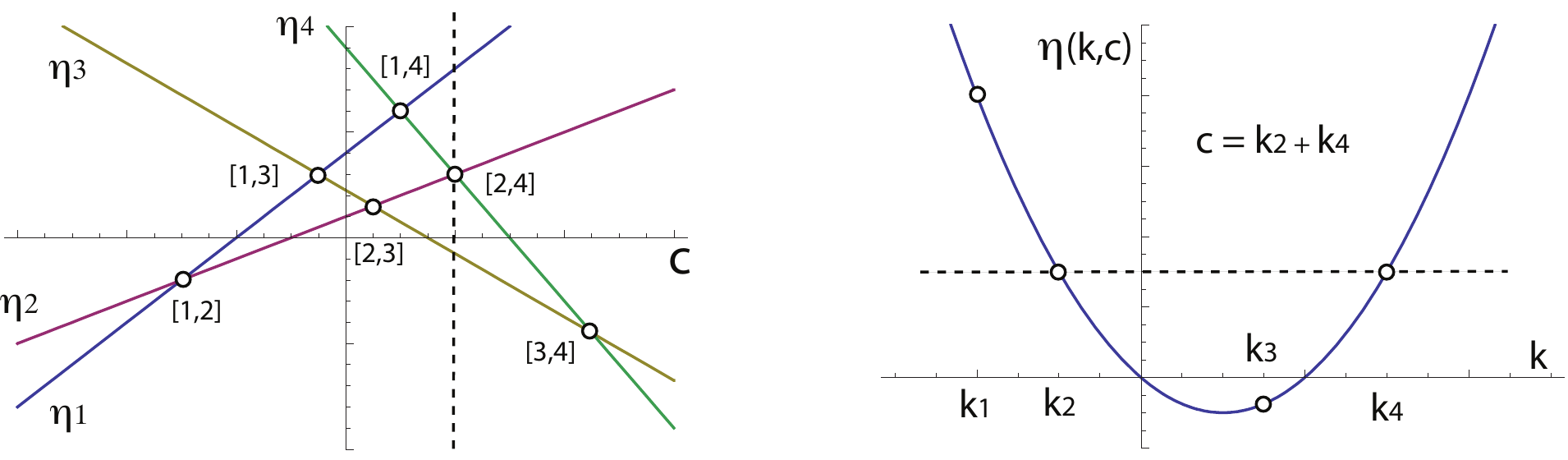}
\caption{The left figure shows $\eta_j(c)=k_j(k_j-c)$ for $j=1,\ldots,4$.
Each $[i,j]$ at the intersection point of $\eta_i(c)=\eta_j(c)$ corresponds to $c=k_i+k_j$.
We assume that the parameters $k_j$ are generic so that there is at most 
one intersection point for each $c$.
The right figure is a plot of $\eta(k,c)$ as a function of $k$ with $c=k_2+k_4$, 
that is, the $\eta$ along the dotted line in the left figure passing through 
the intersection point $\eta_2(c)=\eta_4(c)$.
This figure shows the order $\eta_3<\eta_2=\eta_4<\eta_1$.\label{fig:eta24}}
%\end{center}
\end{figure}
%%%%%%%%%%%%%%%%%%%%%%%%%%%%%
\begin{Example}
Let us consider the $2 \times 4$ matrix,
\begin{equation*}\label{ExA}
A=\begin{pmatrix}
1 & 0 & 0 &-a\\
0 & 1 & b & c
\end{pmatrix}\,,
\end{equation*}
where $a, b, c$ are positive real numbers. In this case, there are 
six maximal minors, five of which are positive, namely,
\[
\xi(1,2)=1,\quad \xi(1,3)=b,\quad\xi(1,4)=c,\quad\xi(2,4)=a,\quad\xi(3,4)=ab\,, 
\]
and $\xi(2,3)=0$. Then from \eqref{tauexp} the $\tau$-function has the form,
\begin{align*}
\tau=&(k_2-k_1)E(1,2)+b(k_3-k_1)E(1,3)+c(k_4-k_1)E(1,4) \\
&+ a(k_2-k_4)E(2,4)+ab(k_4-k_3)E(3,4)\,.
\end{align*}
Proposition \ref{ij} implies that the line solitons are localized along the
lines $x+cy=$constant with $c=k_i+k_j=\tan\Psi_{[i,j]}$. Hence, we look for dominant 
exponential terms in the $\tau$-function along those directions. 
For $y\gg0$, the $c$-values decrease as we sweep clockwise from 
negative to positive $x$-axis starting with the largest value 
$c = k_3+k_4$. We have $\eta_1 , \eta_2 > \eta_3=\eta_4$ from 
the order relations \eqref{dominant} for $c = k_3+k_4$. This
means that $\eta_1+\eta_2$ is the dominant phase combination along this direction. 
Since $\xi(1,2)\neq 0$, $\tau(x,y,t) \approx (k_2-k_1)E(1,2)$ implying
that $u \approx 0$ along the line $[3,4]$, so there is no 
$[3,4]$ line-soliton. By similar reasoning one can verify that
the $[1,4]$- and $[1,2]$-solitons are also impossible. Let us consider the direction 
$c=k_2+k_4$ to check for the $[2,4]$-soliton. From (\ref{dominant}) 
(see also Figure \ref{fig:eta24}), 
$\eta_3<\eta_2=\eta_4<\eta_1$, and since both $\xi(1,2)$ 
and $\xi(1,4)=a$ are nonzero, the $\tau$-function in \eqref{tauexp} corresponds to a dominant 
balance of exponentials:\, $\tau \approx (k_2-k_1)E(1,2) + c(k_4-k_1)E(1,4)$ along the line $[2,4]$.
Therefore $[2,4]$ corresponds to an asymptotic line-soliton as $y \to \infty$. 
The $[1,3]$-soliton also exists by a similar argument. Thus, we have two asymptotic 
line-solitons $[1,3]$- and $[2,4]$-types for $y\gg 0$.
%%%%%%%%%%%%%%%%%%%%%%%%%%%%%%%%%%%%%%%%%%%%
\begin{figure}[t!]
\centering
\includegraphics[scale=0.57]{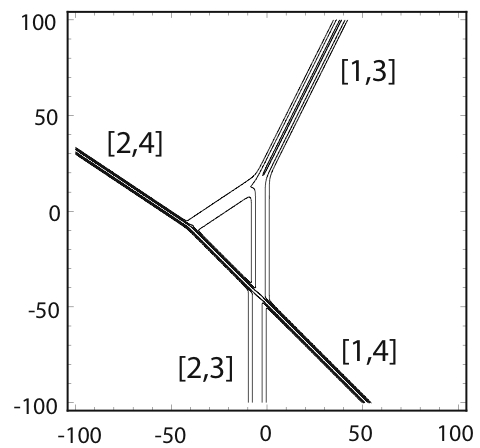}  \hskip 1.5cm
\includegraphics[scale=0.57]{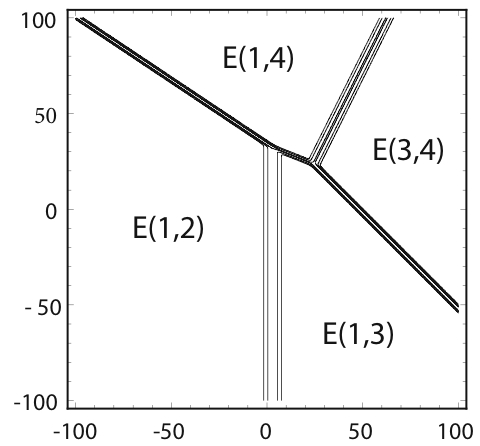} 
\caption{An example of $(2,2)$-soliton solution. The left left figure is
at $t=-16$ and the right one at $t=16$. The parameters $(k_1,\ldots,k_4)$ are
given by $(-1,-0.5,0.5,2)$. The intermediate solitons are $[3,4]$-type for $t=-16$ and
$[1,2]$-type for $t=16$. Note that the middle triangular section at $t=-16$ corresponds to the region with
the dominant exponential $E(2,4)$, and that all five non-zero exponential terms in the $\tau$-function appear
in the figure at $t=16$.}
\label{fig:3421}
\end{figure}
%%%%%%%%%%%%%%%%%%%%%%%%%%%%%%%%%%%%

We next look for the asymptotic solitons for $y\ll 0$ by sweeping from the negative 
$x$-axis to positive $x$-axis. Recall that in this case the dominant exponential 
$E(i,j)$ corresponds to the least value of the sum $\eta_i(c)+\eta_j(c)$.
It is easy to see that $[1,2]$ and $[3,4]$-solitons are impossible since $E(3,4)$ and 
$E(1,2)$ are respectively, the only dominant exponentials along those directions. 
Then consider the $[1,3]$-soliton. Along $c=k_1+k_3$, \eqref{dominant} implies that 
$\eta_2<\eta_1=\eta_3<\eta_4$, and so the exponentials $E(1,2)$ and $E(2,3)$ would give 
the dominant balance. But $E(2,3)$ is not present in the above $\tau$-function because 
$\xi(2,3)=0$. So we conclude that $[1,3]$-soliton does not exist as $y\ll 0$, and
for similar reasons, $[2,4]$-soliton is also impossible. Next, checking for the $[1,4]$-soliton, 
we have $\eta_2, \eta_3<\eta_1=\eta_4$ from \eqref{dominant}.
But as seen earlier, the dominant exponential $E(2,3)$ is {\it not} 
present in the $\tau$-function. However, there does exist a balance between the next dominant 
exponential pairs $\{E(1,3), E(3,4)\}$ or $\{E(1,2), E(2,4)\}$ depending on
whether $\eta_2>\eta_3$ or $\eta_2<\eta_3$. In either case, there exists an asymptotic 
line-soliton along $[1,4]$. A similar argument applies along the line $[2,3]$ which corresponds 
the other asymptotic line-soliton as $y \ll 0$.

In summary, the $\tau$-function corresponding to the $A$-matrix
given above, generates a KP solution with asymptotic line-solitons
$[1,3]$ and $[2,4]$ as $y \gg 0$, and asymptotic line-solitons
$[1,4]$ and $[2,3]$ as $y \ll 0$. This line-soliton solution 
with the parameters $a=b=c=1$ in the $A$-matrix is shown in Figure \ref{fig:3421}.

\end{Example}

We note that the line-solitons associated with the resonant $(1,2)$-
and $(2,1)$-soliton solutions  can be determined
in the same way as the above example by applying for the dominant balance
conditions given by Proposition \ref{TwoDominant} and \eqref{dominant}. We now proceed to
discuss a more general characterization of all line-soliton solutions of
the KP equation whose $\tau$-functions are given in the Wronskian form
\eqref{tau}.

\subsection{Characterization of the line-solitons}
It should be clear from the above examples that 
a dominant exponential term determined by the relations \eqref{dominant}
is actually {\it present} in the given $\tau$-function if 
its coefficient term given by a maximal minor of the $A$-matrix is non-zero. Thus,
in order to obtain a complete characterization 
of the asymptotic line solitons, it is necessary to consider the structure
of the $N \times M$ coefficient $A$-matrix in some detail. 
We consider the matrix $A$ to be in RREF, and we will also assume 
that $A$ is {\it irreducible} as defined below:
\begin{Definition} An $N\times M$ matrix $A$ is irreducible if each 
column of $A$ contains at least one nonzero element, or each row contains at least
one nonzero element other than the pivot once $A$ is in RREF.
\end{Definition}
If an $N\times M$ matrix $A$ is {\it not} irreducible, then 
the corresponding $\tau$-function gives the same KP
solution $u$ which is obtained from another $\tau$-function associated with a
smaller size matrix $\tilde{A}$ derived from $A$. 
One can notice from the determinant expansion in \eqref{tauexp} that
\begin{itemize}
\item[(a)] if the
$m$-th column of $A$ has only zero elements, then $\xi(m_1,\ldots,m_N)=0$ 
if  $m_k=m$ for some $k$, that is, the exponential $E_m$ will 
never appear in the $\tau$-function;  in terms of the chord diagram, this corresponds to
a loop in the lower part of the diagram ($m$ is a non-pivot index),
\item[(b)]
if the $n$-th row of $A$ has the pivot 
as the only non-zero element, then all $\xi(m_1,\ldots,m_N)\ne 0$ contains the index $n$,
that is, the exponential $E_n$ can be factored out from the 
$\tau$-function; in terms of the chord diagram, this corresponds to a loop in the upper part of the
diagram.
\end{itemize}
So the irreducibility implies that we consider only derangements (i.e. no fixed points) of the permutation.

\begin{figure}[t!]
%\begin{center}
\includegraphics[height=5.8cm]{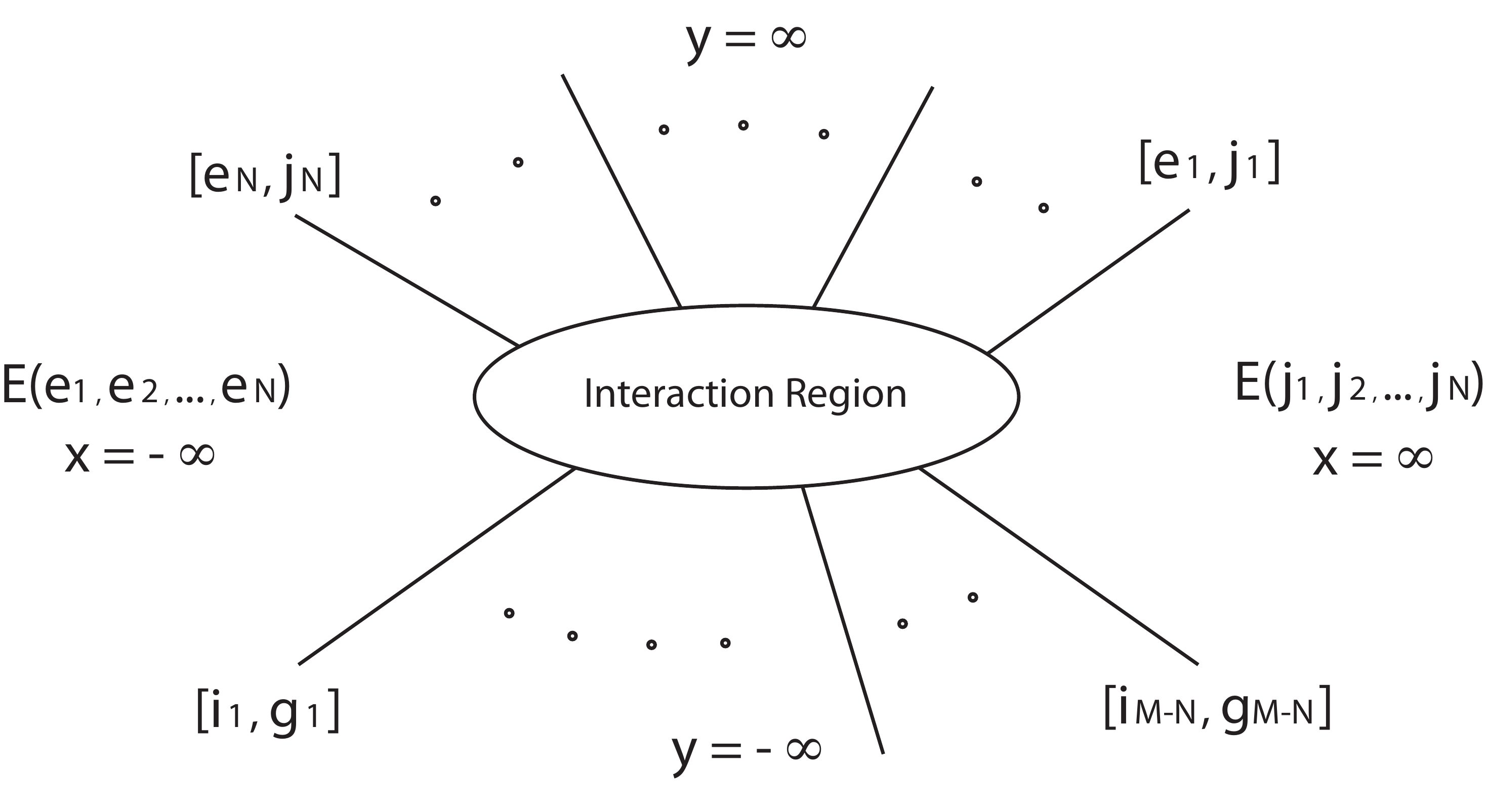}
\caption{$(N_-,N_+)$-soliton solution. The asymptotic line-solitons
are denoted by their index pairs $[e_n,j_n]$ and $[i_m,g_m]$. The sets $\{e_1,e_2,\ldots,e_{N}\}$ and
$\{g_1,g_2,\ldots,g_{M-N}\}$ indicate pivot and non-pivot indices, respectively.
Here $N_-=M-N$ and $N_+=N$ for the $\tau$-function on Gr$(N,M)$, and
$E(\cdot,\ldots,\cdot)$ represents the dominant exponential in that region.\label{NMsoliton}}
%\end{center}
\end{figure}

We now present a classification scheme of the line-soliton solutions
by identifying the asymptotic line-solitons as $y \to\pm \infty$. We denote a
line-soliton solution by $(N_-,N_+)$-soliton whose asymptotic form
consists of $N_-$ line-solitons as $y\to-\infty$ and $N_+$ line-solitons
for $y\to\infty$ in the $xy$-plane as shown in Figure \ref{NMsoliton}.
The next Proposition provides a general result characterizing
the asymptotic line-solitons of the $(N_-,N_+)$-soliton solutions (the proof can be found in \cite{CK:09}):
\begin{Proposition}
Let $\{e_1,e_2,\ldots,e_N\}$ and $\{g_1,g_2,\ldots,g_{M-N}\}$ denote respectively,
the pivot and non-pivot indices associated with
an irreducible, $N\times M$, TNN $A$-matrix. Then the soliton 
solution obtained from the $\tau$-function in (\ref{tauexp}) with this $A$-matrix has
the following structure:
\begin{itemize}
\item[(a)] For $y\gg 0$, there are $N$ asymptotic line-solitons of  $[e_n, j_n]$-type for some $j_n$.
\item[(b)] For $y\ll 0$, there are $(M-N)$ asymptotic line-solitons of $[i_m, g_m]$-type for some $i_m$.
\end{itemize}
\label{engn}
\end{Proposition}

\begin{figure}[t!]
%\begin{center}
\includegraphics[height=7cm]{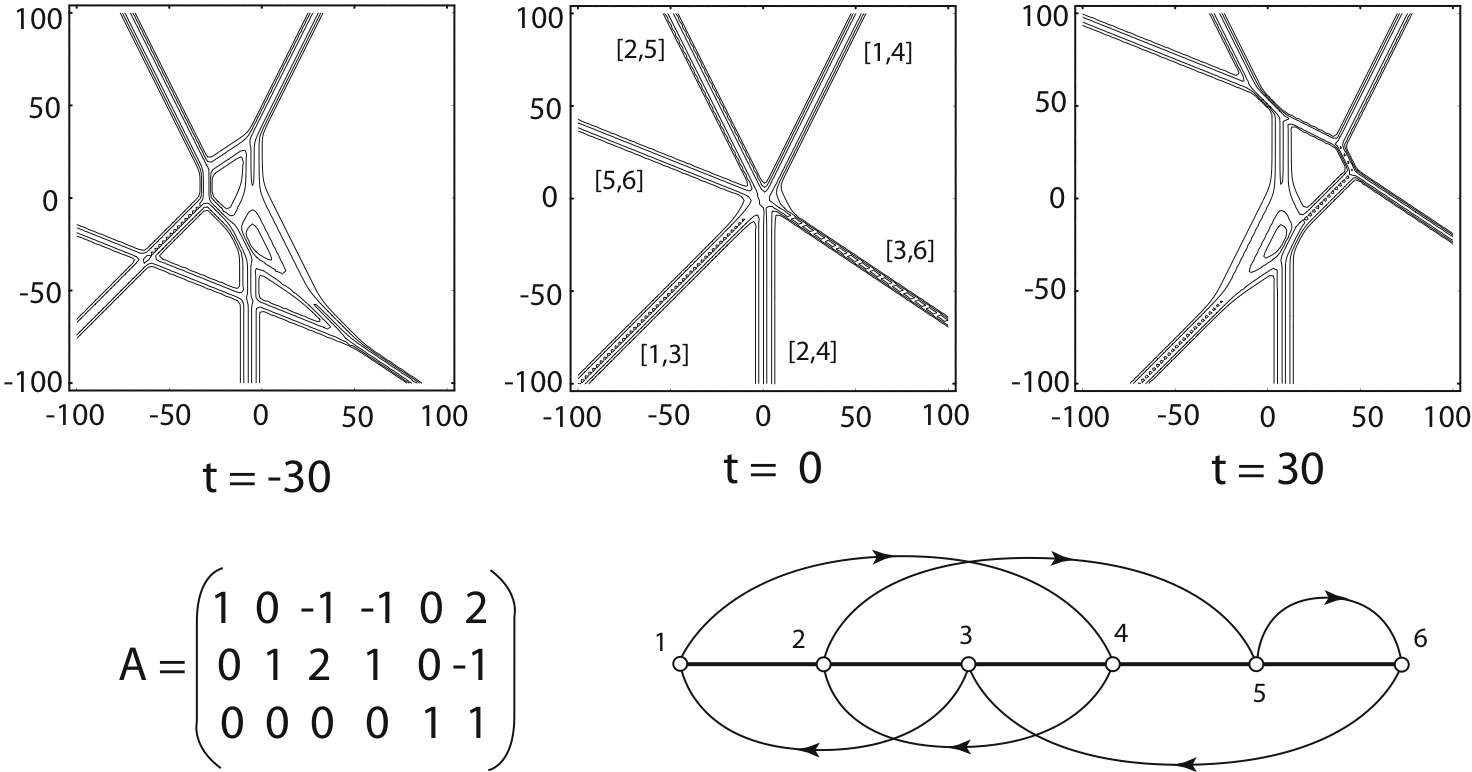}
\caption{An example of $(3,3)$-soliton solution. The permutation of this solution
is $\pi=(451263)$. The $k$-parameters are chosen as $(k_1,k_2,\ldots,k_6)=
(-1,-\sf{1}{2},0,\sf{1}{2},1,\sf{3}{2})$. The dominant exponential for $x\ll 0$ is $E(1,2,5)$, and 
each dominant exponential
is obtained through the derangement representing the solution, e.g. after crossing $[5,6]$-soliton
in the clockwise direction,
the dominant exponential becomes $E(1,2,6)$. That is, $[5,6]$-soliton is given by
the balance of those two exponentials. \label{fig:33soliton}}
%\end{center}
\end{figure}

An important consequence of Proposition \ref{engn} is 
that it defines the {\it pairing} map $\pi: [M] \to [M]$ 
on the integer set $[M] := \{1,2,\ldots,M\}$ according to  
\begin{equation}
\left\{\begin{array}{lll}
\pi(e_n) &= j_n\,,\quad & n=1,2,\ldots,N\,, \\[1.0ex]
\pi(g_m) &= i_m\,, \quad &m=1,2,\ldots,M-N\,.
\end{array}\right.
\label{pi}
\end{equation}
Recall that $\{e_n\}_{n=1}^N$ and $\{g_m\}_{m=1}^{M-N}$ are respectively, 
the pivot and non-pivot indices of the $A$-matrix and form a disjoint partition 
of $[M]$. Then the unique index pairings in Proposition \ref{engn} imply that
the map $\pi$ is a {\it permutation} of $M$ indices. More precisely,
$\pi \in \mathcal{S}_M$
where $\mathcal{S}_M$ is the group of permutations of the index set $[M]$.
Furthermore, since $\pi(e_n)=j_n > e_n,\, n=1,\ldots,N$ and 
$\pi(g_m) = i_m < g_m, \, m=1,\ldots,M-N$, $\pi$ defined by (\ref{pi}) is a 
permutation with no fixed point, i.e. {\em derangements}. Yet another feature of $\pi$ is that 
it has exactly $N$ {\it excedances} defined as follows: an element $l \in [M]$ is an
{\em excedance} of $\pi$ if $\pi(l) > l$. The excedance set of $\pi$ in (\ref{pi})
is the set of pivot indices $\{e_1, e_2, \ldots, e_N\}$. 
The above results can be summarized to deduce the following 
characterization for the line-soliton solution of the
KP equation \cite{CK:09}.
\begin{Theorem}\label{Main}
\label{derangement}
Let $A$ be an $N\times M$, TNN, irreducible matrix which corresponds to
a point in the non-negative Grassmannian Gr$^+(N,M) \subset$ Gr$(N,M)$.
Then the $\tau$-function (\ref{tauexp}) associated with this $A$-matrix 
generates an $(M-N,N)$-soliton solutions. The $M$ asymptotic line-solitons
associated with each of these solutions can be identified via a pairing map $\pi$ 
defined by \eqref{pi}. The map $\pi \in \mathcal{S}_M$ is a derangement of the index 
set $[M]$ with $N$ excedances given by the pivot indices $\{e_1, e_2, \ldots, e_N\}$ 
of the $A$-matrix in RREF.
\end{Theorem}
As explained in Section \ref{sec:Gr},  the derangements $\pi \in \mathcal{S}_M$ are represented by the chord diagrams
with the arrows above the line pointing from $e_n$ to $j_n$ 
for $n=1,2,\ldots,N$, while arrows below the line point from $g_m$ to $i_m$ for 
$m=1,2,\ldots,M-N$. 
Figure \ref{fig:33soliton} illustrates the time evolution of an example of  $(3,3)$-soliton
solution. The chord diagram shows all asymptotic line-solitons for $y\to\pm\infty$.

Theorem \ref{Main} provides a unique parametrization of each TNN Grassmannian cell in terms of
the derangement of $S_M$. This agree with the result obtained by Postnikov et al in \cite{P:06, W:05}.
One should, however, note that
Theorem \ref{Main} does not give us the indices $j_n$ and $i_m$ in the $[e_n, j_n]$ and $[i_m, g_m]$ 
line-solitons. 
The specific conditions that an index pair $[i,j]$ identifies an asymptotic 
line-soliton are obtained by identifying the dominant exponential in each domain in the $xy$-plane.
The example below illustrates
how to apply  Theorem \ref{Main}, and identify
all the asymptotic line-solitons for a given irreducible
TNN $A$-matrix.

\begin{Example}\label{ex:mark}
Let us consider the $3\times 5$ matrix,
\[
A= \begin{pmatrix}
1 &0  &-a &0  & b\\ 
0 &1 &c &0 & -d \\
0 & 0 & 0 & 1 & e 
\end{pmatrix}\qquad {\rm with}\quad ad-bc=0,
\]
where $a,b,c,d$ and $e$ are positive constants, that is, the $A$-matrix marks a point on Gr$^+(3,5)$.
Then the purpose is to find asymptotic line-solitons generated by the $\tau$-function \eqref{tau} 
associated with this $A$-matrix.
From Proposition \ref{engn}, one can see that the
$\tau$-function with this matrix will
produce a $(3,2)$-soliton solution since $N=3$ and $M=5$. Moreover,
the asymptotic line-solitons for this solution are labeled by
$[1,j_1], [2,j_2]$ and $[4,j_3]$ for $y \gg 0$ for some $j_1>1, 
j_2>2$ and $j_3>4$.  Similarly, the line-solitons
for $y \ll 0$ are labeled by $[i_1,3]$ and $[i_2,5]$ for
some $i_1<3$ and $i_2<5$.
The basic idea to determine those indices $j_1,j_2,j_3$ and $i_1,i_2$
is to apply Proposition \ref{engn} and the dominant relations \eqref{dominant}.

Let us first consider the case for $y\gg 0$.
Starting with the last pivot $e_3=4$, it is immediate to find $j_3=5$, because of $j_3>4$
(just Proposition \ref{engn}).  We now take the next pivot $e_2=2$ and find the index $j_2$.
Since the index 5 is already taken as the pair index of $e_3=4$, we need to check only 
the cases $[2,4]$ and $[2,3]$.  For the existence of $[2,4]$-soliton, the dominant relation \eqref{dominant} requires that both $\xi(1,2,5)$ and $\xi(1,4,5)$ are not zero.
Calculating those minors for our $A$-matrix, we have
\[
\xi(1,2,5)=e \ne0,\qquad \xi(1,4,5)=d\ne 0,
\]
and hence $[2,4]$-soliton exists.  Now we consider the case with $e_1=1$, that is, we have
only $[1,2]$ and $[1,3]$ possibility.  In the case of $[1,3]$, we use again the dominant relation \eqref{dominant}, and check the minors $\xi(1,4,5)$ and $\xi(3,4,5)$ which corresponds to
the dominant exponentials.  We then find $\xi(1,4,5)=d\ne 0$ but $\xi(3,4,5)=bc-ad=0$.
This implies that $[1,3]$-soliton is impossible for $y\gg 0$. So the last one is $[1,2]$-type, which
can be confirmed by the condition $\xi(1,4,5)=d\ne 0$ and $\xi(2,4,5)=b\ne 0$.

Now we consider the case for $y\ll 0$.  Theorem \ref{Main} tells us that for the non-pivot index
$g_1=3$, only the pair $[1,3]$ is possible (the index 2 is already taken because [1,2]-soliton exists, i.e. $\pi(1)=2$).
Then the final soliton must be $[3,5]$-type from the non-pivot index $g_2=5$.  The last one can be confirmed by the least condition in \eqref{dominant} with $\xi(2,3,4)=a\ne 0$ and $\xi(2,4,5)=b\ne0$.

Thus we have a $(2,3)$-soliton solution of $\pi=(24153)$-type for the $\tau$-function \eqref{tau}
with the $A$-matrix considered.
The photos in Figure \ref{fig:Mark} show some interacting shallow water waves, which we think a realization of this example.  We demonstrate an exact solution whose parameters are given by 
$(k_1,k_2,\ldots,k_5)=(-2,-1,0,0.5,2)$ and the $A$-matrix  with  $(a,b,c,d,e)=(1,2,1,2,1)$.  
%%%%%%%%%%%%%%%%%%%%%%%%%%%%%%%%%%%%%%%
\begin{figure}[t!]
%\begin{center}
\includegraphics[height=8cm]{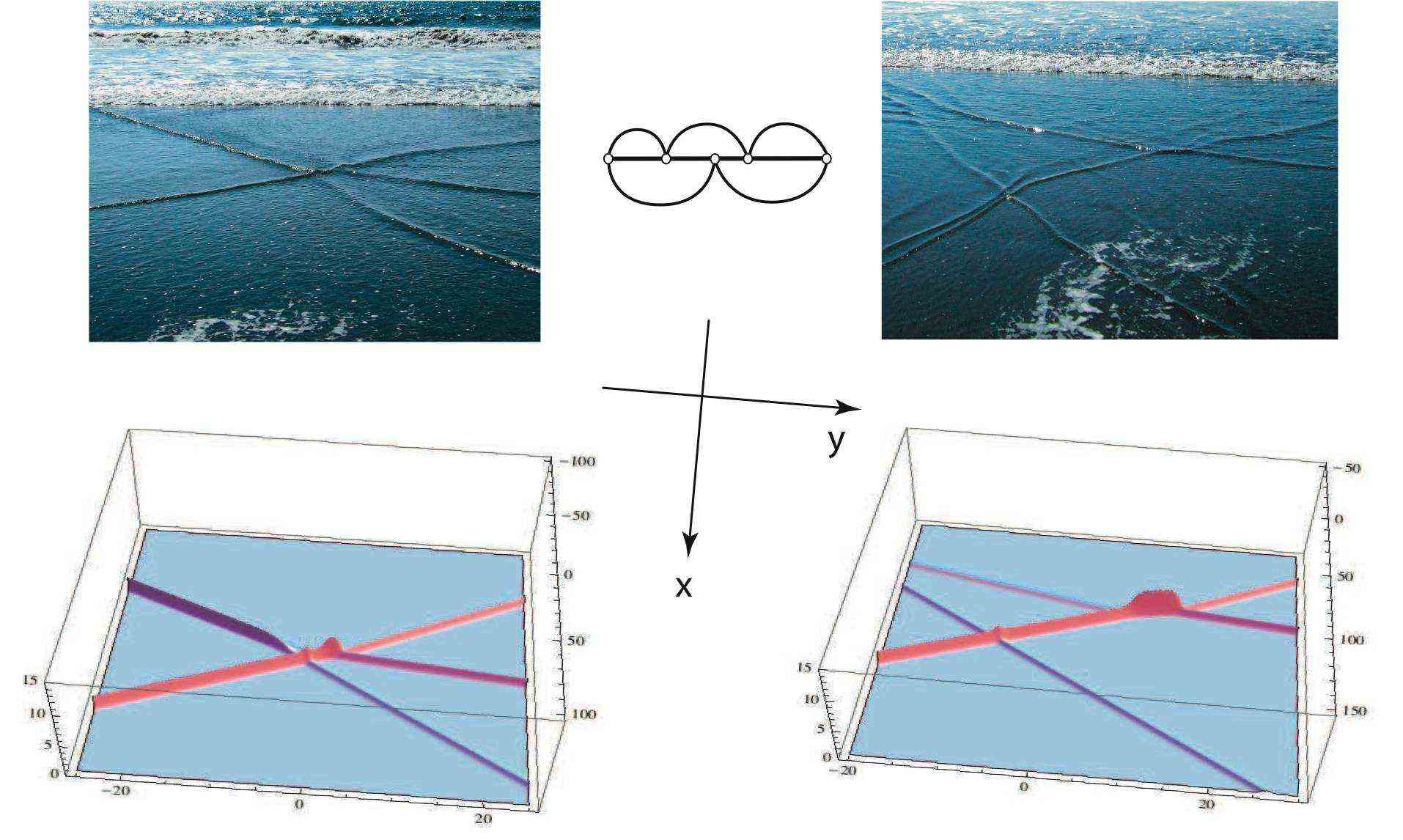}
\caption{Example of shallow water waves. The upper photos are taken at a beach in Mexico. The lower figures show the evolution of the
corresponding exact $(3,2)$-soliton solution of $(24153)$-type shown in the chord diagram
(see Example \ref{ex:mark}):
the left one at $t=1.5$ and the right one at $t=10$. 
(Photographs by courtesy of Mark J. Ablowitz.) \label{fig:Mark}}
%\end{center}
\end{figure}
%%%%%%%%%%%%%%%%%%%%%%%%%%%%%%%%%%%
\end{Example}

\section{$(2,2)$-soliton solutions}\label{sec:2-2S}
\begin{figure}[t]
\centering
\includegraphics[scale=0.5]{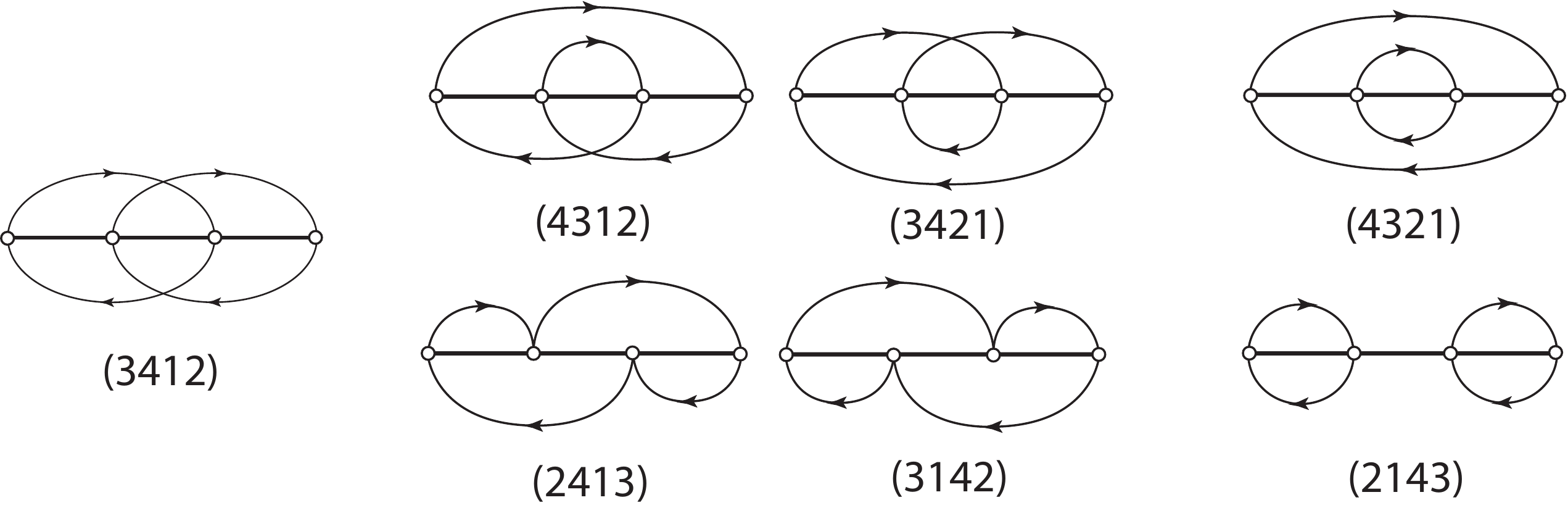}
\caption{The chord diagrams for seven different types
of $(2,2)$-soliton solutions. Each diagram corresponds to a totally non-negative Grassmannian
cell in Gr$(2,4)$.
\label{fig:chords}}
\end{figure}

Here we give a summary of all soliton solutions of the KP equation generated by the $2\times 4$
irreducible, TNN  $A$-matrices. Proposition \ref{engn} implies that each
of the soliton solutions consists of two asymptotic line-solitons as $y\to\pm\infty$.
That is, they are $(2,2)$-soliton solutions.  We outline below the classification scheme 
for the $(2,2)$-soliton solutions, and discuss some of the exact solutions in details for
the applications discussed in the following sections.
First note that for $2\times 4$ matrices, there are only two types  given by
\[
\begin{pmatrix}
1 & 0 & -c & -d \\
0& 1 & a & b
\end{pmatrix}\qquad {\rm and}\qquad
\begin{pmatrix}
1& a & 0 & -c\\
0 & 0 &1 & b
\end{pmatrix}\,,
\]
The fact that $A$ is TNN implies that the constants $a,b,c$ and $d$ must
be non-negative. For the first type, one can easily see that $ad=0$ is impossible because
then either $\xi(3,4)<0$ or $A$ is not irreducible. Then there are 5 possible cases 
with $ad\ne 0$, namely,
\[
\begin{array}{cccc}
(1)~\, ad-bc>0\,, \qquad (2)~\, ad-bc=0\,, \qquad (3)~\, b=0\,, c\ne 0\,,\\[1.0ex]
 (4)~\, c=0, b\ne 0\,, \qquad (5)~\, b=c=0\,.
\end{array}
\]
For the second type, $ab\ne 0$ due to irreducibility. Hence, we have only two cases:
\[
(6)~\, c\ne 0\,, \qquad \qquad (7)~\, c= 0 \,.
\]
Thus we have total seven different types of $A$-matrices, and using Theorem \ref{Main}, we can show that each $A$-matrix gives
a different $(2,2)$-soliton solution which can be enumerated according to the seven
derangements of the index set $[4]=\{1,2,3,4\}$ with two excedances.  Namely, for those cases from (1) to (7) we have
\begin{align*}
&(1)~\, \pi = (3412)\,,  \quad
(2)~\, \pi = (2413)\,, \quad (3)~\, \pi = (4312)\,, \quad (4)~\, \pi = (3421) \,, \\
& (5)~\, \pi = (4321) \,, \quad (6)~\, \pi = (3142) \,, \quad (7)~\, \pi = (2143)\,.
\end{align*}
In Figure \ref{fig:chords}, we show the chord diagrams for all those seven cases. One should note that
any derangement of $\mathcal{S}_4$ with exactly two excedances should be one of the graphs.
This uniqueness in the general case has been used to count the number of totally non-negative
Grassmann cells \cite{P:06, W:05}. 
%%%%%%%%%%%%%%%%%%%%%%%%%%%%%%%%%%%%%%%%%%
%%%%%%%%%%%%%%%%%%%%%%%%%%%%%%%%%%%

Let us now summarize the results for all those seven cases of the $(2,2)$-soliton solutions:

\begin{itemize}
\item[(1)] $\pi=(3412)$: This case corresponds to the {\it T-type} 2-soliton solution
which was first obtained as the solution of the {\it Toda} lattice hierarchy \cite{BK:03}.
This is why we call it ``T-type" (see also \cite{K:04}). The asymptotic line-solitons are
$[1,3]$- and $ [2,4] $-types for $|y|\to\infty$.  The $A$-matrix is given by
\[
A=\begin{pmatrix} 1&0&-c&-d\\0&1&a&b\end{pmatrix}\,,
\]
where $a,b,c,d>0$ are free parameters with $ad-bc>0$. This is the generic solution
on the maximum dimensional cell  of Gr$^+(2,4)$, and the corresponding line-soliton
has the most complicated pattern due to the fully resonant interactions among all line-solitons.
\item[(2)] $\pi=(2413)$: The asymptotic line-solitons are given by
$[1,2]$- and $[2,4]$-solitons for $y\gg 0$, and $[1,3]$- and $[3,4]$-solitons for $y\ll0$.
The $A$-matrix is given by
\[
A=\begin{pmatrix} 1&0&-c&-d\\0&1&a&b\end{pmatrix}\,,
\]
where $a,b,c,d>0$ with $\xi(3,4)=ad-bc=0$. Note the change of the solution structure by imposing  just one constraint $\xi(3,4)=0$ to the previous case (1).
\item[(3)] $\pi=(4312)$: The asymptotic line-solitons for this case are 
$[1,4]$- and $[2,3]$-solitons for $y\gg0$, and $[1,3]$- and $[2,4]$-solitons for $y\ll0$.
The $A$-matrix is given by
\[
A=\begin{pmatrix} 1&0&-b&-c\\0&1&a&0\end{pmatrix}\,,
\]
where $a,b,c>0$ are free parameters. Notice that two line-solitons for $y\ll 0$
are the same as in the T-type solution (see the crossing in the lower chords in Figure \ref{fig:chords}).
\item[(4)] $\pi=(3421)$: The asymptotic line-solitons are given by
$[1,3]$ and $[2,4]$ for $y\gg 0$, and for $y\ll 0$, these are the $[1,4]$- and $[2,3]$-solitons.
The $A$-matrix is given by
\[
A=\begin{pmatrix} 1&0&0&-c\\0&1&a&b\end{pmatrix}\,,
\]
where $a,b,c>0$ are positive free parameters.
This solution can be considered as a dual of the previous case (3), that is,
two sets of line-solitons for $y\gg 0$ and $y\ll 0$ are exchanged (also notice the duality
in the chord diagrams in Figure \ref{fig:chords}).
The example discussed after Proposition \ref{engn} corresponds to this solution (see Figure
\ref{fig:3421}).
\item[(5)] $\pi=(4321)$: The solution in this case is called the {\it P-type} 2-soliton solution 
which has asymptotic line-solitons of $[1,4]$- and $[2,3]$-types as $|y|\to\infty$. This type of solutions 
fits better with the {\it physical} assumption of quasi-two dimensionality with weak $y$-dependence
underlying the derivation of the KP equation. This is why we call it ``P-type" (see \cite{K:04}). 
The $A$-matrix is given by
\[
A=\begin{pmatrix} 1&0&0&-b\\0&1&a&0\end{pmatrix}\,.
\]
The chord diagram indicates that those two line-solitons must have the different amplitudes,
i.e. $A[1,4]>A[2,3]$,
but they  can propagate in the same direction, which correspond to the two soliton solution
of the KdV equation.
\item[(6)] $\pi=(3142)$: The asymptotic line-solitons are given by
$[1,3]$- and $[3,4]$-solitons for $y\gg 0$, and $[1,2]$- and $[2,4]$-solitons for $y\ll 0$.
The $A$-matrix is given by
\[
A=\begin{pmatrix} 1&a&0&-c\\0&0&1&b\end{pmatrix}\,,
\]
where $a,b,c>0$. This solution is dual to the case (2) in the sense that the two
sets of asymptotic line-solitons for $y\gg 0$ and $y\ll 0$ are switched, as well as
the missing minors are switched by $\xi(3,4)\leftrightarrow \xi(1,2)$. 
Also note the duality between the corresponding chord diagrams.
\item[(7)] $\pi=(2143)$: This case is called the {\it O-type} 2-soliton solution.
The asymptotic line-solitons are of
$[1,2]$- and $[3,4]$-types as $|y|\to\infty$. The letter ``O" for this type is due to 
the fact that this solution was {\it originally} found to describe the two-soliton 
solution of the KP equation (see for example \cite{FN:83}). 
The $A$-matrix for the O-type 2-soliton solution is given by
\[
A=\begin{pmatrix} 1&a&0&0\\0&0&1&b\end{pmatrix}\,.
\]
Notice that this $A$-matrix is obtained as a limit $c\to 0$ in the previous one of the case (6), i.e. $(3142)$-soliton solution.
\end{itemize}

Now let us describe the details of some of the $(2,2)$-soliton solutions, which will be important for
an application of those solutions to shallow water problem discussed in the next Section.
In particular, we explain how the $A$-matrix uniquely determines 
the structure of the corresponding soliton solution such as the location of the 
solitons and their phase shifts.

\subsection{O-type soliton solutions}\label{sub:O}
\begin{figure}[t]
\centering
\includegraphics[scale=0.53]{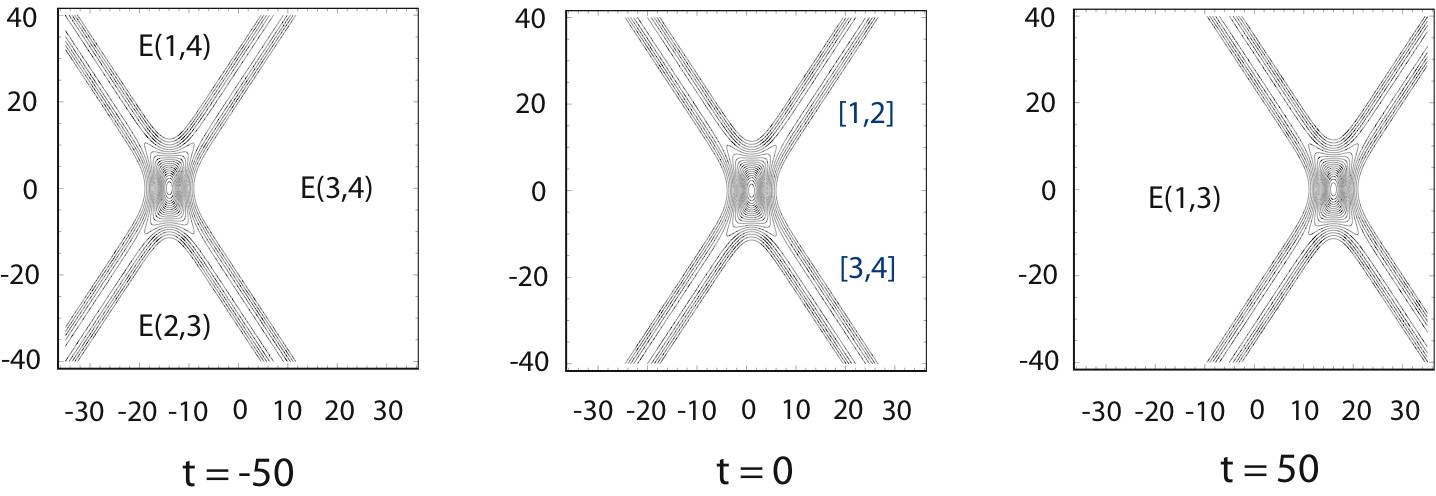} 
\caption{The time evolution of an O-type soliton solution. Each $E(i,j)$ indicates the dominant exponential in that region. The parameters are chosen as $A_{[1,2]}=A_{[3,4]}=0.1$ and
$\Psi_{[3,4]}=-\Psi_{[1,2]}=30^{\circ}$.
\label{fig:V4}}
\end{figure}

This is the original two-soliton solution, and  
the solutions correspond to the chord diagram of $\pi=(2143)$. A solution of this 
type consists of two full line-solitons of $[1,2]$ and $[3,4]$ (see Figure \ref{fig:V4}). 
Note here that they have phase shifts due to their collision.
Let us describe explicitly the structure of the solution of this type:
The $\tau$-function defined in (\ref{tauexp}) for this case is given by
\[
\tau=E(1,3)+bE(1,4)+aE(2,3)+abE(2,4)\,,
\]
where $a,b>0$ are the free parameters given in the $A$-matrix listed above.
As we will show that those two parameters can be used to fix the locations of 
those solitons, that is, they are determined
by the asymptotic data of the solution for large $|y|$.

For the later application of the solution, we assume that $[1,2]$-soliton has a ``negative'' $y$-component in the wave-vector (i.e. $\tan\Psi_{[1,2]}<0$), and $[3,4]$-soliton has a ``positive'' $y$-component, (i.e. $\tan\Psi_{[3,4]}>0$,
see Figure \ref{fig:1soliton}).
Then for the region with large positive $x$, we have $[1,2]$-soliton 
in $y>0$ and $[3,4]$-soliton in $y<0$. 

For $[1,2]$-soliton in $x>0$ (and $y\gg 0$), we have the dominant balance between $E(1,4)$ and $E(2,4)$.
Then the $\tau$-function can be written in the following form,
\begin{align*}
\tau&\approx bE(1,4)+abE(2,4)\\
&=2be^{\theta_4+\sf{1}{2}(\theta_1+\theta_2)}
\cosh\frac{1}{2}\left(\theta_1-\theta_2+\theta_{12}^{+}\right)\,,
\end{align*}
which leads to the $[1,2]$-soliton solution in the region near $\theta_1\approx \theta_2$ for  $x\gg 0$,
\[
u=2{\partial_x^2}\ln\tau\approx 
\frac{1}{2}(k_2-k_1)^2\sech^2\frac{1}{2}\left(\theta_1-\theta_2+\theta_{12}^+\right).
\]
Here the shift $\theta_{12}^+$ ($+$ indicates $x>0$) is related to the parameter $a$ in the $A$-matrix (see below).
%%%%%%%%%%%%%%%%%%%%%%%%%%%%%%%%%%%%
%%%%%%%%%%%%%%%%%%%%%%%%%%%%%%%%%%%%%%%%

For $[3,4]$-soliton in $x>0$ (and $y\ll 0$), from the balance $\tau\approx aE(2,3)+abE(2,4)$, we have
\begin{equation*}
u\approx \frac{1}{2}(k_4-k_3)^2{\rm sech}^2\frac{1}{2}(\theta_3-\theta_4+\theta_{34}^+).
\end{equation*}
The shifts $\theta^+_{12}$ and  $\theta^+_{34}$ are related to the parameters in the $A$-matrix,
\begin{equation}\label{Oshift}
a=\frac{k_4-k_1}{k_4-k_2}e^{-\theta_{12}^+},\qquad
b=\frac{k_3-k_2}{k_4-k_2}e^{-\theta_{34}^+}.
\end{equation}
Thus  the parameters in the $A$-matrix
can be determined by the asymptotic data of the locations of those $[1,2]$- and $[3,4]$-solitons 
for $x\gg 0$ and $|y|\gg 0$.

\begin{figure}[t!]
\centering
\includegraphics[scale=0.4]{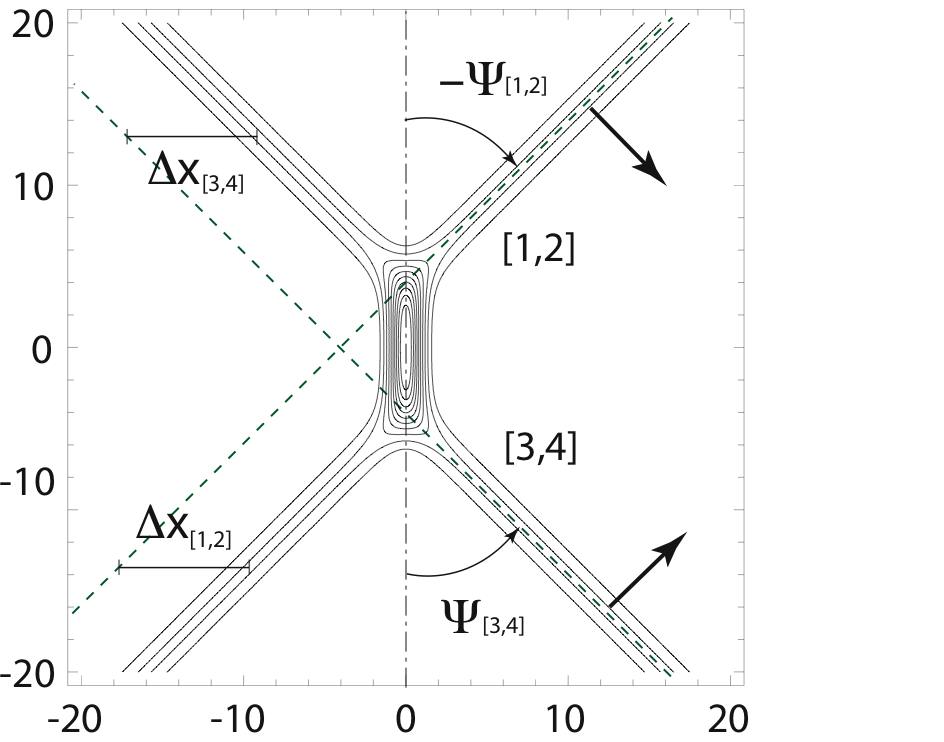} 
\caption{O-type interaction for two equal amplitude solitons.
The parameters $k_i$'s are taken to be $(k_1,k_2,k_3,k_4)=(-1-10^{-4},-10^{-4},10^{-4},1+10^{-4})$,
which give $A_{[1,2]}=A_{[3,4]}=\frac{1}{2}$ and $\tan\Psi_{[3,4]}=-\tan\Psi_{[1,2]}=1+2\times10^{-4}$ (i.e. $\Psi_{[3,4]}\approx
45.0057$).
The constants $a,b$ in the $A$-matrix are chosen so that the center of interaction
point is located at the origin, and $u_{\rm max}=u(0,0,0)\approx 1.96$ and the phase shift
$\Delta x_{[1,2]}=\Delta_{[3,4]}\approx 7.8$.\label{fig:Otype}}
\end{figure}

The most important feature of the O-type solution is the phase shift due to the interaction of
those two oblique line-solitons.
The phase shift for $[i,j]$-soliton is defined by $\theta_{ij}=\theta^-_{ij}-\theta^+_{ij}$
where $\pm$ indicate the values for $x\to\pm\infty$.  The values $\theta_{12}$ and $\theta_{34}$
turn out to be the same (see for example \cite{H:04}),
\[
\theta_{12}=\theta_{34}=-\ln\Delta_{\rm O}.
\]
where we note
\[
\Delta_{\rm O}:=\frac{(k_3-k_2)(k_4-k_1)}{(k_4-k_2)(k_3-k_1)}=1-\frac{(k_2-k_1)(k_4-k_3)}{(k_4-k_2)(k_3-k_1)}<1.
\]
This implies that $\theta_{12}=\theta_{34}>0$, and each $[i,j]$-soliton shifts in $x$ with
\begin{equation}\label{Opshift}
\Delta x_{[i,j]}=\frac{1}{k_j-k_i}\,\theta_{ij}.
\end{equation}
The positive phase shifts $\Delta x_{[1,2]}>0$ and $\Delta x_{[3,4]}>0$ indicate an {\it attractive} force
in the interaction.
Figure \ref{fig:Otype} illustrates an O-type interaction of two solitons which
have the same amplitude, $A_{[1,2]}=A_{[3,4]}=\sf{1}{2}$,  and are symmetric with respect to the $y$-axis,
$\Psi_{[3,4]}=-\Psi_{[1,2]}\approx 45^{\circ}$.  Since the solution is close to the resonance, we have the large phase shifts $\Delta x_{[1,2]}=\Delta x_{[3,4]}\approx 7.8$ and the maximum value of the soltion $u_{\rm max}\approx 1.96$ (almost
four times larger than $A_{[1,2]}$). 

%%%%%%%%%%%%%%%%%%%%%%%%%%%
%%%%%%%%%%%%%%%%%%%%%%%%%%%%%

O-type soliton solution has a steady X-shape with phase shifts in both line-solitons.
One can also find the formula of the maximum amplitude which occurs at the 
center of intersection point (center of the X-shape), which is given by 
\begin{equation}\label{O-center}
u_{\rm max}= \displaystyle{A_{[1,2]}+A_{[3,4]}+
2\,\frac{1-\sqrt{\Delta_{\rm O}}}{1+\sqrt{\Delta_{\rm O}}}\sqrt{A_{[1,2]}A_{[3,4]}}}\,.
\end{equation}
(see for example \cite{CK:09,DSH:04, S:04}.)
Since $0<\Delta_{\rm O}<1$, we have the bound
\[
A_{[1,2]}+A_{[3,4]}<u_{\rm max}<\left(\sqrt{A_{[1,2]}}+
\sqrt{A_{[3,4]}}\right)^2\,.
\]

It is also interesting to note that the formula $\Delta_{\rm O}$ has critical cases at the values 
$k_1=k_2$ or $k_2=k_3$ or $k_3=k_4$. For the case with $k_1=k_2$ or $k_3=k_4$ (i.e. $\Delta_{\rm O}=1$),
one can see that one of the line-soliton becomes small, and the limit consists of just one-soliton
solution. On the other hand, for  the case $k_2=k_3$ (i.e. $\Delta_{\rm O}=0$), 
the $\tau$-function has only three terms, which corresponds to a solution showing
a Y-shape interaction (i.e. the phase shift
becomes infinity and the middle portion of the interaction stretches to infinity). 
This limit has been discussed in \cite{M:77, NR:77} as a resonant interaction of three 
waves to make Y-shape soliton. This limit gives a critical angle
between those solitons which can be found as follows:
First let us express each $k_j$ parameter in terms of the amplitude and the slope,
\begin{align*}
k_{1,2}&=\frac{1}{2}\left(\tan\Psi_{[1,2]}\mp \sqrt{2A_{[1,2]}}\right),\\
k_{3,4}&=\frac{1}{2}\left(\tan\Psi_{[3,4]}\mp\sqrt{2A_{[3,4]}}\right),
\end{align*}
where the angle $\Psi_{[i,j]}$ is measured in the counterclockwise direction from the $y$-axis
(see Figure \ref{fig:Otype}). In particular, we have
\[
\tan\Psi_{[1,2]}=-\sqrt{2A_{[1,2]}}+2k_2, \qquad \tan\Psi_{[3,4]}=\sqrt{2A_{[3,4]}}+2k_3.
\]

For simplicity, let us consider the special case when both solitons
are of equal amplitude and symmetric with respect to the
$y$-axis i.e., $A_{[1,2]}=A_{[3,4]}=A_0$ and  $\Psi_{[3,4]}=-\Psi_{[1,2]}=\Psi_0>0$. 
This corresponds to setting $k_1 = -k_4$ and $k_2 = -k_3$. Then, for fixed
amplitude $A_0$, the angle $\Psi_0$ has a lower bound given by
\[
\tan\Psi_0 = \sqrt{2A_{0}}+2k_3 \geq \sqrt{2A_{0}} := \tan \Psi_c\,.
\]
The lower bound is achieved in the 
limit $k_2=k_3=0$, and the critical angle $\Psi_c$ is given by
\begin{equation}\label{critical}
\Psi_c=\tan^{-1}\sqrt{2A_0}\,.
\end{equation}
In \cite{M:77}, Miles introduced the following parameter to describe the interaction properties 
for O-type solution,
\begin{equation}\label{MilesK}
\kappa:=\frac{\tan\Psi_0}{\sqrt{2A_0}}=\frac{\tan\Psi_0}{\tan\Psi_c}.
\end{equation}
With this parameter, the maximum amplitude of \eqref{O-center} for this symmetric case is given by
\begin{equation}\label{Omax}
u_{\rm max}=\frac{4A_0}{1+\sqrt{\Delta_{\rm O}}}, 
\qquad{\rm with}\quad \Delta_{\rm O}=1-\frac{1}{\kappa^2}.
\end{equation}
Thus, at the critical angle $\Psi_0=\Psi_c$ (i.e., $\kappa=1$), 
we have $u_{\rm max}=4A_0$ and the phase shift $\theta_{[12]}\to\infty$,
leading to the resonant Y-shape interaction (see also \cite{M:77,DSH:04,S:04}).

One should note that if we use the form of the O-type solution even beyond 
the critical angle, i.e.  $k_3<k_2$, then the solution becomes singular (note that 
the sign of $E(2,3)$ changes). In earlier works, this was considered to be an 
obstacle for using the KP equation to describe an interaction of two line-solitons 
with a smaller angle.  On the contrary, the KP equation should give a better 
approximation to describe oblique interactions of solitons with smaller angles.
Thus one should expect to have explicit solutions of the KP equation describing
such phenomena. It turns out that the new types of $(2,2)$-soliton solutions 
discussed above can indeed serve as good models for describing line-soliton 
interactions of solitons with small angles. We will show in Section \ref{sec:SWW2} how
these solutions are related 
to the {\it Mach reflections} in shallow water waves.

\subsection{$(3142)$-type soliton solutions}\label{subsec:3142}
%%%%%%%%%%%%%%%%%%%%%%%%%%%%
\begin{figure}[t]
\centering
\includegraphics[scale=0.55]{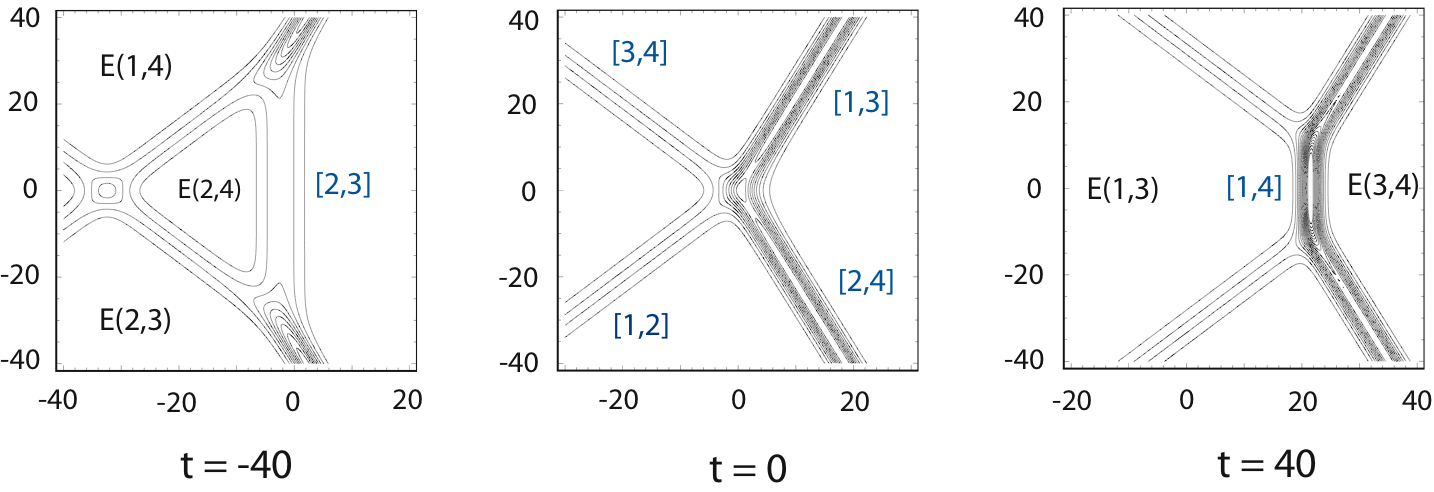} 
\caption{The time evolution of a $(3142)$-type soliton solution. 
The line-solitons have $A_{[1,3]}=A_{[2,4]}=0.5$ and $\Psi_{[2,4]}=-\Psi_{[1,3]}=25^{\circ}$. The critical angle is given by $\Psi_c=\tan^{-1}\sqrt{2A_0}=45^{\circ}$, which also gives $\Psi_c=\Psi_{[3,4]}=-\Psi_{[1,2]}$.  The parameters in the $A$-matrix are given by \eqref{shiftab} with $\theta^+_{13}=\theta^+_{24}=0$ and $s=1$. The amplitude of the intermediate $[1,4]$-soliton approaches asymptotically to $1.075$.
\label{fig:3142}}
\end{figure}
%%%%%%%%%%%%%%%%%%%%%%%%%%%%%%%
We consider a solution of this type which
 consists of two line-solitons for large positive $x$ and two other line-solitons for 
large negative $x$. We then assume that the slopes of two solitons in each region
have opposite signs, i.e. one in $y>0$ and other in $y<0$ (see Figure \ref{fig:3142}). 
The line-solitons for the $(3142)$-type solution are determined from 
the balance between two appropriate exponential terms in its $\tau$-function which has the form,
\[
\tau=E(1,3)+bE(1,4)+aE(2,3)+abE(2,4)+cE(3,4)\,.
\]
The solution contains three free parameters $a,b$ and $c$, which can
be used to determine the locations of three (out of four) asymptotic line-solitons 
(e.g. two in $x\gg0$ and one in $x\ll0$). Thus, the parameters are completely 
determined from the asymptotic data on large $|y|$.

Let us first consider the line-solitons in $x\gg 0$: There are two line-solitons 
which are $[1,3]$-soliton in $y\gg0$ and $[2,4]$-soliton in $y\ll0$. The $[1,3]$-soliton 
is obtained by the balance between the exponential terms $bE(1,4)$ and $cE(3,4)$, 
and the $[2,4]$-soliton is by the balance between $aE(2,3)$ and $cE(3,4)$. Consequently,
the phase shifts of $[1,3]$- and $[2,4]$-solitons for $x\gg0$ are given by 
\begin{equation}\label{shift1324}
\theta_{13}^+=\ln\frac{k_4-k_1}{k_4-k_3}+\ln\frac{b}{c}\,, \qquad \quad
\theta_{24}^+=\ln\frac{k_3-k_2}{k_4-k_3}+\ln \frac{a}{c}\,.
\end{equation}

Now we consider the line-solitons in $x\ll0$: They are $[3,4]$-soliton in $y\gg0$ 
and $[1,2]$-soliton in $y\ll0$. The phase shifts are given respectively by
\begin{align*}
\theta_{34}^-=\ln\frac{k_3-k_1}{k_4-k_1}-\ln b,\qquad
\theta_{12}^-=\ln\frac{k_3-k_1}{k_3-k_2}-\ln a
\end{align*}

\begin{figure}[t]
\centering
\includegraphics[scale=0.54]{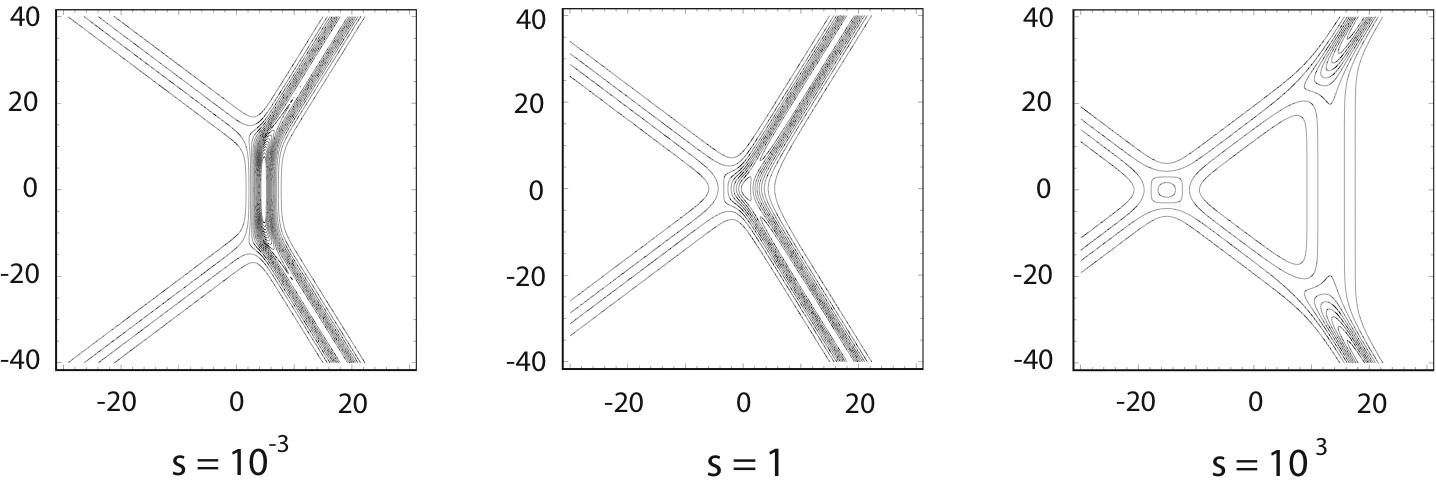} 
\caption{$(3142)$-type soliton solution with the $s$-parameter.
 The line-solitons are given by $A_{[1,3]}=A_{[2,4]}=0.5$ and $\Psi_{[2,4]}=-\Psi_{[1,3]}=25^{\circ}$
 (these then gives the other two solitons uniquely). The parameters in the $A$-matrix
are chosen as (\ref{shiftab}) with $\theta^+_{[1,3]}=\theta^+_{[2,4]}=0$. 
Then at $s=1$, all the solitons
meet at the origin, i.e. the $s$-parameter shifts $[1,2]$- and $[3,4]$-solitons.
\label{fig:phase3142}}
\end{figure}

We then define the parameter $s$ (representing the total phase shifts
$\theta_{13}^++\theta^-_{34}=\theta^+_{24}+\theta^-_{12}$),
\begin{equation}\label{shiftc}
s:=\exp\left(-\theta_{13}^+-\theta_{34}^-\right),
\end{equation}
which leads to
\begin{equation}\label{shiftab}
a=\frac{k_3-k_1}{k_3-k_2}\,se^{\theta_{24}^+}, \qquad b=\frac{k_3-k_1}{k_4-k_1}\,se^{\theta_{13}^+},\qquad
c=\frac{k_3-k_1}{k_4-k_3}\,s.
\end{equation}
The $s$-parameter represents the relative locations of the intersection point of the
$[1,3]$- and $[3,4]$-solitons with the $x$-axis, in particular, 
$\theta_{13}^++\theta_{34}^- = 0$ when $s=1$ (see Figure \ref{fig:phase3142}).
Thus the parameters $a,b$ and $c$ are related to  the locations of $[1,3]$-soliton (with $\theta_{13}^+$),
of $[2,4]$-soliton (with $\theta_{24}^+$), and the intersection point of $[1,3]$- and $[3,4]$-solitons
(with $s$).

%%%%%%%%%%%%%%%%%%%%%%%%%%%%%%%%%%%%%
%%%%%%%%%%%%%%%%%%%%%%%%%%%%%%%%

Now we consider the case where the $[1,3]$- and $[2,4]$-solitons have the same amplitude
($A_{[1,3]}=A_{[2,4]}=A_0$) and they are symmetric with respect to the $x$-axis 
($\Psi_{[2,4]}=-\Psi_{[1,3]}=\Psi_0$). Then in terms of the $k$-parameters, we have
\[
k_3-k_1=k_4-k_2=\sqrt{2A_0}.
\]
Also the symmetry of the wave-vectors, i.e. $\Psi_{[2,4]}=\Psi_0=-\Psi_{[1,3]}$, gives
\[
k_2+k_4=-(k_1+k_3)=\tan\Psi_0.
\]
This implies that we have 
\begin{equation}\label{SymmetryinK}
k_4=-k_1>0,\qquad k_3=-k_2>0.
\end{equation} 
The angle $\Psi_0$ takes the value in $(0,\Psi_c)$, where the critical angle is given by
the condition $k_2=k_3=0$, i.e.
\[
\Psi_c=\tan^{-1}\sqrt{2A_0}.
\]
Notice that this formula is the same as that of the O-type soliton 
solution (see (\ref{critical})), and the $(3142)$-type exists when the $\kappa$-parameter
is less than one, i.e.  for $(3142)$-type, we have
\[
\kappa=\frac{\tan\Psi_0}{\sqrt{2A_0}}<1.
\]

 From (\ref{SymmetryinK}), one can easily deduce the following facts
for $[1,2]$- and $[3,4]$-solitons in $x<0$:
\begin{itemize}
\item[(a)] Those solitons have the same amplitude, i.e. 
\[
A_{[1,2]}=A_{[3,4]}=\frac{1}{2}(k_4-k_3)^2=\frac{1}{2}(k_4+k_2)^2=\frac{1}{2}\tan^2\Psi_0=\kappa^2A_0.
\]
Thus, if the $[1,3]$- and $[2,4]$-solitons in $x>0$ are close to the $y$-axis (i.e.
a small $\Psi_0$), then the amplitudes of the solitons in $x<0$ are small;
whereas at the critical angle
$\Psi_0=\Psi_c$, the solitons $[1,2]$ and $[3,4]$ in $x<0$ 
take the maximum amplitude $A_{[1,2]}=A_{[3,4]}=A_0$.
\item[(b)] The directions of the wave-vectors for the $[1,2]$ and $[3,4]$-solitons
are also symmetric, i.e.
\[
\tan\Psi_{[3,4]}=-\tan\Psi_{[1,2]}=k_3+k_4.
\]
Moreover, the symmetry (\ref{SymmetryinK}) implies that
$\tan\Psi_{[3,4]}=k_4-k_2=\sqrt{2A_{[2,4]}}=\sqrt{2A_0}$, so
\[
\Psi_{[3,4]}=\Psi_c=\tan^{-1}\sqrt{2A_0}.
\]
Thus the directions of the wave-vectors for the $[1,2]$ and $[3,4]$-solitons in $x<0$
depend only on the amplitude of the solitons in $x>0$ but not on their directions
(i.e., angle of their V-shape).
\end{itemize}

Let us choose the parameters in the $A$-matrix for the $(3142)$-soliton
solution appropriately, so that at $t=0$ all the solitons intersect at the 
origin (see Figure \ref{fig:3142}).
Then for $t<0$, the resonant interaction between $[1,3]$- and $[3,4]$-solitons (as well as
$[2,4]$- and $[1,2]$-solitons) generates an intermediate line-soliton (called ``stem" soliton)
which is $[1,4]$ soliton. The amplitude of this soliton is given by
\begin{equation}\label{stemA}
A_{[1,4]}=\frac{1}{2}(k_4-k_1)^2=\frac{1}{2}\left(\sqrt{2A_0}+\tan\Psi_0\right)^2=A_0(1+\kappa)^2.
\end{equation}
Note here that at the critical angle $\Psi_0=\Psi_c$, the amplitude takes the maximum $A_{[1,4]}=4A_0$
(see \cite{TO:07, PTLO:05}).

For $t>0$, the resonant interaction between $[1,3]$- and $[1,2]$-solitons (as well as $[2,4]$- and
$[3,4]$-solitons) generates an intermediate line-soliton of $[2,3]$-soliton. 
The amplitude of $[2,3]$-soliton is given by
\[
A_{[2,3]}=\frac{1}{2}(k_3-k_2)^2=\frac{1}{2}\left(\sqrt{2A_0}-\tan\Psi_0\right)^2=A_0(1-\kappa)^2.
\]
Because of the symmetry (\ref{SymmetryinK}), both $[1,4]$- and $[2,3]$-solitons are
parallel to the $y$-axis, i.e. $\tan\Psi_{[1,4]}=\tan\Psi_{[2,3]}=0$.

\subsection{T-type soliton solutions}\label{Tsoliton}
There are four parameters in the $A$-matrix for T-type soliton
solution. Here we explain that those parameters give the information of
the locations of those line-solitons, the phase shift and on-set of the opening of
a box. Thus three of those four parameters are determined
by the asymptotic data on large $|y|$, and we need an internal data for the other one.
%%%%%%%%%%%%%%%%%%%%%%%%%%%%%%%%%%%%%%%%%%
\begin{figure}[t!]
\centering
\includegraphics[scale=0.52]{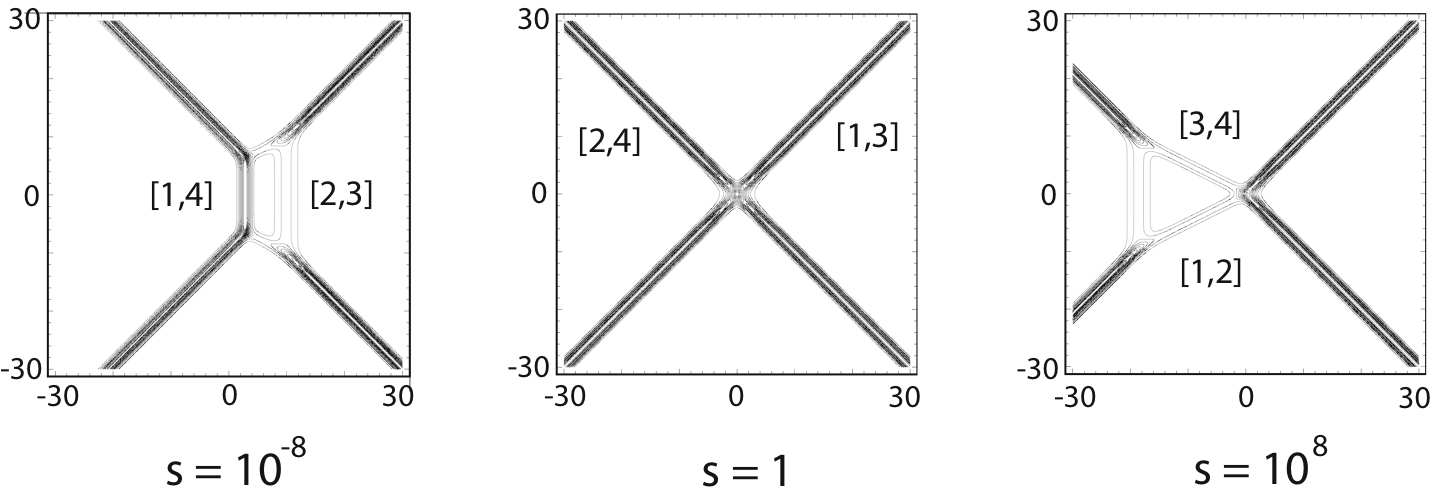} 
\caption{T-type interaction with the $s$-parameter. The $k$-parameters are chosen as
$(k_1,k_2,k_3,k_4)=(-\frac{3}{2},-\frac{1}{2},\frac{1}{2},\frac{3}{2})$. The $A$-matrix is
chosen as \eqref{bcD} and \eqref{r} with $\theta^+_{13}=\theta^+_{24}=0$ and $r=1$.
The $s$-parameter gives the phase shift for the $[1,2]$- and $[3,4]$-solitons in $x<0$.
\label{fig:TS}}
\end{figure}
%%%%%%%%%%%%%%%%%%%%%%%%%%%%%%%%%%%%%%%%

Following the arguments in the previous section, one can find the phase shifts of the
line-solitons of $[1,3]$ and $[2,4]$:
For $[1,3]$-soliton in $x>0$ (and $y\gg0$), the phase shift is calculated as
\[
\theta_{13}^+=\ln\frac{k_4-k_1}{k_4-k_3}-\ln \frac{D}{b}\,,
\]
where $D=ad-bc=\xi(3,4)$. For the same soliton in $x<0$ (and $y \ll0$), we have
\[
\theta^-_{13}=\ln\frac{k_2-k_1}{k_3-k_2}-\ln c\,.
\]
So the total phase shift $\theta_{13}:=\theta^-_{13}-\theta^+_{13}$
 depends on the $A$-matrix unlike the cases of O- and P-types, and it can take
any value.

\begin{figure}[t!]
\centering
\includegraphics[scale=0.52]{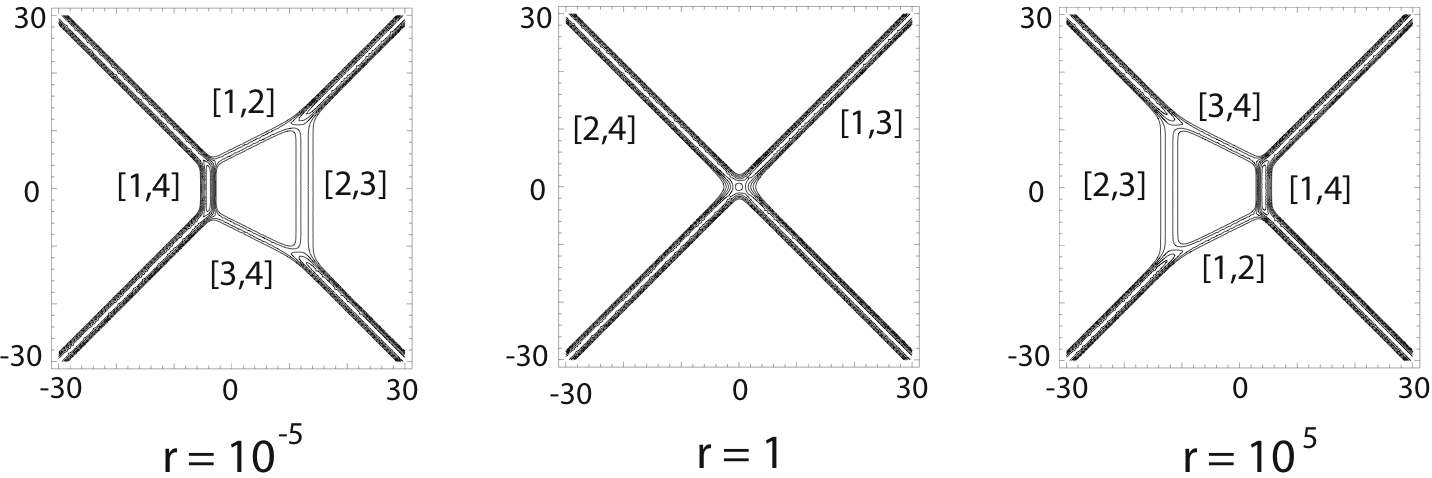} 
\caption{T-type interaction with the $r$-parameter. The $k$-parameters are the same as
those in Figure \ref{fig:TS}.  The $A$-matrix 
are chosen as \eqref{bcD} with $\theta^+_{13}=\theta^+_{24}=0$ and $s=1$. 
The $r$-parameter gives the on-set of the box, and it does not affect the locations of
all four line-solitons, that is, $r$ is an {\it internal} parameter.
\label{fig:T}}
\end{figure}

For $[2,4]$-soliton in $x>0$ (and $y\ll0$), we have
\[
\theta_{24}^+=\ln\frac{k_3-k_2}{k_4-k_3}-\ln\frac{D}{c}.
\]
and for the same one in $x<0$ (and $y\gg0$), we have
\[
\theta_{24}^-=\ln\frac{k_2-k_1}{k_4-k_1}-\ln b.
\]
Note that the total phase shift $\theta_{24}=\theta_{24}^+-\theta_{24}^- $ is the 
same as that for $[1,3]$-soliton, i.e. the phase 
conservation along the $y$-axis $\theta_{13}^++\theta_{24}^-=\theta_{13}^-+\theta_{24}^+$
holds. 
Then as in (\ref{shiftc}) for the case of $(3142)$-type, we define the $s$-parameter,
\begin{equation*}\label{sTtype}
s:=\exp(-\theta_{13}^+-\theta_{24}^-),
\end{equation*}
which represents the intersection point of $[1,3]$- and $[2,4]$-soliton.
With the $s$-parameter, we have
\begin{equation}\label{bcD}
b=\frac{k_2-k_1}{k_4-k_1}\,se^{\theta_{13}^+},\qquad c=\frac{k_2-k_1}{k_3-k_2}\,se^{\theta_{24}^+},\qquad D=\frac{k_2-k_1}{k_4-k_3}\,s.
\end{equation}
Namely, the three parameters $b,c$ and $D=ad-bc$ determine the locations and the phase shift (i.e.
the intersection point of $[1,3]$- and $[2,4]$-solitons). 
One other parameter is then related to an on-set of a box
at the intersection point (see Figure \ref{fig:T}).

In order to characterize this parameter,
let us consider the intermediate solitons of $[1,4]$ and $[2,3]$.
First note that for $t\gg 0$, $[1,4]$-soliton appears as the dominant balance between
$E(1,2)$ and $E(2,4)$. Then one can find the phase shift $\theta_{14}^+$ (here $+$ indicates $t>0$),
\[
\theta^+_{14}=\ln\frac{k_2-k_1}{k_4-k_2}-\ln\, d .
\]
Similarly one can get the phase shift $\theta_{14}^-$ for $t\ll 0$ as
\[
\theta^-_{14}=\ln\frac{k_3-k_1}{k_4-k_3}-\ln\frac{D}{a}.
\]
 Now consider the sum of $\theta_{14}^{\pm}$, i.e.
\begin{align*}
\theta_{14}^++\theta_{14}^-&=\ln\frac{(k_2-k_1)(k_3-k_1)}{(k_4-k_2)(k_4-k_3)}-\ln\frac{dD}{a}.
\end{align*}
Also, for the $[2,3]$-soliton, one can get
\begin{align*}
\theta_{23}^++\theta_{23}^-&=\ln\frac{(k_2-k_1)(k_4-k_2)}{(k_3-k_1)(k_4-k_3)}-\ln\frac{aD}{d}.
\end{align*}
%%%%%%%%%%%%%%%%%%%%%%%%%%%%%%%%%%%%%%%%%%
%%%%%%%%%%%%%%%%%%%%%%%%%%%%%%%%%%%%%%%%

  Now we introduce a parameter $r$ in the form,
\begin{equation}\label{r}
\frac{a}{d}=r\frac{k_4-k_2}{k_3-k_1}\,,
\end{equation}
so that we have
\begin{align*}
\theta_{14}^++\theta_{14}^-=\ln\,\frac{r}{s},\qquad \theta_{23}^++\theta_{23}^-=-\ln\,(rs).
\end{align*}
Suppose that at $t=0$, $[1,3]$- and $[2,4]$-solitons in $x>0$ are placed so that they meet at the origin,
that is, we choose $\theta_{13}^+=\theta_{24}^+=0$. Also if there is no phase shifts for those solitons,
i.e. $s=1$. then the sums become
\[
\theta_{14}^++\theta_{14}^-=\ln\,r=-(\theta_{23}^++\theta_{23}^-)\,.
\]
This implies that at $t=0$ (and $s=1$) if $r=1$, then the T-type soliton solution has an exact 
shape of ``X" without any opening of a box at the intersection point on the origin.
Moreover, at $t=0$ if $r>1$, then $[1,4]$-soliton appears in $x>0$ and $[2,3]$-soliton in $x<0$;
whereas if $0<r<1$, then $[1,4]$-soliton appears in $x<0$ and $[2,3]$-soliton in $x>0$.
Figure \ref{fig:T} illustrates those cases with $s=1$. The parameter $r$ determines the
exponential term that is dominant in the region inside the box. When
$r<1$, $E(2,4)$ is the dominant exponential term, and when $r>1$ the dominant
exponential is $E(1,3)$. One should note that the parameter $r$ cannot be
determined by the asymptotic data, that is, $r$ is considered as an ``internal" parameter.

\section{Numerical simulation and the stability of the soliton solutions}\label{sec:NS}

In this section, we present some numerical simulations of the KP equation with
``V-shape'' initial wave form related to a physical situation (see for examples \cite{PTLO:05, TO:07, F:80}). The main purpose of the numerical simulation is to study the interaction properties of
line-solitons, and we will show that the solutions of the initial value problems with V-shape incident
waves  approach asymptotically to some of the exact soliton solutions of the KP equation discussed in the 
previous section.  This implies a stability of those exact solutions under the influence
of certain deformations (notice that the deformation in our cases are not so small).

The initial value problem considered here is essentially an infinite energy problem in the sense
that each line-soliton in the initial wave is supported asymptotically in either $y\gg 0$ or $y\ll 0$,
and the interactions occur only in a finite domain in the $xy$-plane.  In the numerical scheme,
we consider the rectangular domain $D=\{(x,y): |x|\le L_x, |y|\le L_y\}$, and each line-soliton
is matched with a KdV soliton at the boundaries $y=\pm L_y$.  
The details of the numerical scheme and the results can be found in \cite{KK:10}.

We consider the initial data given in the shape of ``V'' with the amplitude $A_0$ and the oblique
angle $\Psi_0>0$,
\begin{equation}\label{initialdata}
u(x,y,0)=A_0\sech^2 \sqrt{\frac{A_0}{2}}\left(x-|y|\tan\Psi_0\right)\,.
\end{equation}
Note here that two semi-infinite line-solitons are propagating toward each other into the positive $x$-direction,
so that they interact strongly at the corner of the V-shape.
At the boundaries $y=\pm L_y$ of the numerical domain, those line-solitons are
patched to the KdV one-soliton solutions given by 
\[
u(x,\pm L_y,t)=A_0\sech^2\sqrt{\frac{A_0}{2}}\,(x\mp L_y\tan\Psi_0-\nu t)\,,
\]
with $\nu=\frac{3}{4}\tan^2\Psi_0+\frac{1}{2}A_0$. Note here that these solitons correspond to  the exact one-soliton solution
of the KdV equation with the velocity shift due to the oblique propagation of the line-soliton, i.e.
$\partial^2 u/\partial y^2=\tan^2\Psi_0\partial^2 u/\partial x^2$. The numerical simulations are based on a spectral method with window-technique similar to the method used in \cite{TO:07}  (see \cite{KK:10} for the details).
The V-shape initial wave was first considered by Oikawa and Tsuji (see for example \cite{PTLO:05,
TO:07}) in order to study  the  generation of freak (or rogue) waves. They noticed  generations of
different types of asymptotic solutions depending on the initial oblique angle $\Psi_0$, and found the resonant interactions which create localized high
amplitude waves.
%%%%%%%%%%%%%%%%%%%%%%%%%%%%%%%%%%%%%%%%%%
\begin{figure}[t!]
\centering
\includegraphics[scale=0.4]{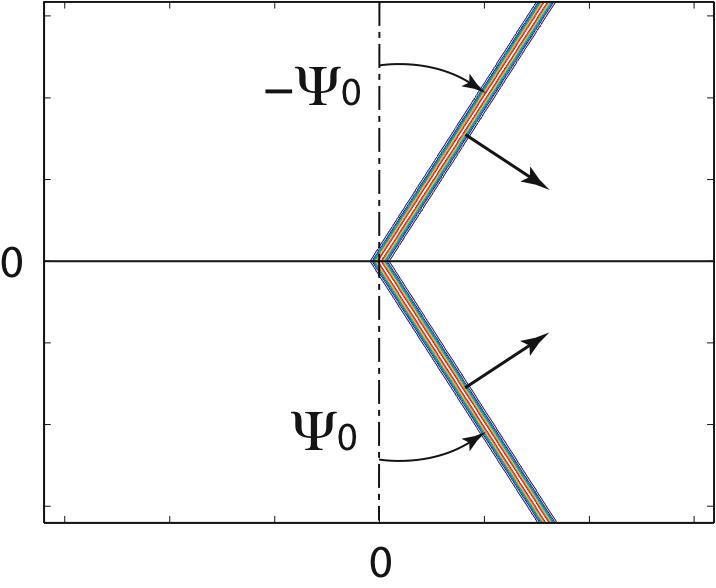} 
\caption{Initial data with V-shape wave. Each line of the V-shape is locally a line-soliton solution.
We set those line-solitons to meet at the origin. 
\label{fig:IV}}
\end{figure}
%%%%%%%%%%%%%%%%%%%%%%%%%%%%%%%%%%%%%%%%
In this section, we present the results for the cases corresponding to $A_0=2$ and two different angles,
$\Psi_1$ and $\Psi_2$ with $\Psi_1<\Psi_c<\Psi_2$.
where the critical angle is given by $\Psi_c=\tan^{-1}\sqrt{2A_0}\approx 63.4^{\circ}$.
Then we explain these results in terms of certain $(2,2)$-soliton solutions discussed in the 
previous section, and in particular,
we describe the connection with the Mach reflection (this will be further discussed in Section \ref{sec:SWW2}).

The main idea here is to consider the V-shape initial wave as the part of some $(2,2)$-soliton solutions
listed in the previous section.  In order to identify those soliton solutions from the V-shape initial wave form, let us first denote
them as $[{i_1},{j_1}]$-soliton for $y\gg 0$ and $[{i_2},{j_2}]$-soliton for $y\ll 0$.
Then using the relations, $k_j-k_i=\sqrt{2A_0}=2$ and $k_j+k_i=\tan\Psi_0$, for $[i,j]$-soliton
and the Miles parameter $\kappa=\tan\Psi_0/\sqrt{2A_0}$ of \eqref{MilesK}, we have
\begin{equation}\label{NumericalK}
\left\{\begin{array}{llll}
\displaystyle{k_{i_1}=-(1+\kappa),\qquad } &k_{j_1}=1-\kappa,\\[1.0ex]
\displaystyle{k_{i_2}=-(1-\kappa),\qquad}  &k_{j_2}=1+\kappa.
\end{array}\right.
\end{equation}
Notice that $k_{j_2}=-k_{i_1}$ and $k_{i_2}=-k_{j_1}$ because of the symmetry in the initial wave.
Moreover, 
at the critical angle $\Psi_0=\Psi_c$  (i.e. $\kappa=1$),  we have $k_{i_2}=k_{j_1}=0$. 
We also note $k_{i_1}$ as the smallest parameter and $k_{j_2}$ as the largest one, so that depending on the angle $\Psi_0$, we obtain the following ordering in the $k$-parameters:

For $0<\Psi_0<\Psi_c$ (i.e. $\kappa<1$), we have
\[
k_{i_1}<k_{i_2}<0<k_{j_1}<k_{j_2},
\]
implying that the corresponding chords of the $[i_1,j_1]$- and the $[i_2,j_2]$-solitons overlap.
That is, $[1,3]$ chord appears on the upper side of the diagram,
and $[2,4]$ chord on the lower side. 
This means that the two solitons can be identified as part of either the $(3412)$-type (T-type) or the $(3142)$-type
solution (see the chord diagrams in Figure \ref{fig:chords}).  

For $\Psi_c<\Psi_0<\sf{\pi}{2}$ (i.e. $\kappa>1$), we have
\[
k_{i_1}<k_{j_1}<0<k_{i_2}<k_{j_2}.
\]
In this case, the corresponding chords are separated, and the two
solitons form part of either $(2413)$- or $(2143)$-type (O-type) solution. Here 
 $[1,2]$- and $[3,4]$-chords appear on the upper and lower sides of the chord diagram, respectively.

Then the numerical simulations show that we have  the following types of the asymptotic solutions depending on the values  $\Psi_0$:
\begin{itemize}
\item[(a)] If the angle satisfies $\Psi_0<\Psi_c$ (i.e. $\kappa<1$), then the
solution converges asymptotically to $(3142)$-type soliton solution (not T-type)
\item[(b)] If the angle satisfies $\Psi_c<\Psi_0$ (i.e. $\kappa>1$), then the solution converges
asymptotically to an O-type soliton solution (not $(2413)$-type).
\end{itemize}
The convergence here is in a locally defined $L^2$-sense with the usual norm,
\[
\|f\|_{L^2(D)}:=\left(\iint_D|f(x,y)|^2dxdy\right)^{\sf{1}{2}}.
\]
where $D\subset \mathbb{R}^2$ is a compact set which covers the main structure of the interactions in the solution.
To confirm the convergence statements, we define the (relative) error function,
\begin{equation}\label{error}
E(t):=\|  u^t - u_{\rm exact}^t\|^2_{L^2(D^t_r)}\big/ \|u_{\rm exact}^t\|^2_{L^2(D^t_r)},
\end{equation}
with the solution $u^t(x,y):=u(x,y,t)$ and an exact solution $u^t_{\rm exact}(x,y)$, where
 $D_r^t$ is the circular disc given by
\[
D_r^t:=\left\{(x,y)\in\mathbb{R}^2: (x-x_0(t))^2+(y-y_0(t))^2\le r^2\right\}.
\]
The center $(x_0(t),y_0(t))$ of the circular domain $D^t_r$  is chosen as the intersection point of two lines determined from the corresponding exact solution.
We find the exact solution $u^t_{\rm exact}(x,y)$ by minimizing $E(t)$ at certain large time $t=T_0$:
In the minimization process, we assume that the $k$-parameters remain the same as those given by
\eqref{NumericalK}, and vary the corresponding $A$-matrix to adjust the solution pattern
(recall that the $A$-matrix determines the locations of the line-solitons in the solution,
see Section \ref{sec:CL}).
After minimizing $E(t)$, that is, finding the corresponding exact solution, we check
that $E(t)$ further decreases for a larger time $t>T_0$ up to a time $t=T_1>T_0$, just before the effects
of the boundary enter the disc $D_r^t$ (those effects include the periodic condition in $x$ and a mismatch on the boundary patching).
We take the radius $r$ in $D_r^t$ large enough so that the main interaction
 area is covered for all $t<T_0$, but $D_r^t$ should be kept away  from the boundary
 to avoid any influence coming from
 the boundaries. The time $T_1>T_0$ gives an
 optimal time to develop a pattern close to the corresponding exact solution, but it is also limited
 to avoid any disturbance  from the boundaries for $t<T_1$.
 Thus, our convergence implies the separation of the radiations from the soliton solution,
 just like the case of the KdV equation (see the end of Section \ref{sec:KP}).
 
 We also note that the convergence here implies a {\it completion} of the partial chord diagram
 consisting of only two chords which corresponds to the semi-infinite solitons in the initial
 V-shape wave. Namely, the asymptotic solution of the initial value problem with
 V-shape initial wave is given by an exact solution parametrized by a unique chord diagram,
 and the initial (partial) chord diagram is completed by adding two other solitons (chords) generated by
 the interaction.  The completion may not be unique, and in \cite{KOT:09}, we proposed
 a concept of {\it minimal} completion in the sense that the completed diagram has the minimum
 total length of the chords and the corresponding TNN Grassmannian cell has the minimum dimension.
 However, this problem is still open, and we need to make a precise statement of the minimal completion
 of partial chord diagram given by the initial wave profile.

\subsection{Regular reflection: $\kappa>1$}
\begin{figure}[t!]
\begin{centering}
\includegraphics[scale=0.63]{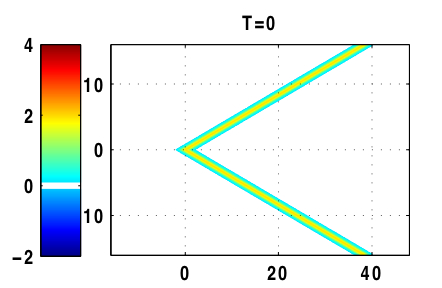}\includegraphics[scale=0.6]{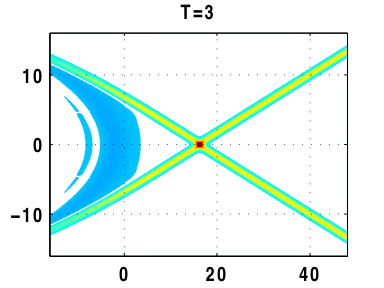}\includegraphics[scale=0.6]{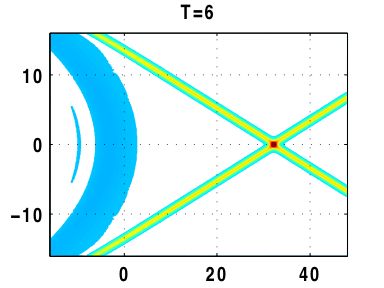}
\par\end{centering}
\begin{centering}
\includegraphics[scale=0.63]{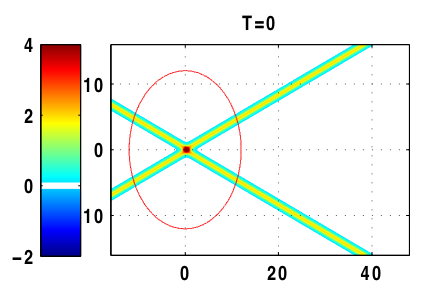}\includegraphics[scale=0.6]{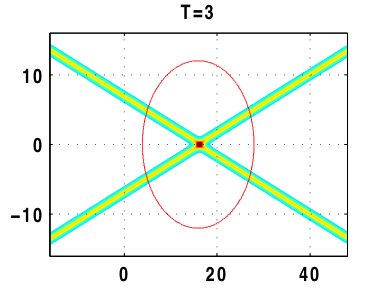}\includegraphics[scale=0.6]{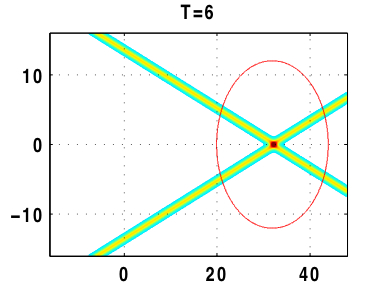}
\par\end{centering}
\begin{centering}
\includegraphics[scale=0.7]{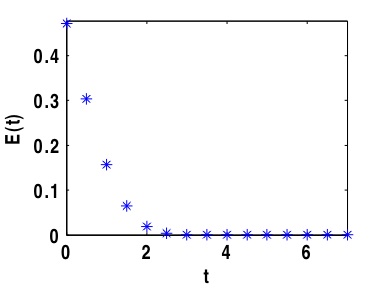} \hskip 1cm \raisebox{0.3in}{\includegraphics[scale=0.4]{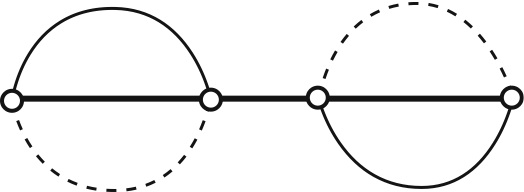}}
\par\end{centering}
\caption{\label{fig:ON} Numerical simulation of V-shape initial wave  for $\kappa>1$ (regular reflection).
The initial wave consists of $[1,2]$-soliton in
$y>0$ and $[3,4]$-soliton in $y<0$, with $A_{[1,2]}=A_{[3,4]}=2$ and
$\Psi_0\approx 67.3^{\circ}$ ($\Psi_c=63.4^{\circ}$. The upper figures show the
result of the direct simulation.  Notice a large wake behind the interaction point which extends the initial
solitons. The middle figures show the
corresponding exact solution of O-type. The circle in these figures show the domain $D^t_r$
with $r=12$.  About $t=3$, the wakes seem to be out of the domain.
 The bottom graph shows the error function $E(t)$ which is
minimized at $t=6$. The solid chords in the diagram indicates the incident solitons, and the dotted
ones show the reflected solitons (i.e. a completion of the chord diagram \cite{KOT:09}).}
\end{figure}

We consider the V-shape initial wave with $A_0=2$ and $\tan\Psi_0=\frac{12}{5}, (\Psi_0\approx 67.3^{\circ})$ which gives $\kappa=1.2$.
Here the critical angle is $\Psi_c=\tan^{-1}(2)\approx 63.4^{\circ}$, and we expect 
 asymptotically an O-type soliton solution. The corresponding $k$-parameters are
obtained from (\ref{NumericalK}), i.e.
\[
(k_1,k_2,k_3,k_4)=\left(-\sf{11}{5},-\sf{1}{5},\sf{1}{5},\sf{11}{5}\right).
\]
Figure \ref{fig:ON} illustrates the result of the numerical
simulation. The top figures show the direct simulation of the KP
equation. The wake behind the interaction point has a large negative
amplitude, and it disperses and decays in the negative $x$-direction.
This shows a separation of the radiations from the exact solution similar to the case of KdV soliton.
The steady pattern left after shedding the radiations can be identified as an O-type solution.
 The middle figures show the corresponding O-type exact
solution whose $A$-matrix is determined by minimizing the error
function $E(t)$ at $t=6$,
\[
A=\begin{pmatrix}
1 & 1.91 & 0 & 0\\
0 & 0 & 1 &  0.17
\end{pmatrix}
\]
Using (\ref{Oshift}), we obtain the shift of the initial line-solitons,
\[
x_{[1,2]}=x_{[3,4]}=-0.020.
\]
(Note here that because of the symmetric profile, the shifts for initial solitons 
are the same.)
The negative shifts imply the slow-down of the incidence waves due
to the generation of the solitons extending the initial solitons in
the negative $x$-direction.  The phase shifts $\Delta
x_{[i,j]}$ for the O-type exact solution are
calculated from (\ref{Opshift}), and they are
\[
\Delta x_{[1,2]}=\Delta x_{[3,4]}=0.593.
\]
The positivity of the phase shifts is due to the attractive force
between the line-solitons, and this explains the slow-down of the
initial solitons, i.e. the small negative shifts of $x_{[i,j]}$.
The bottom graph in Figure \ref{fig:ON} shows $E(t)$ of \eqref{error}, where we take $r=12$ for the domain $D_r^t$. One
can see a rapid convergence of the solution to the O-type exact
solution with those parameters. One should however remark that when $\Psi_0$ is close
to the critical one, i.e. $k_2\approx k_3$, there exists a large
phase shift in the soliton solution, and  the convergence is very
slow. Note that in the limit $k_2=k_3$  the
amplitude of the intermediate soliton generated at the intersection
point reaches {\it four} times larger than the initial solitons. 
This large amplitude wave generation has been considered
as the Mach reflection problem of shallow water wave \cite{M:77,
PTLO:05, CK:09, KOT:09} (see also Section \ref{sec:SWW2}).
The chord diagram in Figure \ref{fig:ON} shows a {\it completion} of the (partial) chord diagram:
The solid chords indicate the initial solitons forming V-shape, and the dotted chords corresponds
to the solitons generated by the interaction (see \cite{KOT:09} for further discussion).

%%%%%%%%%%%%%%%%%%%%%%%%%%%%%%%%%%%%%%%%%%%%
%%%%%%%%%%%%%%%%%%%%%%%%%%%%%%%%%%%%%%%%%%%%%%%%

\subsection{The Mach reflection: $\kappa<1$}\label{sub:M}
We consider the initial V-shape wave with $A_0=2$ and $\Psi_0=45^{\circ}$
(i.e. $\kappa=0.5$).
The angle $\Psi_0$ is now less than the critical angle $\Psi_c\approx 63.4^{\circ}$.
The asymptotic solution is expected to be of $(3142)$-type whose $k$-parameters are
obtained from \eqref{NumericalK}, i.e.
\[
(k_1,k_2,k_3,k_4)=\left(-\sf{3}{2},-\sf{1}{2},\sf{1}{2},\sf{3}{2}\right).
\]

%%%%%%%%%%%%%%%%%%%%%%%%%%%%%%%%%%%%%%%%%%%%%%%%%%%%%%%%%%%%%%%%%%%%%
\begin{figure}[t!]
\begin{centering}
\includegraphics[scale=0.63]{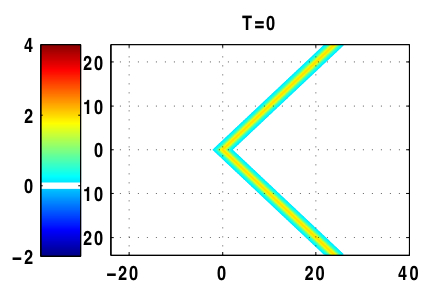}\includegraphics[scale=0.6]{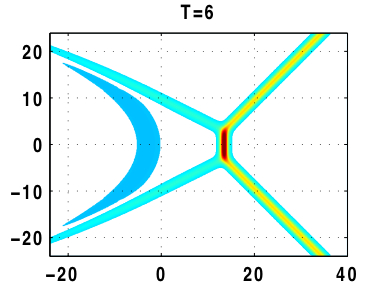}\includegraphics[scale=0.6]{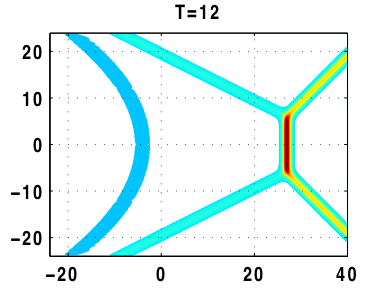}
\par\end{centering}
\begin{centering}
\includegraphics[scale=0.63]{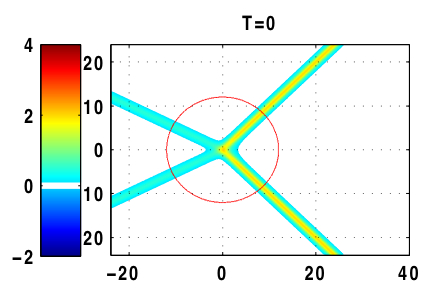}\includegraphics[scale=0.6]{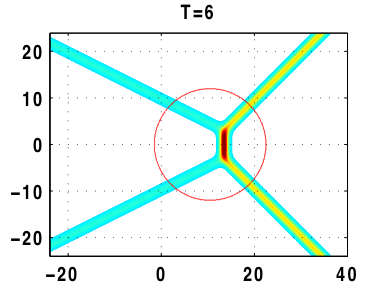}\includegraphics[scale=0.6]{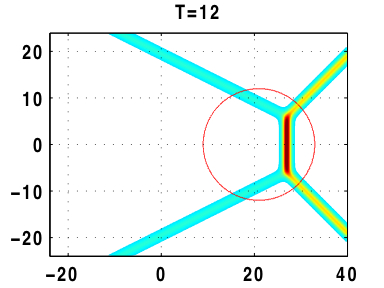}
\par\end{centering}
\begin{centering}
\includegraphics[scale=0.7]{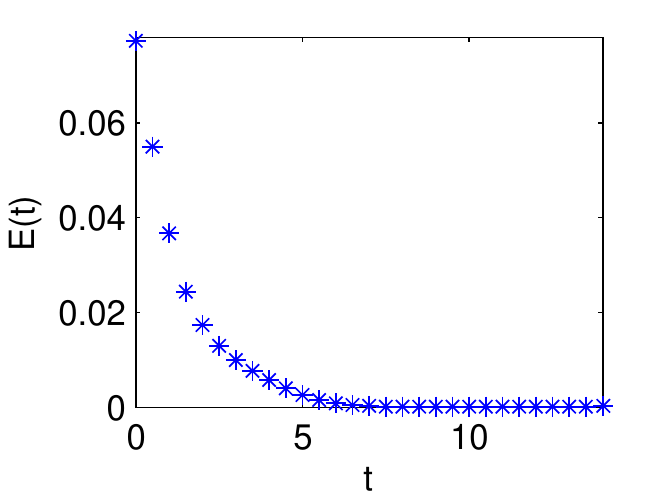} \hskip 1cm \raisebox{0.3in}{\includegraphics[scale=0.4]{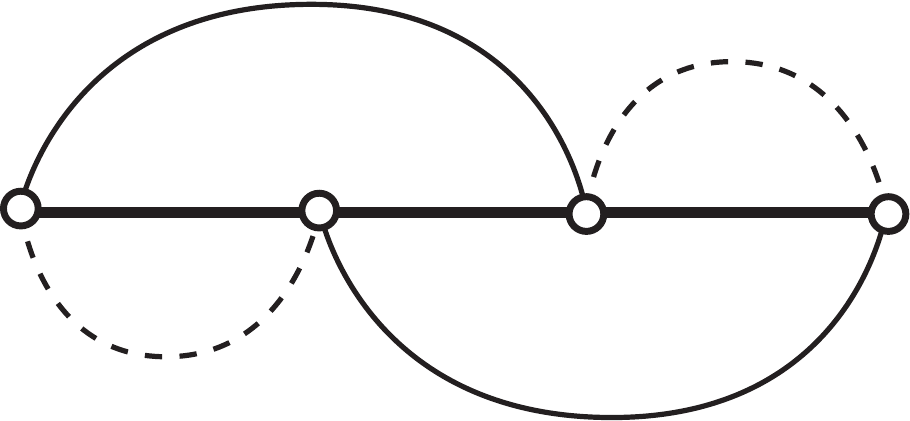}}
\par\end{centering}\caption{\label{fig:3142N}  Numerical simulation of V-shape initial wave for $\kappa<1$ (Mach reflection):   The initial wave consists of  $[1,3]$-soliton in $y>0$ and $[2,4]$-soliton in $y<0$ with $A_0=2$ and $\Psi_0=45^{\circ}$. The top figures show the result of the direct simulation, and the middle figures show the corresponding exact solution of $({{3142}})$-type.
Notice that a large amplitude intermediate soliton is generated at the intersection point, and it corresponds
to the $[1,4]$-soliton with the amplitude $A_{[1,4]}=4.5$.  The circles in the middle figures 
are $D^t_r$ with $r=12$, which cover well the main part of the interaction regions up to $t=12$.
The bottom graph of the error function $E(t)$ which is minimized at $t=10$.  The solid chords
in the diagram indicate the incident solitons, and the dotted ones show the reflected solitons.
}
\end{figure}
%%%%%%%%%%%%%%%%%%%%%%%%%%%%%%%%%%%%%%%%%

Figure \ref{fig:3142N} illustrates the result of the numerical
simulation. The top figures show the direct simulation of the KP
equation. We again observe a bow-shape wake behind the interaction point.
The wake expands and decays, and then we see the appearance of new
solitons which form resonant interactions with the initial solitons.
One should note that the solution generates a large amplitude
intermediate soliton at the interaction point, and this soliton is
identified as $[1,4]$-soliton with the amplitude $A_{[1,4]}=4.5.$
This $[1,4]$-soliton is called the Mach stem in the Mach reflection \cite{CK:09, KOT:09} (see also Section \ref{sec:SWW2}).

The middle figures in Figure \ref{fig:3142N} show the corresponding
exact solution of $(3142)$-type whose $A$-matrix is found by
minimizing $E(t)$ at $t=10$,
\[
A=\begin{pmatrix}
1 & 1.92  &0&  -1.96\\
0& 0 &1 & 0.64
\end{pmatrix}
\]
Using
(\ref{shiftab}), we obtain the phase shifts $x_{[i,j]}$ for the initial solitons of $[1,3]$- and $[2,4]$-type in $x>0$, and the $s$-parameter,
\[
 x_{[1,3]}=x_{[2,4]}=-0.01,\qquad s=0.980.
 \]
Those values indicate that the solution is very close to the exact solution for all the time.
The negative value of the shifts $x_{[i,j]}$ is due to the generation of  a large amplitude
soliton $[1,4]$-type (i.e. the initial solitons slow down), and $s<1$ implies that the $[1,4]$-soliton is  generated after $t=0$.
Also note that the
$[1,4]$-soliton now resonantly interact with $[1,3]$- and
$[2,4]$-solitons to create new solitons $[1,2]$- and
$[3,4]$-solitons (called the reflected waves in the
Mach reflection problem \cite{M:77, CK:09, KOT:09}).  This process
then seems to compensate the shifts of incident waves, even though
we observe a large wake behind the interaction point. 

The bottom graph in Figure \ref{fig:3142N} shows a rapid convergence
of the initial wave to $(3142)$-type soliton solution with those
parameters of the $A$-matrix and $k$ values given above, and for the
error function $E(t)$,  we minimize it at $t=10$ for $D^t_r$ with $r=12$.

 %%%%%%%%%%%%%%%%%%%%%%%%%%%%%%%%%%%%%%%%%%%%%%
\begin{figure}[t!]
\begin{centering}
\includegraphics[scale=0.63]{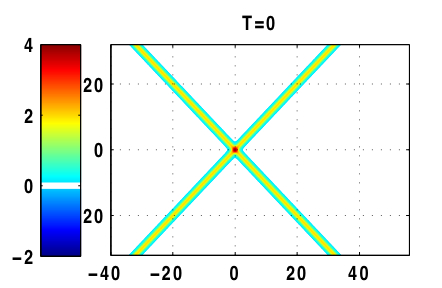}\includegraphics[scale=0.6]{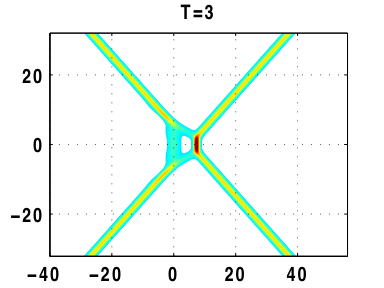}\includegraphics[scale=0.6]{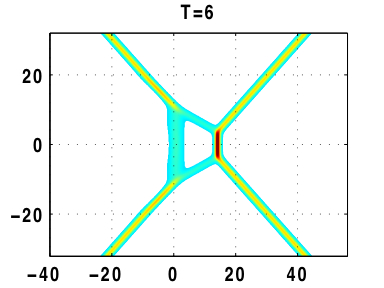}
\par\end{centering}
\begin{centering}
\includegraphics[scale=0.63]{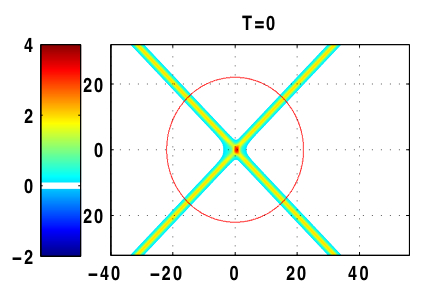}\includegraphics[scale=0.6]{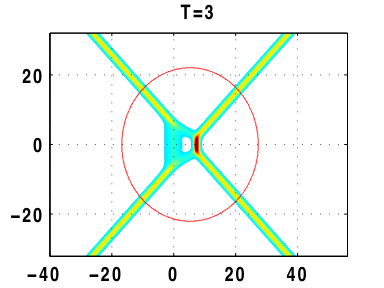}\includegraphics[scale=0.6]{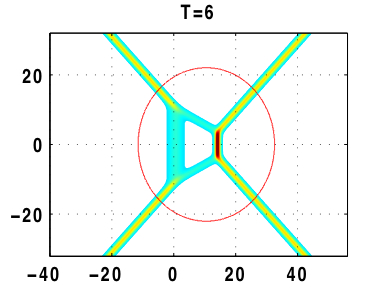}
\par\end{centering}
\begin{centering}
\includegraphics[scale=0.7]{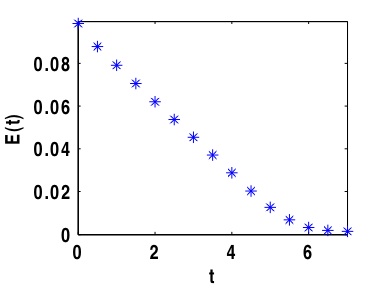} \hskip 1cm \raisebox{0.3in}{\includegraphics[scale=0.4]{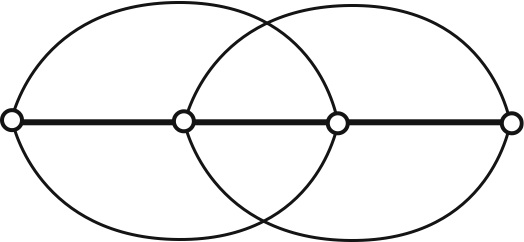}}
\par\end{centering}\caption{\label{fig:X-h} Numerical simulation for X-shape initial wave with
$A_0=2$ and $\Psi_0=45^{\circ}$ (i.e. $\kappa=0.5<1$):
The initial wave is the sum of $[1,3]$- and $[2,4]$-solitons.  The top figures show the numerical
simulation, and the middle figures show the corresponding exact solution of $(3412)$-type, i.e. T-type.
Notice that the circle showing $D^t_r$ with $r=22$ covers well the box generated by the resonant
interaction up to $t=7$.
The bottom graph shows the error function $E(t)$ of \eqref{error} which is minimized at $t=6$.
The four solid chords in the diagram show the asymptotic solitons in the incident wave, and they form a T-type soliton solution.}
\end{figure}
%%%%%%%%%%%%%%%%%%%%%%%%%%%%%%%%%%%%%%%%%%%%%%%
\subsection{T-type interaction with X-shape initial wave}
In this example, we consider an X-shape initial wave given by
the sum of two line-solitons.  For simplicity, we consider a symmetric initial wave
with $A_0=2$ and $ \Psi_0=45^{\circ}$ (i.e. extend the initial wave in  Figure \ref{fig:3142N}
into the negative $x$-region).  Since $\kappa=0.5<1$, the corresponding chord diagram
shows the T-type with
 the $k$-parameters
$(k_1,k_2,k_3,k_4)=(-\frac{3}{2}, -\frac{1}{2},\frac{1}{2},\frac{3}{2})$.
Although T-type soliton appears for smaller angle $\Psi_0$, one should not take so small value.
For an example of the symmetric case, if we take $\Psi_0=0$ giving twice higher
amplitude than one-soliton case, we obtain KdV 2-soliton solution
with different amplitudes (as can be shown by the method of IST).
So for the case with very small angle $\Psi_0$, we expect to see those KdV solitons near the intersection point.
However, the solitons expected from the chord diagram have almost the same amplitude
as the incidence solitons for the case with a small angle.
The detailed study also shows that near the intersection point for T-type solution at the time when
all four solitons meet at this point (i.e X-shape), the solution at the intersection point 
has a small amplitude due to the
repulsive force similar to the KdV solitons. Then the initial X-shape wave with small angle
generates a large soliton at the intersection point. This then implies that our initial wave given by
the sum of two line-solitons creates a large
dispersive perturbation at the intersection point, and one may need to wait a long time to see the convergence.

The top figures in Figure \ref{fig:X-h} illustrate the numerical
simulation, which clearly shows an opening of a resonant box as
expected by the feature of T-type. The corresponding exact
solution is illustrated in the middle figures, where the $A$-matrix
of the solution is obtained by minimizing the error function $E(t)$
of \eqref{error} at $t=6$,
 \[
 A=\begin{pmatrix}
 1 & 0 &  -0.368  & -0.330 \\
 0 & 1 & 1.198  & 0.123
 \end{pmatrix}
 \]
In the minimization process, we take $x_{[1,3]}=x_{[2,4]}$ due to the symmetric profile of the solution,
and adjust the on-set of the box (see subsection \ref{Tsoliton}).  Note that the symmetry reduces the number of
free parameters to three.  We obtain
\[
x_{[1,3]}= x_{[2,4]}=0.025, \qquad r=3.63, \qquad s=0.350.
\]
The positive shifts of those $[1,3]$- and $[2,4]$-solitons in the wavefront indicate also the positive shift
of the newly generated soliton of $[1,4]$-type at the front. This is due to the repulsive force which exists in
the KdV type interaction as explained above, that is,
the interaction part in the initial wave has a larger amplitude than that of the exact solution, so that
this part of the solution moves faster than that in the exact solution. This difference may result in a shift
of the location
of the $[1,4]$-soliton. The relatively large value $r>1$ indicates that the onset of the box is actually much earlier than $t=0$,
and $s<1$ shows the positive phase shifts as calculated from \eqref{bcD}.

The bottom graph in Figure \ref{fig:X-h} shows the evolution of the
error function $E(t)$ of \eqref{error} which is minimized at $t=6$.
Note here that the circular domain $D_r^t$ with $r=22$ covers well
the main feature of the interaction patterns for all the time
computed for $t\le 7$.  The chord diagram in the figure shows four asymptotic solitons in the initial
wave which form a T-type soliton solution (see \cite{KK:10} for further discussion).

\section{Shallow water waves: The Mach reflection}\label{sec:SWW2}
\begin{figure}[t]
\centering
\includegraphics[scale=0.45]{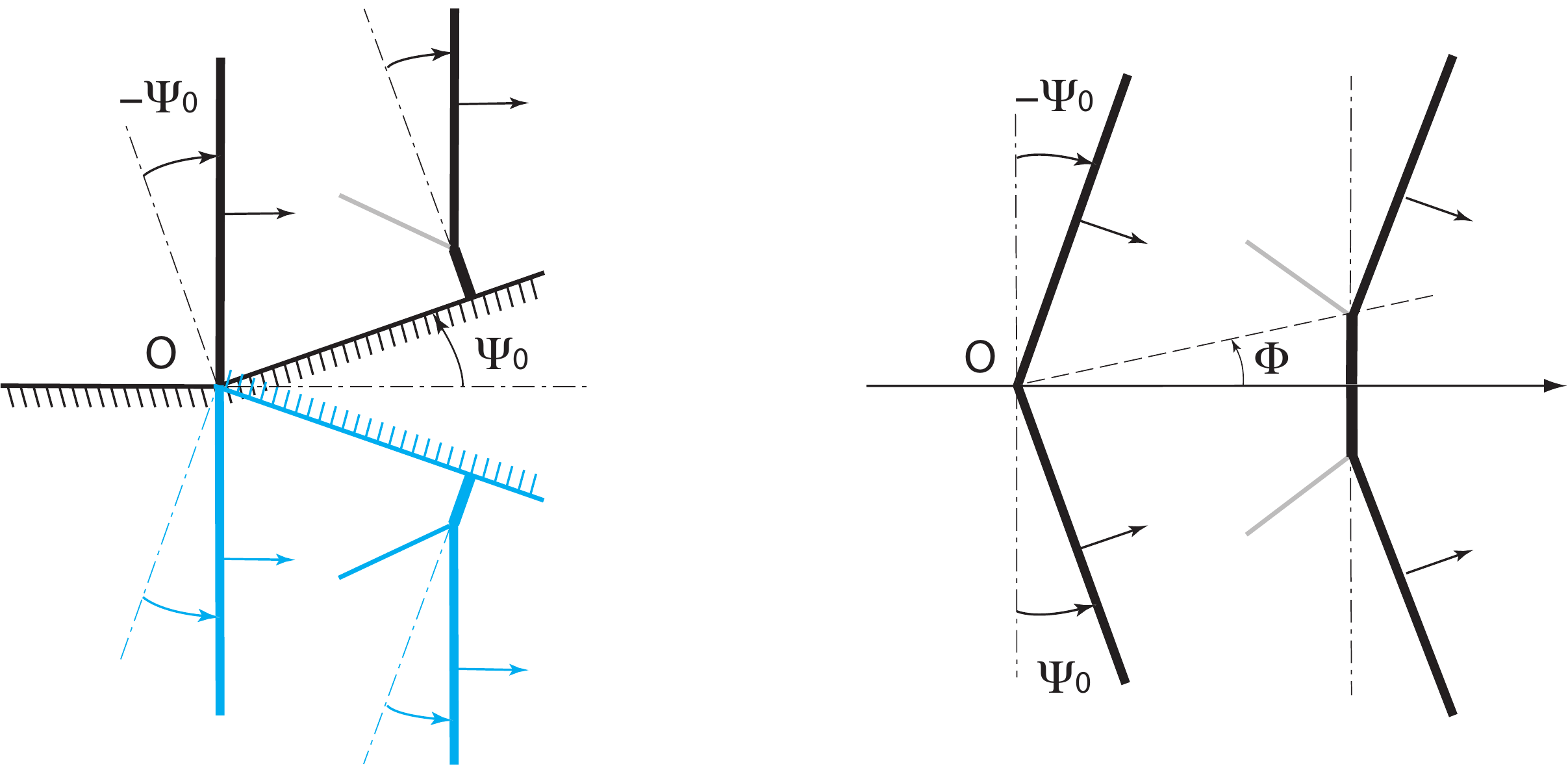} 
\caption{The Mach reflection. The left panel illustrates a semi-infinite line-soliton (incidence wave) propagating
parallel to the wall with the mirror image. The right panel is an equivalent system to the left one when we ignore
the viscous effect on the wall. The incident wave then forms a V-shape wave at $t=0$ as
discussed in Section \ref{sec:NS}.  The resulting wave pattern shown here  is a $(3142)$-soliton solution.
\label{fig:MR}}
\end{figure}

In this last section, we discuss a real application of the exact soliton solutions of the KP equation
described in the previous sections to the Mach reflection phenomena in shallow water.
In \cite{M:77}, J. Miles considered an oblique interaction of two line-solitons  using O-type solutions.
He observed that resonance occurs at the critical angle $\Psi_c$, and when the initial oblique angle $\Psi_0$
is smaller than $\Psi_c$, the O-type solution becomes
singular (recall  that at the critical angle $\Psi_c$, one of the exponential term in the $\tau$-function vanishes, see subsection \ref{sub:O}). 
He also noticed a similarity between this resonant interaction and the Mach reflection
found in shock wave interaction (see for example {CF:48,Wh:64}). This is illustrated by the left figure of Figure \ref{fig:MR},
where an incidence wave shown by the vertical line is propagating to the right, and it hits a rigid wall
with the angle $-\Psi_0$ measured counterclockwise from the axis perpendicular to the wall
(see also \cite{F:80}).
If the angle $\Psi_0$ (equivalently the inclination angle of  the wall) is large, 
the reflected wave behind the incidence
wave has the same angle $\Psi_0$, i.e. a regular reflection occurs. However, if the angle is small,
then an intermediate wave called the Mach stem appears as illustrated in Figure \ref{fig:MR}.
The Mach stem, the incident wave and the reflected wave interact resonantly, and
those three waves form
a resonant triplet. The right panel in Figure \ref{fig:MR} illustrates the wave propagation which is  equivalent to that in the left panel, if one ignores the effect of viscosity on the wall (i.e. no boundary layer). At the point $O$, the initial wave has V-shape
with the angle $\Psi_0$, which forms the initial data for our simulation discussed in the previous section.
Then as we presented, the numerical simulation describes the reflection 
of line-soliton with an inclined wall, and these results explain well the Mach
reflection phenomena in terms of the exact soliton solutions of the KP equation.
%%%%%%%%%%%%%%%%%%%%%%%%%%%%%%%%%%%%%%%%%%
%%%%%%%%%%%%%%%%%%%%%%%%%%%%%%%%%%%%%%%%

\subsection{Previous numerical results of the Boussinesq-type equations}
One of the most interesting things of the Mach reflection is that the KP theory predicts 
an extraordinary four-fold amplification of the stem wave at the critical angle \cite{M:77}.
We recall the formulae of the maximum amplitudes which are
given by (\ref{Omax}) for the O-type solution ($\kappa>1$) and (\ref{stemA}) for $(3142)$-type solution ($\kappa<1$).  Let $\alpha$ denote the amplification
factor in terms of the Miles parameter $\kappa$ of \eqref{MilesK}, i.e.
\begin{equation}\label{alpha}
\alpha=\left\{\begin{array}{lll}
\displaystyle{(1+\kappa)^2},\quad &{\rm for}\quad \kappa<1, \\[1.0ex]
\displaystyle{\frac{4}{1+\sqrt{1-\kappa^{-2}}}},\quad &{\rm for}\quad \kappa>1.
\end{array}
\right.
\end{equation}
Several laboratory and numerical experiments tried to confirm the formula \eqref{alpha}, in particular,
 the four-fold amplification at the critical value $\kappa=1$ (see for example \cite{P:57, Me:80, F:80, T:93, YCL:10}).  In \cite{F:80}, Funakoshi made a numerical simulation
of the Mach reflection problem using the system of
equations,
\[\left\{\begin{array}{lll}
\displaystyle{\eta_t+\Delta\psi+\alpha\nabla\cdot(\eta\nabla\psi)-\frac{\beta}{6}\Delta^2\psi=0},\\[1.5ex]
\displaystyle{\left(\psi-\frac{\beta}{2}\Delta\psi\right)_t+\eta+\frac{\alpha}{2}|\nabla\psi|^2=0,}
\end{array}\right.
\]
which is equivalent to the Boussinesq-type equation \eqref{Boussinesq} up to this order.
He considered the initial wave to be the KdV soliton with  higher order corrections up to $\mathcal{O}(\epsilon)$.
In his paper, he mainly presented the results for the incidence waves with the amplitude
$a_i=0.05=\hat{a}_0/h_0$ and the angles $\frac{\pi}{40}\le \Psi_0\le \frac{\pi}{3}$.
He concluded that his results agree very well with the resonantly interacting solitary
wave solution predicted by Miles. However his results on the amplification parameter
$\alpha$ are slightly shifted to the lower values of the Miles $\kappa$-parameter.
Tanaka in \cite{T:93} then re-examined Funakoshi's results for higher amplitude incidence waves
with $a_i=0.3$ using the high-order spectral method.
He noted that  the effect of large amplitude tends to prevent the Mach reflection to occur, and
all the parameters such as the critical angle $\Psi_c$ are shifted toward the values corresponding
to the regular reflection (i.e. O-type). For example, he obtained the maximum amplification 
$\alpha=2.897$ at $\kappa=0.695$.

\begin{figure}[t!]
\centering
\includegraphics[scale=0.43]{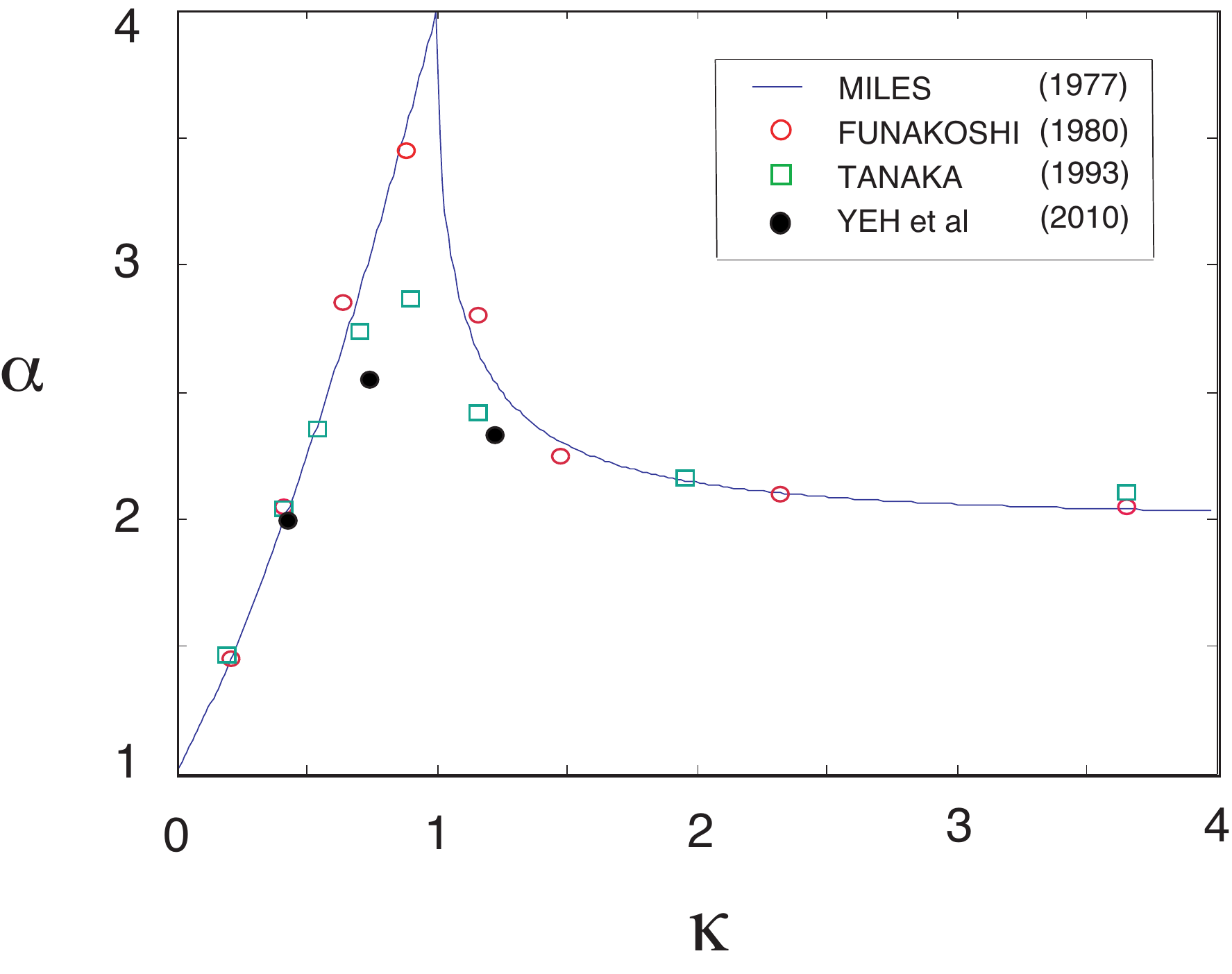}
\caption{Numerical results of the amplification factor $\alpha$ versus the $\kappa$-parameter.
The circles show Funakoshi's result \cite{F:80}, the squares show Tanaka's result \cite{T:93}.
The black dots shows the experimental results by Yeh et al \cite{YCL:10}.\label{fig:FTY}}
\end{figure}

However, we claim in our recent paper \cite{YLK:10} that
 those previous results did not properly interpret  their comparisons with the theory, and 
 in fact  their results are in good agreement with the the predictions given by the KP theory
 except for the cases near $\kappa=1$.  One should emphasize that the KP equation is derived under the assumptions of quasi-two dimensionality and weak nonlinearity.  Thus 
 the key ingredient  is to include higher order corrections to those assumptions when we compare
the numerical  or experimental results with the theory.
 In particular, the quasi-two dimensionality can be corrected by comparing the KP soliton
 with the KdV soliton in the propagation direction as mentioned in Section \ref{sec:KP}.
 More precisely, we have the amplitude correction \eqref{anglecorrection}, i.e.
 \[
 \hat{a}_0=\frac{a_0}{\cos^2\Psi_0}=\frac{2h_0A_0}{3\cos^2\Psi_0},
 \]
 where $\hat{a}_0$ is the amplitude observed in the numerical computation of the Boussinesq-type
 equation (which has rotational symmetry in $\mathbb{R}^2$).   This then suggests that 
 the $\kappa$-parameter should be evaluated by the following formula using the experimental
 amplitude $\hat{a}_0$,
 \begin{equation}\label{ampcorrection}
 \kappa:=\frac{\tan\Psi_0}{\sqrt{2A_0}}=\frac{\tan\Psi_0}{\sqrt{3(\hat{a}_0/h_0)}\,\cos\Psi_0}.
 \end{equation}
  Because of the quasi-two dimensional approximation, i.e. $|\Psi_0|\ll 1$, Miles in his paper
  \cite{M:77} replaced $\tan\Psi_0$ by $\Psi_0$, and then in \cite{F:80,T:93}, the authors continued on to use this replacement. Then their computations
  with rather large values of $\Psi_0$ gave significant shifts of the $\kappa$-parameter.
We then re-evaluate their results with our formula \eqref{ampcorrection}, and the
new results are shown in Figure \ref{fig:FTY}.  Since Funakoshi's simulations are based on small amplitude incidence waves,
his results agree quite well with the KP predictions.  Tanaka's results are also in good agreement
with the KP theory except for the cases near the critical angle (i.e. $\kappa=1$), where the amplification
parameter $\alpha$ gets close to 3. This region clearly violates the assumption of the weak nonlinearity.   Although one needs to
make higher order corrections to weak nonlinearity, the original plots of Tanaka's
are significantly improved with the formula \eqref{ampcorrection}.
%%%%%%%%%%%%%%%%%%%%%%%%%%%%%%%%%%%%
%%%%%%%%%%%%%%%%%%%%%%%%%%%%%%%%%%%%%%%%%%%%
 The black dots in Figure \ref{fig:FTY} indicate the results of recent laboratory experiments done by  Yeh
 and his colleagues \cite{YCL:10}.
 We will discuss their experimental results in the next section.

 Before closing this section, we remark on the length of the Mach stem 
 (i.e. the intermediate soliton of $[1,4]$-type).  From $(3142)$-soliton solution, one can find the point 
 $(x_*,y_*)$ of the interaction of the triplet with $[1,3]$-, $[3,4]$- and $[1,4]$-solitons \cite{CK:09} ,
\[\left\{\begin{array}{lll}
\displaystyle{x_*=\frac{1}{4}\left(\tan\Psi_c+\tan\Psi_0\right)^2 t=\frac{A_0}{2}(1+\kappa)^2t,}  \\[1.5ex]
\displaystyle{ y_*=\frac{1}{2}\left(\tan\Psi_c-\tan\Psi_0\right)t=\sqrt{\frac{A_0}{2}}(1-\kappa)t,}
\end{array}\right.
\]
(see Figure \ref{fig:MR}).
In the physical coordinates with $(\tilde{x}_*,\tilde{y}_*, \tilde{t})$, we have
\[
\tilde{x}_*=c_0\left(1+\frac{\hat{a}_0\cos^2\Psi_0}{h_0}(1+\kappa)^2\right)\tilde{t},\qquad
\tilde{y}_*=c_0\sqrt{\frac{\hat{a}_0\cos^2\Psi_0}{3h_0}}\,(1-\kappa)\,\tilde{t}.
\]
The angle $\Phi$ in Figure \ref{fig:MR} is then given by  $\tan\Phi=\tilde{y}_*/\tilde{x}_*$,
which is approximated in \cite{M:77, F:80} by
\[
\tan\Phi\approx \sqrt{\frac{\hat{a}_0}{3}}(1-\kappa).
\]
Using the corrected formula $\tan\Phi=\tilde{y}_*/\tilde{x}_*$, one can see again
a good agreement with the KP theory (see Figure 8 in \cite{F:80}).

\subsection{Experiments}
Recently, Yeh and his colleagues \cite{YCL:10} performed several laboratory experiments on the Mach reflection phenomena using 7.3 m long and 3.6 m wide  wave tank with a water depth of 6.0 cm.  Here we briefly describe their results
and show that our KP theory can predict very well the evolution of the waves observed in the experiments.

The wave tank is equipped with 16 axis directional-wave maker system along the 3.6 m long side wall
marked by $x=0$. 
An oblique incident solitary wave is created by driving those 16 paddles synchronously along
the sidewall, and the wave maker is designed to generate a KdV soliton with any heights
before the breaking.  The temporal and spatial variations of water-surface profiles are 
measured by the Laser Induced Fluorescent (LIF) method (a highly  accurate measurement technique).
The water dyed with fluorescein (green) fluoresces when excited by the laser sheet.
The illuminated image of the water-profiles are recorded by a high-speed and high-resolution video camera.

\subsubsection{The Mach reflection}
%%%%%%%%%%%%%%%%%%%%%%%%%%%%%%%
\begin{table}[t]
\centering
\caption{Amplification factor $\alpha$ for different  values of  
$\kappa=\tan\Psi_0/\sqrt{2A_0}$:
$\alpha_{\tilde x=71.1}$(Exp.) are the laboratory data at $x=71.1$ ($\tilde x=4.27$ m),
$\alpha_{x=71.1}({\rm KP})$ are calculated from the corresponding KP exact solutions at $t=41.05$, and
$\alpha_{x=\infty}$(KP) are
from \eqref{Omax} and \eqref{stemA}.  In the row of $A_0=0.413$ with $\Psi_0=30^{\circ}$, the values of $\alpha$ in the brackets are obtained at
 $x = 50.8$, because of the wave breaking immediately after this point; hence, the greater amplification cannot be realized \cite{YLK:10}. \label{Table1}}
\begin{tabular}{ccccccc}
\hline
\multicolumn{1}{c}{$\kappa$}&\multicolumn{1}{c}{$A_0$} &\multicolumn{1}{c}{$\Psi_0$}& \multicolumn{1}{c}{$\alpha_{x=71.1} ({\rm Exp.})$} 
& \multicolumn{1}{c}{$\alpha_{x=71.1}({\rm KP})$} & \multicolumn{1}{c}{$\alpha_{x = \infty}({\rm KP})$}\\
\hline
1.392 &0.086 & $30^{\circ}$ & 2.10 & 2.36&   2.36 \\
1.242 &0.108 & $30^{\circ}$ & 2.13 &  2.51 &          2.51\\
1.017 &0.161 & $30^{\circ}$& 2.24 &    3.38 &      {3.38} \\
\hline
0.887 &0.212 & $30^{\circ}$& 2.33 &    2.43 &     {3.56} \\
0.731 &0.312 & $30^{\circ}$& 2.52 &    2.61 &        2.99 \\
0.722 & 0.127 & $20^{\circ}$ & 1.89& 1.84 & 2.96 \\
0.635 &0.413 & $30^{\circ}$&   (2.48)     &     (2.54)       &   2.67 & \\
0.591 & 0.189 & $20^{\circ}$ & 1.95  &  1.93 &  2.53  \\
0.516 & 0.249 & $20^{\circ}$ & 1.99 & 2.08 & 2.30  \\
0.425 & 0.367 & $20^{\circ}$ & 2.01 & 1.99 & 2.03\\
\hline
\end{tabular}
\end{table}
%%%%%%%%%%%%%%%%%%%%%%%%%%%%%%%%%%%%%%%
 In Table \ref{Table1}, their experimental results of the amplification factors $\alpha$ are compared with those obtained from the exact solutions of the KP equation
(i.e. O-type for $\kappa>1$ and $(3142)$-type for $\kappa<1$).  
The waves were measured at $\tilde x=4.27$ m ($x=\tilde x/h_0=71.1$) which is the farthest measuring location in the experiments, except for the case with $A_0=0.413$.
In the later case, the $\alpha$ values in the brackets are measured at
$x=50.8$ because of the wave-breaking (notice that at this point $\alpha=2.48$ implies $a_i= 1.02$). We calculate the corresponding KP exact solution at $t=41.05$ (recall here that the relations \eqref{realvariables} gives $\tilde x-c_0\tilde t=h_0 x$ and $c_0\tilde t=\frac{3h_0}{2}t$).
The amplification factor $\alpha$ is still growing along the propagation direction, and the values obtained from the exact
solutions are in good agreement with the measurements.  
We note here that near the critical case (i.e. $\kappa=1$) the growth of the stem amplitude is
very slow and at $x=71.1$ ($\tilde x=4.27$ m) the amplification factor is only achieved about $65\%$.
Also for the cases with small oblique angle, i.e. $\Psi_0=20^{\circ}$ in Table \ref{Table1},  the amplification factor $\alpha$ grows slower
when the incidence wave amplitude is smaller, that is, at $x=71.1$, $\alpha$
is almost constant (slightly decreases) as $\kappa$ increases.  However the asymptotic value
of  $\alpha$ for large $x$ increases as $\kappa$ increases.  This means that the observed waves are
still in the transient stage, and a longer tank is necessary to observe further growth
of the amplification factor (see \cite{YLK:10} for the details).
%%%%%%%%%%%%%%%%%%%%%%%%%%%%%%%%%%%%%%
\begin{figure}[t]
\centering
\includegraphics[width=2in,height=2in]{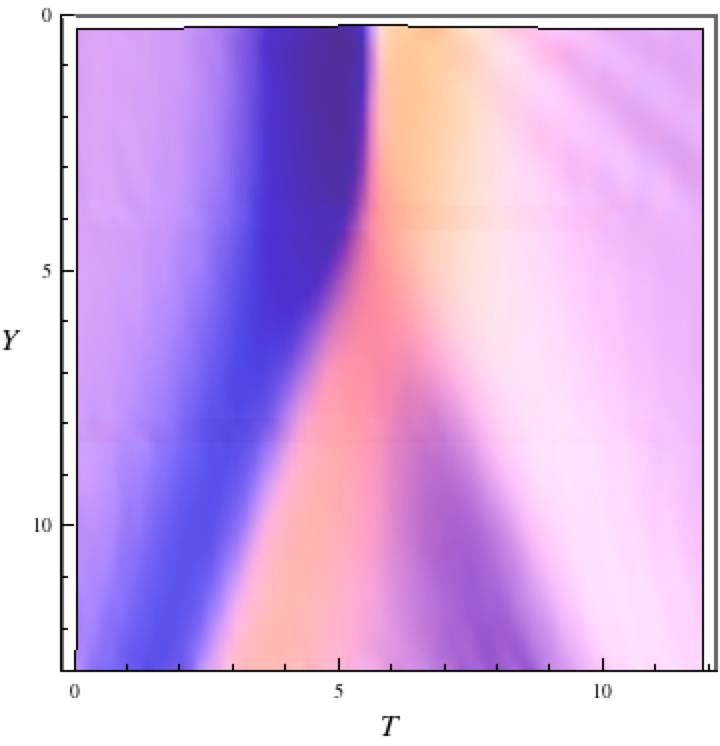} \hskip 1.5cm 
\includegraphics[width=2in,height=1.8in]{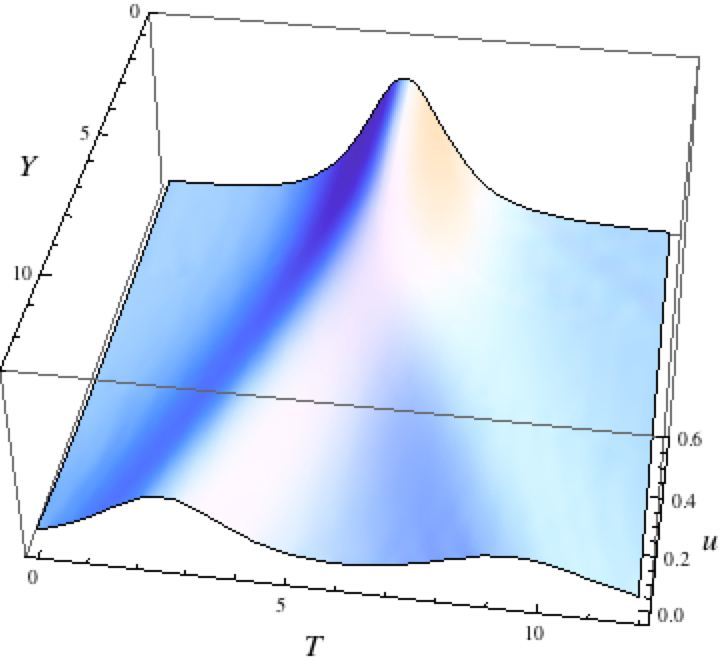}
\label{fig:Fig7a}\\[2.0ex]
\includegraphics[width=2in,height=2in]{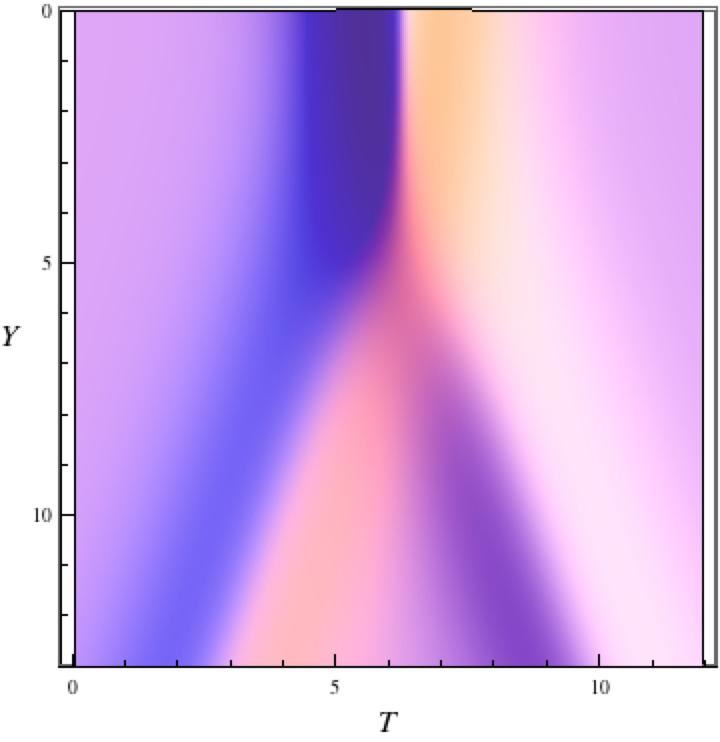} 
\hskip 1.5cm \includegraphics[width=2in,height=1.8in]{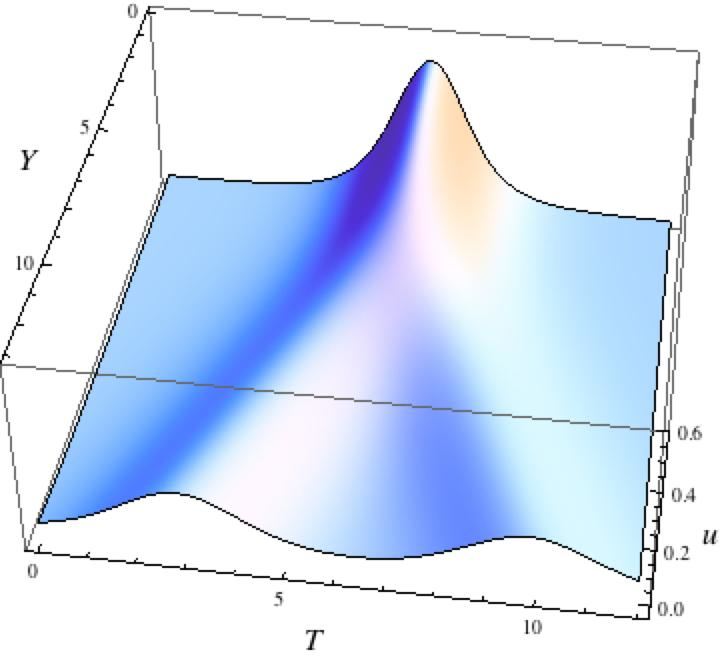} 
\label{fig:Fig7b}
\caption{Two views of the temporal variation of the water-surface profile in the $y$-direction (perpendicular to the wall) at $x= 71.1$ \cite{YLK:10}: The top panels show the experimental result, and the bottom ones show the
corresponding $(3142)$-type exact soliton solution of the KP equation. The incident wave amplitude $A_0 = 0.212$, and the angle $\Psi_0 = 30^{\circ}$.
The amplification factors obtained from the experiment and the exact solutions are
close, and they are $\alpha_{x=71.1}({\rm Exp.})=2.33$ and $\alpha_{x=71.1}(\rm KP)=2.43$
(see Table \ref{Table1}).\label{fig:Fig7}}
\end{figure}
%%%%%%%%%%%%%%%%%%%%%%%%%%%%%%%%%%%%%%%%%
In Figure \ref{fig:Fig7}, we show the image of the wave-profile at $x=71.1$
for the case when the incident wave amplitude have $A_0=0.212$ and $\Psi_0=30^{\circ}$.
The corresponding exact solution with those parameters is of (3142)-type (i.e. $\kappa=0.887<1$).
The upper panels show two views in different angles  of the temporal variation of the wave profile of the experiment at $x=71.1$, which is made from 250 slices of the spatial profiles (100 slices per second) with approximately 3000 pixel resolution n the $y$-direction. As expected form the (3142)-type exact solution,
the stem-wave formation is realized, in which the incident and reflected waves separate
away from the wall by the stem-wave.
The lower panels  show the corresponding $(3142)$-type exact solution at $t=41.05$ (the $x$-coordinate
is converted to the $t$-coordinate, using \eqref{realvariables}).
  Here  the $k$-parameters are $(k_1,k_2,k_3,k_4)=(-0.614,-0.037,0.037,0.614)$ from $A_0=0.212$ and $\Psi_0=30^{\circ}$.  Then 
 we calculate the $A$-matrix using \eqref{shiftab} with $s=1$, and take
 \[
 A=\begin{pmatrix}
 1 & 8.797 & 0 & -1.128 \\
 0 & 0 & 1 & 0.530
 \end{pmatrix}.
 \]
This choice of the $A$-matrix places the incidence wave crossing at the origin at $t=0$, i.e.
$\theta^+_{[1,3]}=\theta^+_{[2,4]}=0$ in \eqref{shiftab}.
We see a good agreement between the experiment and the KP theory. At $x=71.1$, the wave-profile 
observed in the experiment is close to the corresponding exact solution of $(3142)$-type, that is,
the radiations generated at the beginning stage dispersed and well separated from the main
part of the wave-profile as predicted in the numerical simulation (see subsection \ref{sub:M}).

\subsubsection{T-type interaction}
%%%%%%%%%%%%%%%%%%%%%%%%%%%%%%%%%%%%%
\begin{figure}[t!]
\centering
\includegraphics[scale=0.55]{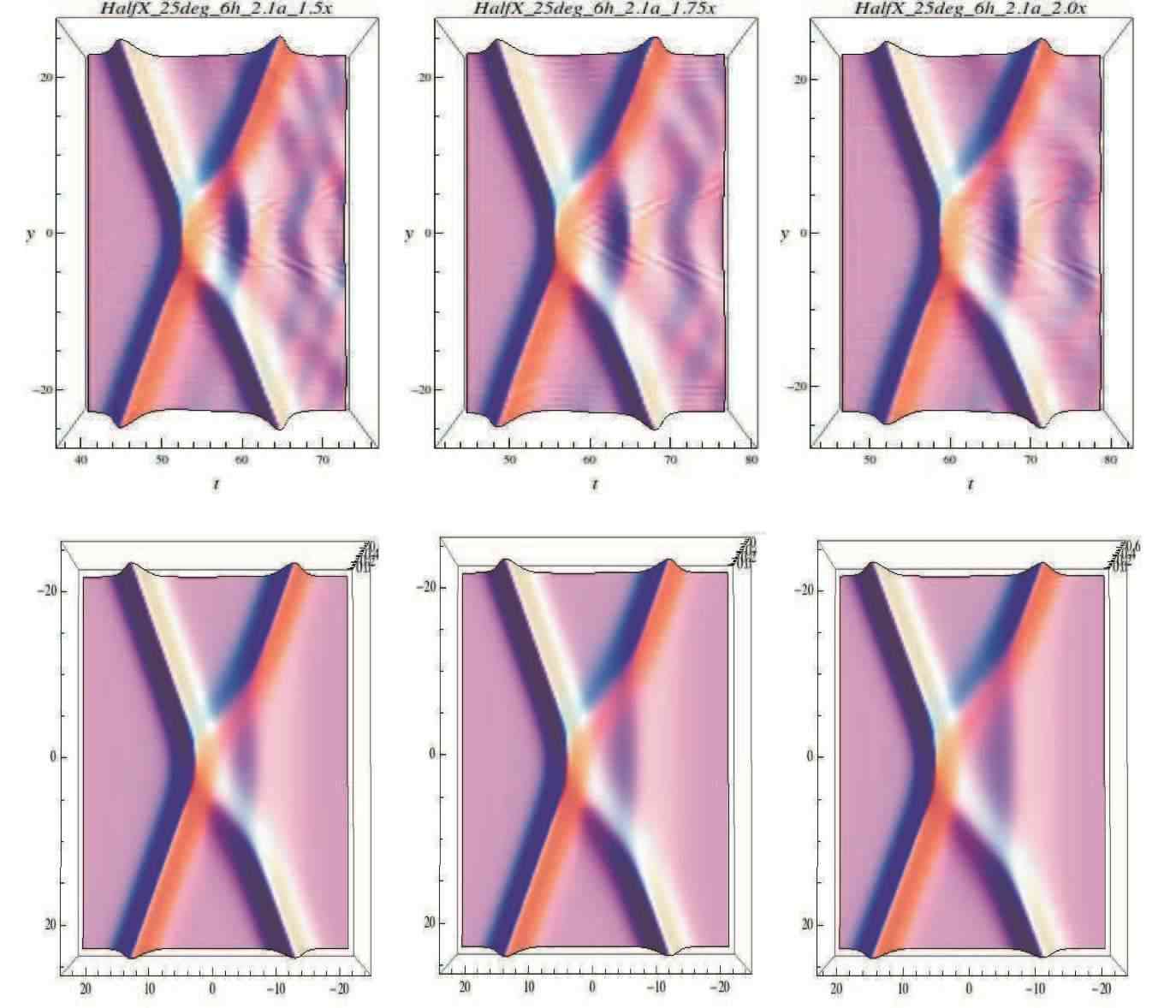}
\caption{T-type solutions generated in the water tank \cite{YCL:10}.
The experimental results are shown in the upper panels.  Those figures are made by combining
the real image of the wave profiles with their mirror images of the wall at $y=0$ (the center horizontal line).  The lower panels show the corresponding T-type exact solution of the KP equation. The initial wave has $a_i=2.1$ cm ($A_0=0.43$) and $\Psi_0=25^{\circ}$ (i.e. $\kappa=0.503$).  The exact solution is plotted in the $xy$-plane, and the solitons are propagating to the left. (The experimental figures by courtesy of Harry Yeh.) 
\label{fig:Ttype12}}
\end{figure}
%%%%%%%%%%%%%%%%%%%%%%%%%%%%%%%%%%%%%%%%%%%
Figure \ref{fig:Ttype12} shows a preliminary result for T-type interaction pattern generated in 
the same tank by Yeh and his collaborators.  The T-type solution is the most complex
and interesting soliton solution associated with the $\tau$-function on Gr$^+(2,4)$.
The initial wave has the V-shape (half of the X-shape), then 
other half of the X-shape is generated by the line-soliton with opposite angle.
The upper panels of Figure \ref{fig:Ttype12} show the evolution of the wave pattern with $A_0=0.431$ and $\Psi_0=25^{\circ}$.  
Behind the crossing wave form (the right side in the figure), the large wakes are generated at the early stage of the evolution, 
but they eventually separate from the main pattern of the T-type interaction,
as we observed in the numerical simulation.
The figures clearly show the formation of a box pattern as expected by the KP theory.
The lower panels show the corresponding T-type exact soliton solution whose parameters are
$(k_1,k_2,k_3,k_4)=(-0.697,-0.231,0.231,0.697)$ and  the $A$-matrix given by
\eqref{bcD} and \eqref{r} with $\theta^+_{[1,3]}=\theta^+_{[2,4]}=0, s=40$ and $r=1$, i.e.
\[A=
\begin{pmatrix}
1 & 0 & -40.34 & -24.07 \\
0 & 1 & 24.07 & 13.37
\end{pmatrix}
\]
The large value of the $s$-parameter indicates large phase shift of the incident line-solitons 
(see Figure \ref{fig:TS}), that is, the parts of line-solitons in the right side (i.e. behind the interaction point)
were created with some delayed time.
In finding those parameter values, we did not make a precise minimization of certain error function,
like the one given in \eqref{error}. In a future communication, we hope we will be able to develop a method to determine the exact solution from the experimental data, that is, the inverse problem of the KP equation.

\vskip 1cm

\leftline{\bf Acknowledgements}

\vskip 0.5cm
\noindent
I would like to thank Harry Yeh for letting me to use his excellent and important experimental data 
before publication and for making extremely fruitful collaboration on the Mach reflection problem. 
I am grateful to Sarbarish Chakravarty for his many excellent suggestions which made
a tremendous improvement on the paper and for
many valuable discussions throughout our intensive but very pleasant collaboration.
I would also like to thank my colleagues, Chiu-Yen Kao, Masayuki Oikawa and Hidekazu Tsuji for many useful discussions related to the subjects presented in this paper.
Special thanks to Mark J. Ablowitz who provides his excellent photos in Figure \ref{fig:Mark}.
My research is partially supported by NSF grant DMS-0806219.

\medskip

\end{document}